\documentclass[12pt]{article}
\usepackage{geometry}
\usepackage{a4}
\usepackage{graphicx,subfigure}
\usepackage{epsf}
\usepackage{amsmath}
\usepackage{amssymb}
\usepackage{cite}
\usepackage{tensor}
\begin{document}
\begin{titlepage}
\title{On Holographic R\'enyi Entropy in Some Modified Theories of Gravity}
\author{}
\date{
Anshuman Dey, Pratim Roy, Tapobrata Sarkar
\thanks{\noindent E-mail:~ deyanshu, proy, tapo@iitk.ac.in}
\vskip0.4cm
{\sl Department of Physics, \\
Indian Institute of Technology,\\
Kanpur 208016, \\
India}}
\maketitle\abstract{
\noindent
We perform a detailed analysis of holographic entanglement R\'enyi entropy in some modified theories of gravity with four dimensional 
conformal field theory duals. First, we construct perturbative black hole 
solutions in a recently proposed model of Einsteinian cubic gravity in five dimensions, and compute the R\'enyi entropy as well as the scaling dimension of 
the twist operators in the dual field theory. Consistency of these results are verified from the AdS/CFT correspondence, via a corresponding computation of 
the Weyl anomaly on the gravity side. Similar analyses are then carried out for three other examples of modified gravity in five dimensions that include a chemical 
potential, namely Born-Infeld gravity, charged quasi-topological gravity and a class of Weyl corrected gravity theories with a gauge field, with the last example being
treated perturbatively. Some interesting bounds in the dual conformal field theory parameters in quasi-topological gravity are pointed out. We also provide 
arguments on the validity of our perturbative  analysis, whenever applicable.}
\end{titlepage}

\section{Introduction}

Entanglement in quantum mechanical systems is a topic that has intrigued and engaged physicists ever since the discovery of quantum mechanics. 
Two quantum subsystems that make up a composite system are said to be entangled when measurements performed on the subsystems are correlated, 
i.e the states of the individual subsystems are not independent. Broadly speaking, a measure of entanglement for a subsystem can be obtained by tracing 
out the complementary degrees of freedom from the total density matrix. Applications of the idea of entanglement to quantum many body systems and quantum field theories \cite{Cardy}, \cite{Cardy1} are relatively recent, but these have witnessed a lot of activity in the recent past, in the context of condensed matter physics, quantum information theory, etc \cite{Kitaev}, \cite{Levin}, \cite{Hsu}.

In the context of the AdS/CFT correspondence \cite{Maldacena}, \cite{Witten}, the calculation of the entanglement entropy of a strongly coupled
quantum field theory holographically
from a dual gravitational description arose out of the seminal work by Ryu and Takayanagi \cite{RyuT}. The Ryu-Takayangi prescription 
and its generalisations \cite{Rangamani}, \cite{Myers1}, \cite{Lewkowycz}, \cite{Fursaev}, \cite{Dong}, \cite{Mohammadi} 
(see also \cite{Naseh}, \cite{Naseh1} for related discussions) have been at the forefront of research over the last decade and promises to yield a deep 
understanding of entanglement in strongly coupled 
quantum systems. Following up on this line of research, in recent times, there has been a surge of interest in investigating higher curvature theories of gravity, like 
Gauss-Bonnet gravity, Lovelock gravity \cite{Lovelockmain}, and quasi-topological gravity \cite{Ray}, \cite{Robinson}, \cite{Ray1}. 
We will call such theories generically as modified theories of gravity (in the sense of being modified from the standard Einstein gravity). 
It is well known that such theories often suffer from the problem of negative energy states in their spectrum, but nonetheless might prove
important clues towards the quantization of gravity. Viewed from a string theory perspective, these theories might be viewed as higher order corrections to 
the standard Einstein-Hilbert action, the latter being a low energy effective theory, and the corrections becoming important at large energy scales. 
Indeed, CFT duals to these modified theories of gravity have been extensively studied by holographic methods by now 
(see, for example, \cite{Brigante1}, \cite{Brigante}, \cite{Buchel}, \cite{Boer}, \cite{Sinha}). 

Understanding entanglement in modified theories of gravity is important and interesting. Apart from providing insights into the nature of entanglement in
novel examples of strongly coupled quantum field theories, these often provide us with extra tuneable physical parameters that substantially enrich the phase 
structure of the system, and is expected to be of use to model realistic situations. Further, these theories often provide useful insights in terms of novel physical 
bounds on the parameters of the dual CFT. With these motivations, the purpose of this paper is to understand the nature of entanglement
in some examples of modified theories of gravity. In particular, we focus on the analysis of the entanglement R\'enyi entropy (ERE) \cite{Renyi}, \cite{Renyi1} and study its various features. The tools needed for our analysis is well established, by now. Hence, in order to set the notations and conventions to be used in the rest of the paper, we briefly recapitulate these in section 2. 

The first case that we study in this paper in section 3 is ERE in the recently proposed Einsteinian cubic gravity \cite{Bueno}. To this end, we first note that 
most of the higher derivative theories of gravity that exist in the literature have the characteristic that the coupling constants associated with different terms 
in the Lagrangian are dimension dependent. In Einsteinian cubic gravity on the other hand, the coupling constants are dimension independent. 
Unlike, say, Gauss-Bonnet gravity, the extra terms in the Lagrangian are not topological in four spacetime dimensions. 
Motivated by the holographic studies of higher derivative gravity theories, in the first part of this work, we 
undertake an investigation of the entanglement R\'enyi entropy of the CFT dual to Einsteinian cubic gravity, after constructing a black hole solution 
perturbatively, for a particular choice of parameters in the theory. We also compute the scaling dimensions of the twist operators (to be elaborated upon in the
next section) and verify the central charges of the dual CFT via a computation of the Weyl anomaly. 

In sections 4, 5 and 6, similar analyses are performed in extended theories of gravity with a chemical potential, and we study various features of their
ERE and twist operators. In section 4, we consider Born-Infeld gravity \cite{Tanay}, \cite{Cai}, and elaborate upon the behavior of the charged ERE both as a function of the chemical 
potential for fixed values of the Born-Infeld parameter, as well as a function of the Born-Infeld parameter with fixed chemical potential. In section 5, we perform 
a similar analysis for charged quasi-topological gravity \cite{Mann}. In this case, we are able to provide interesting bounds on the central charges as well as other parameters of the dual CFT via our analysis. Finally, in section 6, we briefly study Einstein's gravity with a Weyl-corrected gauge field \cite{Dey}, \cite{Sarkar}. Sections 3 and 6 contain perturbative analysis of the
ERE, and in order to justify the numerical values of the coupling constant that we have chosen, we have to check that the second order corrections to
our linear order results in this paper are indeed small. For ERE, this has been done in the paper, and the relevant second order black hole solutions for 
Einsteinian cubic gravity are summarised in Appendix A. For Weyl-corrected gravity, the second order expressions 
are detailed in Appendix B, but the analysis become tedious, and the second order corrections could not be performed satisfactorily, 
due to numerical issues, as we explain towards the end of section 6. 

A word about the notation used : we consider four different theories and consider similar physical quantities in each of them. However, using different
symbols for different theories will unnecessarily clutter the notation and we avoid doing it. It has to be thus remembered that a particular symbol used in 
a section remains relevant for that section only. 

\section{Holographic Entanglement R\'enyi Entropy}

In this section, we will briefly review the relevant details of the computation of various quantities associated with holographic ERE. Since the material
is quite well established by now, we will be brief here, and point the reader to the references herein for further details. 

Let us first briefly review the idea of entanglement entropy (EE) for a field theory which is a composite
system of two subsytems, $A$ and its complement $B$. Consider the reduced density matrix of $A$, given by $\rho_A$, obtained by tracing over the 
degrees of freedom corresponding to $B$ out of the density matrix of the full system. The entanglement entropy of the subsystem $A$ is given by,
\begin{equation}
S_{EE} = -\text{tr}\, (\rho_A\, \text{log}\, \rho_A)\, ,
\end{equation}
and is by definition, the von Neumann entropy of subsystem $A$. 
The entanglement R\'enyi entropy (ERE) \cite{Renyi}, \cite{Renyi1}, which will be the main focus of this paper is another useful measure of entanglement, and is given by,
\begin{equation}
S_n = \frac{1}{1-n}\, \text{log}\,  \text{tr}\, (\rho_{\text{A}}^n)\, .
\label{EREformula}
\end{equation}
Here, $n$, called the order of the ERE is a positive real number. By construction, $S_{EE} = \lim_{n \to 1} S_n$, and hence we can obtain the EE
from the ERE. 

The calculation of the R\'enyi entropy in CFTs depends on the replica trick \cite{Cardy}, \cite{Cardy1} which involves computing a path integral over $n$-fold cover consisting of replicas of the original CFT. The different copies are separated by twist operators \cite{Caraglio} inserted at branch points. 
Recently, holographic methods have been used to compute the ERE for a variety of CFTs with different bulk duals. Extending the idea of \cite{Casini}, the authors of \cite{Myers} derived a formula for holographic ERE with spherical entangling surfaces and calculated the ERE explicitly for Einstein gravity as well as some higher derivative theories like Gauss-Bonnet gravity and quasi-topological gravity. In a related work, \cite{Galante} computed the ERE for four dimensional $\mathcal{N}=4$ super-Yang-Mills holographically, using the same method. This proposal of ERE was then generalized by \cite{Belin} to include a chemical potential in the field theory. They constructed the charged entanglement R\'enyi entropy in a grand canonical ensemble where one has to compute the Euclidean path integral by inserting a Wilson line encircling the entangling surface. For some recent interesting works on holographic  R\'enyi entropy we refer the reader to \cite{Matsuura}, \cite{Matsuura1}, \cite{Czinner}, \cite{Czinner1}, \cite{Miao}.

Let us briefly recapitulate the computation of the ERE, closely following \cite{Casini} and \cite{Myers}. 
We consider a spherical entangling surface $S^{d-1}$ in a flat $d$-dimensional CFT. From \cite{Casini}, it can be shown that the CFT 
can be mapped to a hyperbolic cylinder $R \times H^{d-1}$ by a conformal transformation. So, effectively the correlators in the vacuum CFT are 
mapped to a thermal ensemble on $R \times H^{d-1}$, with the temperature being fixed by the curvature scale $\mathcal{R}$ of the cylinder 
(which is also the radius of the original sphere $S^{d-1}$), given by,
\begin{equation}
T_0 = \frac{1}{2\pi \mathcal{R}}\, .
\end{equation} 
Now, using the AdS/CFT correspondence, we can infer that the CFT on the hyperbolic cylinder $R \times H^{d-1}$ has a bulk dual which is a black hole 
with a hyperbolic horizon. The above argument implies a simple relation between the thermal density matrix and the reduced density matrix $\rho_A$ as,
\begin{eqnarray}
 \rho_A=U^{-1}\, \frac{e^{-H / T_0}}{Z(T_0)}\, U\, ,
\end{eqnarray}
where $U$ is a unitary transformation and $Z(T_0)=\text{tr}\, (e^{-H / T_0})$ and hence we have,
\begin{eqnarray}
 \text{tr}(\rho_A^n)=\frac{Z(T_0 / n)}{Z(T_0)^n}\, .
\end{eqnarray}
Now making use of the expression of free energy $F$ in thermodynamics, $F(T)=-T\, \text{log}\, Z(T)$, the ERE of the CFT can be calculated as,
\begin{equation}
S_n = \frac{n}{1-n}\frac{1}{T_0}\bigg[F(T_0)-F\bigg({T_0\over n}\bigg)\bigg]\, .
\end{equation}
Since $S= - \frac{\partial F}{\partial T}$, the above relation can be rewritten as,
\begin{equation}
S_n = \frac{n}{n-1}\frac{1}{T_0} \int_{T_0/n}^{T_0} S\, dT\, ,
\label{Sneqn}
\end{equation}
where, $S$ denotes the thermal entropy of the CFT living on $R\times H^{d-1}$. Using the gauge/gravity duality, we can identify this thermal entropy with the Wald entropy $S_{\text{Wald}}$ \cite{Wald} of the hyperbolic black hole and hence we can compute the R\'enyi entropy using eq.(\ref{Sneqn}) \footnote{Recently, \cite{Mahapatra} considered corrections to Wald entropy of topological black holes and found logarithmic corrections to R\'enyi entropy.}.

Since we will be considering some modified  theories of gravity dual to CFTs with a conserved global charge, we have to understand how the entanglement 
between the two subsystems $A$ an $B$ depends on the charge distribution among themselves. For a grand canonical ensemble, the charged ERE is defined \cite{Belin}, \cite{Pastras} by 
simply generalizing eq.(\ref{EREformula}),
\begin{equation}
S_n(\mu_c) = \frac{1}{1-n}\, \text{log}\, \text{tr}\, \bigg[\rho_{\text{A}}\, \frac{e^{\mu_c\, Q_{A}}}{n_A(\mu_c)}\bigg]^n\, ,
\label{ChargedEREformula}
\end{equation}
where, $\mu_c$ is the chemical potential conjugate to the charge $Q_A$ and $n_A(\mu_c)= \text{tr}\, \big(\rho_{\text{A}}\, e^{\mu_c\, Q_{A}}\big)$ is introduced to normalize the new density matrix with unit trace. In a similar way, one can derive the formula for charged ERE which resembles eq.(\ref{Sneqn}) with $S_n$ being a function of $\mu_c$,
\begin{equation}
S_n (\mu_c) = \frac{n}{n-1}\frac{1}{T_0} \int_{T_0/n}^{T_0} S(T,\mu_c)\, dT\, .
\label{Sneqnmuc}
\end{equation}

The R\'enyi entropy satisfies the following inequalities \cite{Myers}, \cite{Zyczkowski},
\begin{eqnarray}
\frac{\partial S_n}{\partial n} \leq 0\,, \ \ \ \ \frac{\partial}{\partial n}\left(\frac{n-1}{n} S_n \right) \geq  0\,, \nonumber \\ 
\frac{\partial}{\partial n}\bigg((n-1)S_n\bigg)\geq 0\,, \ \ \ \ \ \frac{\partial^2}{\partial n^2}\bigg((n-1)S_n\bigg)\leq 0\,. 
\label{EREinequalities}
\end{eqnarray}
where it can be shown from the gravity side that satisfying the inequalities in the first line of eq.(\ref{EREinequalities}) automatically leads
to the ones in the second line of that equation \cite{Pastras}. 
As was argued in \cite{Myers}, these inequalities will hold for any CFT, as long as we have a stable thermal ensemble. Nevertheless, in 
all the examples considered in this paper, we will explicitly check the above inequalities. This is particularly important for the perturbative solutions that we consider. 

Now, as we have mentioned, the replica trick for computing the R\'enyi entropy can be understood as the insertion of a surface operator, known as the 
twist operator $(\sigma_n)$, at the entangling surface \cite{Myers}, \cite{Belin}, \cite{Smolkin}. The scaling dimension of this operator is determined from the leading singularity in the 
correlator $\langle T_{\mu\nu}\sigma_n\rangle$. The leading singular behavior is fixed by the symmetry, tracelessness and the conservation properties 
of $T_{\mu\nu}$.

Let us consider a four dimensional CFT where a planar twist operator is positioned at $x^1 = x^2 = 0$ in flat Euclidean space. The twist operator extends 
along the other two directions $x^3$ and $x^4$. Now, one inserts a stress tensor at $(y^1,y^2,y^3,y^4)$ and thus the orthogonal distance between the 
twist operator and the stress tensor operator is given by, $l = \sqrt{(y^1)^2+(y^2)^2}$. With this notation, the leading singularity in the correlator of the 
twist operator and the stress tensor can be calculated as \cite{Myers}, \cite{Belin},
\begin{eqnarray}
\langle T_{i j}\, \sigma_n \rangle&=& -\frac{h_n}{2\pi} \,\frac{\delta_{i
j}}{l^4} \,,\qquad \langle T_{i a}\, \sigma_n
\rangle=0\,,\nonumber \\
\langle T_{a b}\, \sigma_n \rangle&=& \frac{h_n}{2\pi} \,\frac{3\,\delta_{a
b}-4\,n_{a}\,n_{b}}{l^4}\ \,.
\label{leadingsingularbehavior}
\end{eqnarray}
where, $(a, b)=(1, 2)$ are the normal directions and $(i, j)=(3, 4)$ are the tangential directions to the twist operator. Here $n^{a} = \frac{y^{a}}{l}$ 
is the orthogonal unit vector from the twist operator $\sigma_n$ to the point where the stress tensor $T_{\mu\nu}$ has been inserted. From 
eq.(\ref{leadingsingularbehavior}), one should notice that the leading singularity is now fixed up to a constant $h_n$, known as the scaling dimension 
of $\sigma_n$. Following \cite{Myers}, \cite{Belin}, we quote the final expression of the scaling dimension $h_n$ (in the four dimensional boundary CFT)
in terms of the thermal energy density,
\begin{eqnarray}
 h_n(\mu_c)={n\over 3}{\mathcal{R}^3\over T_0} \bigg[\mathcal{E}(T_0,\, 0)-\mathcal{E}\bigg({T_0\over n},\, \mu_c\bigg)\bigg]\, ,
 \label{hnformula}
\end{eqnarray}
where, $\mathcal{E}(T,\, \mu_c)$ is the thermal energy density given by,
\begin{eqnarray}
\mathcal{E}(T,\, \mu_c) = \frac{E(T,\, \mu_c)}{\mathcal{R}^3\, V_{\Sigma}}\, .
\end{eqnarray}
$E(T,\, \mu_c)$ being the total thermal energy and $V_{\Sigma}$ is the regulated volume of the hyperbolic plane $H^3$.
We should point out here that the first term in eq.(\ref{hnformula}) appears due to the anomalous contribution from the stress tensor under 
conformal transformations. One can also express eq.(\ref{hnformula}) in terms of the thermal entropy density $\mathcal{S}$ using the 
first law of thermodynamics,
\begin{eqnarray}
 d\mathcal{E}=Td\mathcal{S}+T_0\, \mu_c\, d \rho\, ,
 \label{1stlaw}
\end{eqnarray}
where, $\mathcal{S}$ and $\rho$ are respectively the thermal entropy density and the charge density of the dual CFT.

The conformal dimension $h_n(\mu_c)$ possesses an interesting universal property, when expanded around $n=1$ and $\mu_c=0$,
\begin{eqnarray}
 h_n(\mu_c) = \sum_{\lambda,\delta}\frac{1}{\lambda!\delta!} h_{\lambda \delta}(n-1)^\lambda \mu_c^\delta, \ \  \mbox{where} \ \  h_{\lambda\, \delta} =  (\partial_n)^\lambda(\partial_{\mu_c})^\delta h_n(\mu_c)|_{n=1,\ \mu_c=0}.
  \label{hnexpansion}
\end{eqnarray}
Note that $h_{00}=0$, while the authors of \cite{Myers}, \cite{Belin} showed that the quantity $h_{10}$ is related to the central charge 
$\tilde{C}_T$ for any $d$-dimensional CFT 
\begin{eqnarray}
h_{10} = \frac{2}{d-1}\pi^{1-\frac{d}{2}}\Gamma\bigg(\frac{d}{2}\bigg)\tilde{C}_T\, .
\label{h10universal}
\end{eqnarray}
As mentioned in \cite{Belin}, \cite{Pastras}, since we are discussing charged R\'enyi entropy, one may think about the correlator of the twist operator $\sigma_n$ and the current operator $J_\mu$. Here also one can extract the leading singular behavior of this correlator using the conservation of the current $J_{\mu}$ as \cite{Belin}, \cite{Pastras}, 
\begin{eqnarray}
 \langle J_a\sigma_n(\mu_c)\rangle = -\frac{i k_n(\mu_c)}{2\pi}\frac{\epsilon_{a\, b}n^b}{l^3}, ~~~ \langle J_{a}\sigma_n\rangle = 0,
\end{eqnarray}
where $\epsilon_{a\, b}$ stands for the volume form and $ k_n(\mu_c)$ represents the magnetic response characterizing the response of the current to the magnetic flux. Finally, one can derive the magnetic response $k_n(\mu_c)$ using a conformal mapping \cite{Belin}, \cite{Pastras} as,
\begin{eqnarray}
 k_n(\mu_c) = 2 \pi n \mathcal{R}^3 \rho(n, \mu_c).
  \label{knformula}
\end{eqnarray}
In a similar way as argued with the scaling dimension, here one may expand the magnetic response $ k_n(\mu_c)$ around $n=1$ and $\mu_c=0$,
\begin{eqnarray}
 k_n(\mu_c) = \sum_{\lambda,\delta}\frac{1}{\lambda!\delta!} k_{\lambda \delta}(n-1)^\lambda \mu_c^\delta, \ \  \mbox{where} \ \  k_{\lambda\, \delta} =  (\partial_n)^\lambda(\partial_{\mu_c})^\delta k_n(\mu_c)|_{n=1,\ \mu_c=0}.
 \label{knexpansion}
\end{eqnarray}
In this work, we will broadly study the concepts introduced above. Having set up the notations and conventions, we will now proceed to the main body of this paper. 

\section{ERE for Einsteinian Cubic Gravity \label{ERE}}

Our first example is the analysis of ERE in Einsteinian cubic gravity. This is an interesting proposal recently put forward in \cite{Bueno}. As stated in the
introduction, modified theories of gravity often suffer from ghost modes in the spectrum. The work of \cite{Bueno} on the other hand attempts to 
formulate a theory which at the linearized level has no ghost modes, and where the coupling constants are independent of spacetime dimensions.
This theory has the following action 
\begin{eqnarray}
\label{cubicaction}
&& S=\frac{1}{2\ell _p ^3}\int d^5 x \bigg[R-2\Lambda + \frac{\lambda }{2}L^2 \mathcal{\chi}_4 
+ \beta_1\, L^4\big(\tensor{R}{_{\mu}^{\rho}_{\nu}^{\sigma}}\tensor{R}{_{\rho}^{\gamma}_{\sigma}^{\delta}}\tensor{R}{_{\gamma}^{\mu}_{\delta}^{\nu}}+\frac{1}{12}\tensor{R}{_{\mu\nu}^{\rho\sigma}}\tensor{R}{_{\rho\sigma}^{\gamma\delta}}\tensor{R}{_{\gamma\delta}^{\mu\nu}}\nonumber \\
&& -\tensor{R}{_{\mu\nu\rho\sigma}}\tensor{R}{^{\mu\rho}}\tensor{R}{^{\nu\sigma}}+\frac{2}{3}\tensor{R}{_{\mu}^{\nu}}\tensor{R}{_{\nu}^{\rho}}\tensor{R}{_{\rho}^{\mu}}\big) + \beta_2 L^4 \big(\frac{14}{3}\tensor{R}{_{\mu\nu}^{\rho\sigma}}\tensor{R}{_{\rho\sigma}^{\gamma\delta}}\tensor{R}{_{\gamma\delta}^{\mu\nu}}- 24\tensor{R}{_{\mu\nu\rho\sigma}}\tensor{R}{^{\mu\nu\rho}_{\gamma}}\tensor{R}{^{\sigma\gamma}} \nonumber \\
&& + 3\tensor{R}{_{\mu\nu\rho\sigma}}\tensor{R}{^{\mu\nu\rho\sigma}}R +16\tensor{R}{_{\mu\nu\rho\sigma}}\tensor{R}{^{\mu\rho}}\tensor{R}{^{\nu\sigma}} +\frac{64}{3}\tensor{R}{_{\mu}^{\nu}}\tensor{R}{_{\nu}^{\rho}}\tensor{R}{_{\rho}^{\mu}}-12\tensor{R}{_{\mu\nu}}\tensor{R}{^{\mu\nu}}R+R^3\big)\bigg]
\label{actionECG}
\end{eqnarray}
where, $R$ is the Ricci scalar and $L$ is the AdS length, related to the cosmological constant $\Lambda$ by, $\Lambda = -{6\over L^2}$. 
$\mathcal{\chi}_4$ is the standard four-derivative Gauss-Bonnet term,
\begin{eqnarray}
 \mathcal{\chi}_4 = \tensor{R}{_{\mu\nu\rho\lambda}}\tensor{R}{^{\mu\nu\rho\lambda}}-4\tensor{R}{_{\mu\nu}}\tensor{R}{^{\mu\nu}}+R^2\, ,
 \label{GBtermECG}
\end{eqnarray}
while $\beta_1$ and $\beta_2$ are the coupling constants of two new sets of six derivative terms. 
As stated earlier, our goal is to study the ERE of Einsteinian cubic gravity using holographic methods. 
As in any other case of extended theories of gravity, this provides a novel dual CFT, which is interesting to analyze.  

It is difficult (if not impossible) to analytically solve the Einstein equations for this cubic theory taking into account all the three parameters 
$\lambda$, $\beta_1$ and $\beta_2$. We will be interested in the cubic terms in the Lagrangian, and hence exclude the Gauss-Bonnet coupling from
our analysis (the latter was studied in \cite{Myers}). We will thus focus on one of the new coupling terms $\beta_1$ and $\beta_2$. To simplify the
computation, we will set $\beta_1={3\over 2}\beta$ and $\beta_2=0$ (the numerical factor of ${3 \over 2}$ is just a choice which simplifies some 
constants that appear in our analysis in what follows. Any other choice will not alter the essential physics). Of course, one could alternatively set 
$\beta_1=0$ to begin with, as well. Now, even with a single coupling, we found it difficult to solve the system exactly. So, we proceed to solve the system 
up to linear order in the new coupling constant $\beta$. 

In order to compute the holographic R\'enyi Entropy dual to this cubic gravity, we need to construct a hyperbolic black hole solution which 
can be done by choosing the following metric ansatz,
\begin{eqnarray}
ds^2 = -\bigg({r^2\over L^2}f_0(r)-1\bigg) \big(1+ \beta\mathcal{F}_1(r)\big) \mathcal{N}(r)^2 dt^2 +{dr^2 \over \big({r^2\over L^2}f_0(r)-1\big) 
\big(1+  \beta\mathcal{F}_2(r)\big)} + r^2 d\Sigma_{3} ^{2}
 \label{metricECG}
\end{eqnarray}
where, $d\Sigma_{3} ^{2}$ is the line element for the three dimensional hyperbolic plane $H^3$ having unit curvature. We will set 
\begin{eqnarray}
 \mathcal{N}(r)=\frac{\tilde{L}}{\mathcal{R}}\, ,
 \label{NeomECG}
\end{eqnarray}
where $\tilde{L}$ is the AdS curvature scale which is to be determined and ${\mathcal R}$ is the curvature of the hyperbolic spatial slice. 

Here $f_0(r)$ is the leading order solution which represents a hyperbolic Schwarzschild black hole in five dimensional AdS space, and has the form,
\begin{eqnarray}
 f_0(r)=1-{\omega^4 \over r^4}\, .
 \label{feomECG}
\end{eqnarray}
The position of the horizon of this unperturbed solution can be obtained by demanding, $f_0(r_h)={L^2\over r_h^2}$. This in turn, fixes the constant $\omega$ in 
the leading order solution as, 
\begin{eqnarray}
\omega=(r_h^4-L^2\, r_h^2)^{1\over 4}
\label{omegaeom}
\end{eqnarray}

Now varying the action with respect to $\mathcal{F}_1(r)$ and $\mathcal{F}_2(r)$ yield the following differential equations at order $\beta^2$ (the 
absence of $\mathcal{O}(\beta)$ terms signifies that the ansatz satisfies the Einstein's equations at linear order)
\begin{eqnarray}
r^9 \left(2 r F_2(r) \left(L^2-2 r^2\right)+F_2'(r) \left(L^2 r^2-r^4+\omega^4\right)\right)-4{\mathcal A}(r) = 0\nonumber\\
r^9 \left(2 r F_2(r) \left(L^2-2 r^2\right)+F_1'(r) \left(L^2 r^2-r^4+\omega^4\right)\right)-4 {\mathcal B}(r)= 0~,
\label{FeomECG}
\end{eqnarray}
where we have denoted
\begin{eqnarray}
{\mathcal A}(r) &=& -72 L^2 r^2 \omega ^8+r^{12}+69 r^4 \omega ^8-74\omega^{12}\nonumber\\
~~{\mathcal B}(r) &=& -48 L^2 r^2 \omega ^8+r^{12}+45 r^4 \omega ^8-50\omega^{12}
\end{eqnarray}
from which we obtain the following analytical solutions 
\begin{eqnarray}
\mathcal{F}_1(r) &=& \frac{36 L^2 r^2 \omega ^8+r^{12}-57
r^4 \omega ^8+25 \omega ^{12}}{r^8 \left(L^2 r^2-r^4+\omega ^4\right)}-\frac{C_1}{L^2 r^2-r^4+\omega ^4}+C_2\nonumber\\
\mathcal{F}_2(r) &=& \frac{48 L^2 r^2 \omega ^8+r^{12}-69 r^4
   \omega ^8+37 \omega ^{12}}{r^8 \left(L^2 r^2-r^4+\omega ^4\right)} - \frac{C_1}{L^2 r^2-r^4+\omega ^4}
\label{FsolutionsECG}
\end{eqnarray}
where $C_1$ and $C_2$ are constants of integration, to be determined using appropriate boundary conditions.
At this point it is important to note that we have inserted an extra factor of $1\over \mathcal{R} ^2$ in the metric component $g_{tt}$. 
With this inclusion, we make sure that the boundary CFT dual to this black hole geometry resides on $R \times H^3 $, 
where the hyperbolic spatial slice $H^3$ has curvature $\mathcal{R}$ as mentioned before,
\begin{eqnarray}
 ds_{\infty}^2=-dt^2 + \mathcal{R}^2 \ d\Sigma_{3} ^{2} \,.
 \label{BoundaryMetricECG}
\end{eqnarray}
The constant $C_2$ can be set to zero since it simply produces a rescaling of the time coordinate, and we demand that the metric of
eq.(\ref{metricECG}) be conformally equivalent to the one of eq.(\ref{BoundaryMetricECG}). Then 
$C_1$ can be fixed by demanding that the  position of the horizon $r_h$ remains the same as with the unperturbed solution, and by 
the fact that the potential singularities in $\mathcal{F}_1(r_h)$ and $\mathcal{F}_2(r_h)$ need to be avoided. It can be checked that
this can be obtained by setting 
\begin{equation}
C_1 = \frac{11L^6}{r_h^2} -54L^4 + 75L^2r_h^2 - 31r_h^4 
\label{firstc1}
\end{equation}

Since the asymptotic form of the bulk metric of eq.(\ref{metricECG}) should represent the background metric for the dual boundary CFT given 
by eq.(\ref{BoundaryMetricECG}), we now determine the AdS curvature scale $\tilde{L}$ as,
\begin{eqnarray}
\tilde{L}={L\over \sqrt{1-\beta}} \equiv {L \over \sqrt{f_{\infty}}}\, ,
\label{LtildeECG}
\end{eqnarray}
where we have used the fact that 
\begin{eqnarray}
 (1+ \beta \mathcal{F}_1(r)) \left(f_0(r)-\frac{L^2}{r^2}\right)|_{r\rightarrow \infty}=1-\beta \equiv f_{\infty}\, .
\end{eqnarray}

After determining the perturbed black hole solution up to $\mathcal{O}(\beta)$, we proceed to compute the Hawking temperature of the black hole. 
At this stage, it is convenient to make a change of variables and use a new dimensionless variable, $x={r_h\over {\tilde L}}$, instead of using $r_h$. 
The Hawking temperature of the black hole then can be expressed as,
\begin{equation}
T = \frac{2 x^2-1}{2 \pi  x \mathcal{R}}-\beta \frac{ \left(x^2-1\right)^3 }{\pi x^5
\mathcal{R}}
\label{TempECG}
\end{equation}
Note that there is a $\mathcal{O}(\beta)$ correction in the temperature and as we take the limit $\beta \rightarrow 0$, we recover the expression 
of the temperature with pure Einstein gravity \cite{Myers}, as expected.
 
To compute the thermal entropy of the hyperbolic black hole, we proceed to evaluate the Wald entropy following \cite{Wald},
\begin{eqnarray}
S_{Wald}=-2\pi \int_{horizon} d^3 x \sqrt{h} \frac{\partial \mathcal{L}}{\partial R_{\mu\nu\rho\lambda}} \epsilon_{\mu\nu} \epsilon_{\rho\lambda}
\label{WaldEntropyFormula}
\end{eqnarray}
This  proposal was put forward to compute the black hole entropy in higher derivative gravity theories. Here, $\mathcal{L}$ represents the 
Lagrangian of the particular higher derivative gravity, and $\epsilon_{\mu\nu}$ is the binormal Killing vector normalized as $\epsilon_{\mu\nu}\epsilon^{\mu\nu} = -2$.
The final expression of the Wald entropy turns out to be 
\begin{equation}
S_{Wald} = 2 \pi \big({L\over \ell_p}\big)^3\, V_{\Sigma}\big[x^3+{3\over 2}\beta \big(25 x^3-48 x+\frac{18}{x}\big)\big]\, .
\label{WaldEntropyECG}
\end{equation}
The first term in eq.(\ref{WaldEntropyECG}) represents the usual thermal entropy in case of a hyperbolic $AdS$ black hole in pure Einstein gravity, 
while the second term stands for the correction in entropy due to the presence of the cubic term in the action.
Here $V_{\Sigma}=\int_{H^3} d\Sigma_3$ is the volume of the hyperbolic plane. As shown in \cite{Myers},  $V_{\Sigma}$ is divergent and this 
behavior mimics the UV divergence of the  R\'enyi Entropy in the dual CFT. Hence, one needs to regularize the entropy and this is done by integrating
out the hyperbolic volume element up to a maximum radius ${\mathcal{R}\over \delta}$, where $\delta$ is the short-distance cut-off of the dual CFT.
Now after expanding $V_{\Sigma}$ in powers of ${\mathcal{R}\over \delta}$, it is straightforward to single out the universal contribution coming
from the sub-leading terms,
\begin{eqnarray}
 V_{\Sigma, univ}=-2\pi \log\bigg({\mathcal{R}\over \delta}\bigg) \,.
\end{eqnarray}
 
Now, writing eq.(\ref{Sneqn}) in terms of the dimensionless variable $x$ , the entanglement R\'enyi entropy can be expressed as,
\begin{eqnarray}
 S_n &=& \frac{n}{n-1} \frac{1}{T_0}\, \int_{x_n}^{x_1} S(x)\,\frac{dT}{dx}\,dx \nonumber \\
&=& \frac{n}{n-1} \frac{1}{T_0}\,\left[S(x)\,T(x)|^{x_1}_{x_n} -\int_{x_n}^{x_1} \frac{dS}{dx}\,T(x)\,dx \right] \,.
\label{RenyiEntropyFormula}
\end{eqnarray}
where in the second line we have done an integration by parts.

It follows that $x_1$ and $x_n$ (the upper and lower limits, 
respectively, in eq.(\ref{RenyiEntropyFormula})) are the only two remaining quantities that we need for the computation of the ERE. These two quantities 
can be determined by solving the following equation,
\begin{eqnarray}
 T(x_n)={T_0\over n}
 \label{xneqnECG}
\end{eqnarray}
where, $T_0={1\over 2\pi\mathcal{R}}$. From eqs.(\ref{TempECG}) and (\ref{xneqnECG}), note that at the ${\beta\rightarrow 0}$ limit, $x_1$ and $x_n$ are given by,
\begin{eqnarray}
 (x_1)_{\beta\rightarrow 0} &=& 1 \nonumber \\
 (x_n)_{\beta\rightarrow 0} &=& {1\over 4n}(1+\sqrt{1+8n^2})\, \equiv \tilde{x}_n .
 \label{xnunperturbed}
\end{eqnarray}
Now substituting eq.(\ref{TempECG}) in eq.(\ref{xneqnECG}) yields a sixth order algebraic equation for $x_n$ which can not be solved exactly for 
arbitrary $n$. However, with $n=1$, we find $x_1=1$ is still a solution of the equation. To determine $x_n$ for arbitrary $n$, we solve $x_n$ up to linear order in $\beta$,
\begin{eqnarray}
x_n = \tilde{x}_n - 2\beta \frac{(1-{\tilde x}_n^2)^3} {{\tilde x}_n^3(1+2 {\tilde
x}_n^2)}
\label{xnECG}
\end{eqnarray}
where, $\tilde{x}_n$ is defined in eq.(\ref{xnunperturbed}). We construct the solution in such a way that $x_n$ agrees with eq.(\ref{xnunperturbed}) 
when $\beta$ vanishes. 

Finally, we compute the ERE,
\begin{eqnarray}
S_n &=& \frac {n}{n-1} T_0 \int_{x_1}^{x_n}\, S_{Wald}\, T'(x)\, dx\, \nonumber \\
&=& {n\over n-1} \pi V_{\Sigma} \bigg(\frac{L}{\ell_p}\bigg)^3\bigg[\big(2-{\tilde
x}_n^2-{\tilde x}_n^4\big)
+{3\beta \over 2 {\tilde x}_n^2}(1-{\tilde x}_n^2)(14-56{\tilde x}_n^2+27{\tilde
x}_n^4)\bigg]
\label{SnECG}
\end{eqnarray}

We can also calculate the entanglement entropy from the above expression of ERE by taking the $n \rightarrow 1$ limit, which yields,
\begin{equation}
S_1 = 2 \pi  \bigg(\frac{L}{\ell_p}\bigg)^3 V_{\Sigma} \big(1-{15\over 2}\beta\big)\, .
\label{EEECG}
\end{equation}
Notice that, the above expression in the $\beta \rightarrow 0$ limit corresponds to the entanglement entropy for any CFT in four space-time dimensions dual to a five dimensional pure Einstein gravity. The second term in the above expression represents the correction in entanglement entropy due to the presence of the cubic term in the bulk gravity. This correction in entanglement entropy is related to the central charge $a$ for a four dimensional CFT. As we will show in section \ref{Anomaly},
this matches with a corresponding calculation from the gravity side, via a computation of the Weyl anomaly. 

\subsection{Numerical Analysis and Results}

In this subsection, we present the results for the ERE in Einsteinian cubic gravity, plotted against two quantities, namely the order of 
the ERE $n$ and the coupling $\beta$. Since $S_n$ contains the volume factor $V_{\Sigma}$ involving the short-distance cut-off 
of the boundary CFT, we plot the ratio of R\'enyi entropy $S_n$ to the entanglement entropy $S_1$, instead of plotting $S_n$.

\begin{figure}[h!]
 \centering
 \subfigure[]{
 \includegraphics[width=2.5in,height=2.3in]{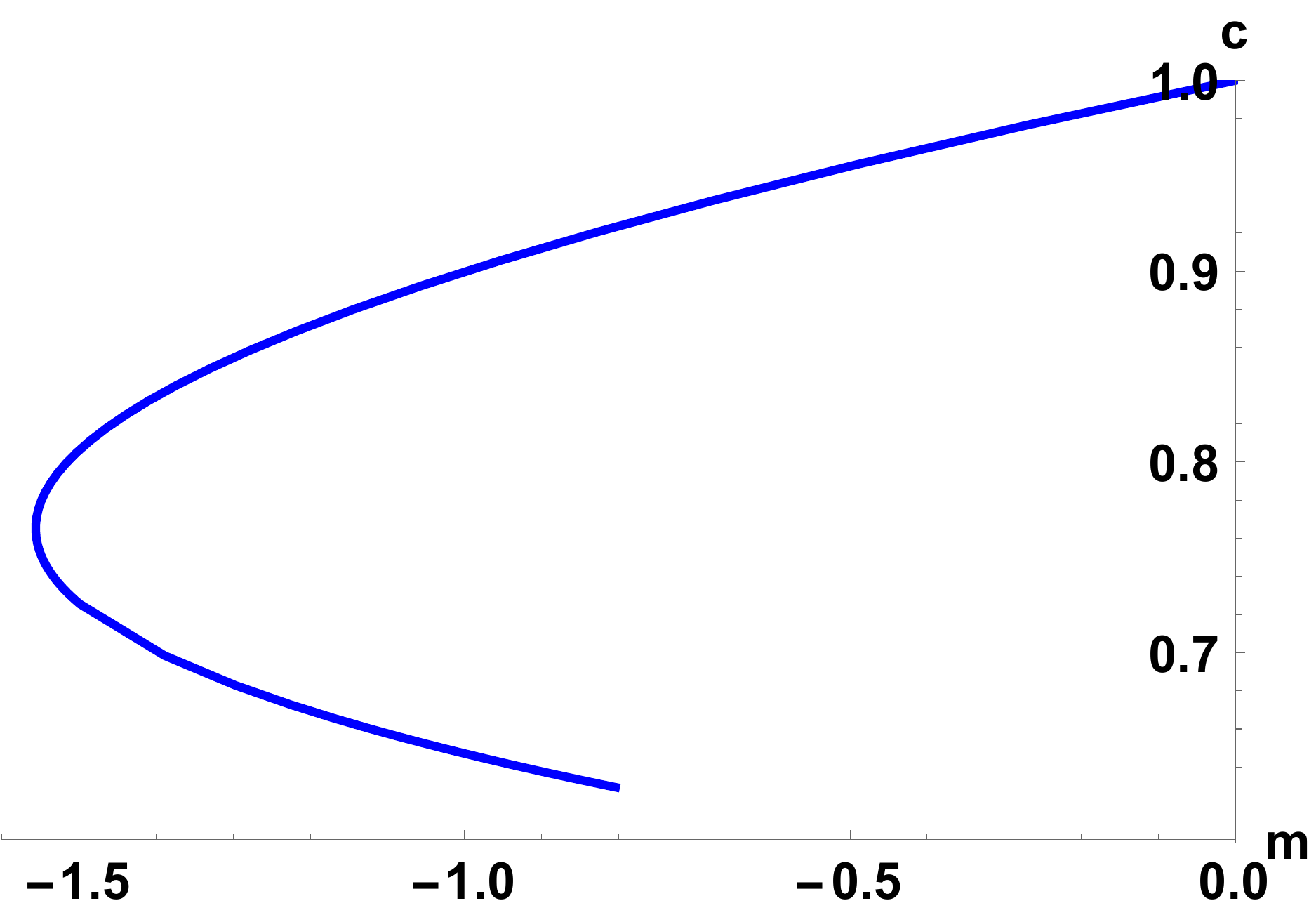}
 \label{SnbyS1vsbeta} } 
 \subfigure[]{
 \includegraphics[width=2.5in,height=2.3in]{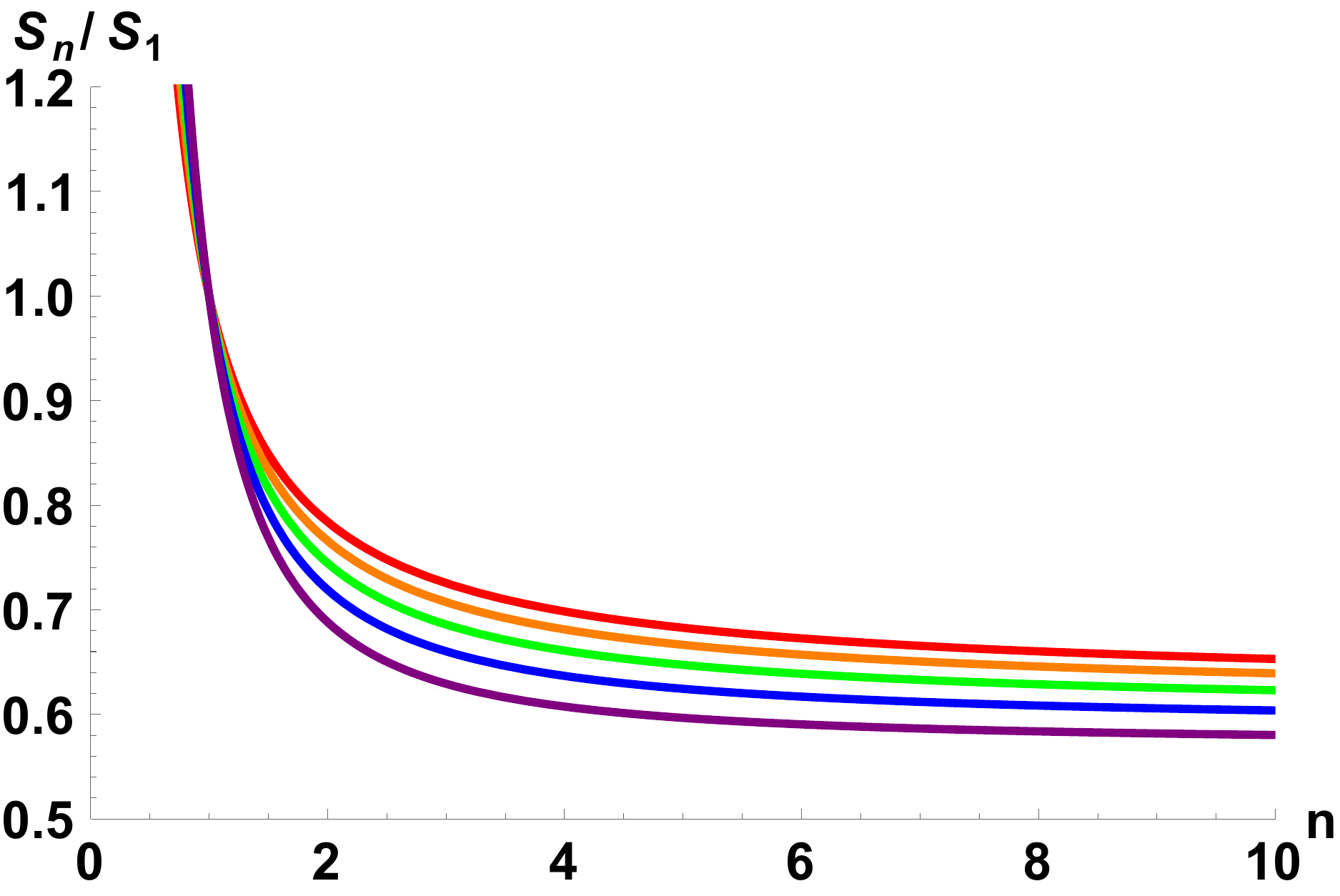}
 \label{SnbyS1vsn} }
  \caption{\small{(a)\, Plot of $c$ vs $m$ when ${S_n\over S_1}$ is fitted to $m\beta + c$, where the different points on the line correspond to different values
$n$. (b)\, ${S_n\over S_1}$ is plotted as a function of $n$, where the red, orange, green, blue and purple lines denote $\beta=0$, $0.01$, $0.02$, $0.03$ and $0.04$, respectively.}}
 \label{SnbyS1ECG}
 \end{figure}
Let us first consider the behavior of the ERE as a function of the coupling $\beta$ for different values of $n$ as shown. This is shown (in a slightly
different way) in figure \ref{SnbyS1vsbeta}. Since we have treated the problem perturbatively, we consider small values of $\beta$ 
up to $\beta = 0.05$ (these numerical values will be justified in a while, towards the end of this section) and fitted
${S_n\over S_1}$ for different $n$ to $m\beta + c$. In the figure, we have plotted $c$ vs $m$, where the different points on the curve correspond to different 
values of $n$. Expectedly, for $n=1$, we have $c=1$ and $m=0$, while for larger values of $n$, the slope remains negative and the intercept 
remains positive. 
In figure \ref{SnbyS1vsn}, we have shown the variation of ${S_n\over S_1}$ 
with $n$ for different values of $\beta$, and the red, orange, green, blue and purple lines denote $\beta=0$, $0.01$, $0. 02$, $0.03$ and $0.04$, respectively.
The slope of all the lines are negative, i.e.,  $\frac{\partial S_n}{\partial n} \leq 0$. 
Hence, we notice that the first inequality of eq.(\ref{EREinequalities}) is obeyed for any small value of the coupling constant $\beta$ in Einsteinian cubic gravity.
\begin{figure}[h!]
 \centering
 {
 \includegraphics[width=3in,height=2.8in]{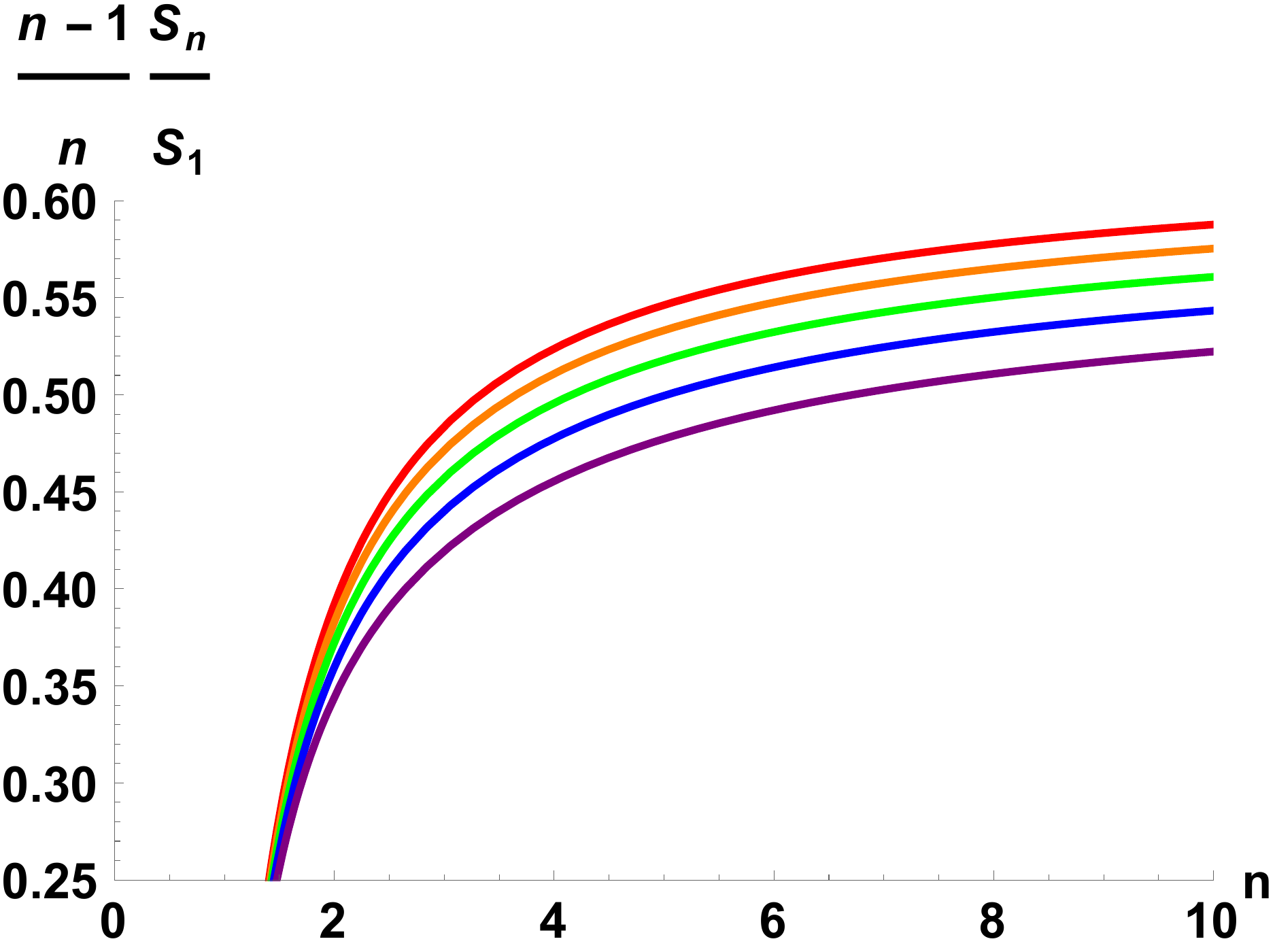}}
  \caption{\small{${n-1\over n}{S_n\over S_1}$ is plotted as a function of $n$ for different values of $\beta$, where we follow the same color coding for $\beta$ as used in figure \ref{SnbyS1vsn}.}}
   \label{SecondInequalityECG}
 \end{figure}

In figure \ref{SecondInequalityECG} we show the variation of the quantity ${n-1\over n}{S_n\over S_1}$ with $n$ for different values of the coupling $\beta$. Here again we follow the same color coding for $\beta$ as in figure \ref{SnbyS1vsn}. Note that the slope of the lines are positive for all chosen values of $\beta$, i.e., $\frac{\partial}{\partial n}\left(\frac{n-1}{n} S_n \right) \geq  0$, which supports the second inequality obeyed by R\'enyi entropy in eq.(\ref{EREinequalities}). 
It should be mentioned that choosing a relatively higher value of $\beta$ violates the inequalities obeyed by  R\'enyi entropy. This is an artifact of our
perturbative analysis. 

\subsection{Scaling Dimension of Twist Operators \label{Twist Operators}}
Recall that as mentioned in section 2, the field theoretic method for calculating entanglement entropy involves performing a path integral over $n$ replicas 
of the original CFT, and that twist operators open branch cuts between these copies. In this example, there is no chemical potential, and hence using eq.(\ref{1stlaw}), eq.(\ref{hnformula})  reduces to the result\cite{Myers},
\begin{eqnarray}
h_n = \frac{n}{3 T_0 V_\Sigma} \int_{x_n}^{x_1} dx T(x) S'(x)\, .
 \label{hnformulamuc0}
\end{eqnarray}
Using the above equation, the dimension of the twist operator can be computed up to linear order in $\beta$ as,
\begin{eqnarray}
&& h_n = -{\pi\over n^3} \bigg({L\over \ell_p}\bigg)^3\bigg[{1\over
32}{(1+\sqrt{1+8n^2})^3\over 3+\sqrt{1+8n^2}}(1-n^2)\bigg]\nonumber \\
&&-{\pi\over n^3} \bigg({L\over \ell_p}\bigg)^3 
{\beta \over 64 \sqrt{1+8 n^2} }\bigg[\sqrt{1+8 n^2}(88 n^4-340
n^2+81)+432n^2+81\bigg] + \mathcal{O}(\beta^2)\, .\nonumber \\
\end{eqnarray}
As expected, the leading order term in the expression corresponds to a CFT with a bulk Einstein dual. Also note that, $h_n$ at linear order vanishes 
if one takes the limit $n\rightarrow 1$. We also find that 
\begin{eqnarray}
 \partial_n h_n |_{n=1}=\frac{2}{3} \pi \bigg({L\over \ell_p}\bigg)^3 \big(1+{9\over 2}\beta+\mathcal{O}(\beta^2)\big)={2\over 3 \pi}\, c\,
\end{eqnarray}
where $c=\pi ^2 \big({L\over \ell_p}\big)^3 \big(1+{9\over 2}\beta+\mathcal{O}(\beta^2)\big)$ is the central charge of the four dimensional CFT. The above relation between the central charge $c$ and the first order coefficient of the expansion of the scaling dimension $\partial_n h_n |_{n=1}$ is a general property and we show that this holds for Einsteinian cubic gravity, as expected. We will compute the central charge $c$ in the next section by studying holographic Weyl anomaly and show that it matches
with the result above. 

\subsection{Weyl Anomaly and Central Charges\label{Anomaly}}

In this section, we compute both the central charges $c$ and $a$ characterizing a four dimensional CFT dual to Einsteinian cubic gravity using the 
gauge/gravity duality. It is well known that, although the trace of the energy momentum tensor $\langle T_{\mu}^{\mu}\rangle$ vanishes in flat space, it is 
no longer zero if we place the CFT in a curved background. This is known as Weyl anomaly or conformal anomaly \cite{Duff}, \cite{Henningson}, 
which relates the CFT parameters to the coupling constants of the dual gravity theory. The trace anomaly for a four dimensional CFT is given by,
\begin{equation}
\langle T_{\mu}^{\mu}\rangle =\frac{c}{16\pi ^{2}}I_{4}-\frac{a}{16\pi ^{2}}E_{4},  
\label{Weyl Anomaly}
\end{equation}
where, 
\begin{equation}
I_{4}=R_{ijkl}R^{ijkl}-2R_{ij}R^{ij}+\frac{1}{3}R^{2}\, ,
\end{equation}%
is the square of the Weyl tensor $C_{ijkl}$, while $E_4$ represents the Euler density in four dimensions,
\begin{equation}
E_{4}=R_{ijkl}R^{ijkl}-4R_{ij}R^{ij}+R^{2}  \label{E4I4}\, .
\end{equation}
Following the prescription of \cite{Henningson}, one can obtain the central charges $c$ and $a$ via the gauge/gravity duality. Starting with the 
gravity action and employing  the Fefferman-Graham expansion \cite{Fefferman} one can write down the bulk metric as,
\begin{eqnarray}
ds^{2} =\frac{\tilde{L}^{2}}{4\rho ^{2}}d\rho ^{2}+\frac{\bar{g}_{ij}}{\rho }
dx^{i}dx^{j},  
\label{FeffermanGraham} 
\end{eqnarray}
where,
\begin{equation}
 \bar{g}_{ij} =\bar{g}_{(0)ij}+\rho \bar{g}_{(1)ij}+\rho ^{2}\bar{g}
_{(2)_{ij}}+...~,
\label{BoundaryFG}
\end{equation}
and $\bar{g}_{(0)ij}$ represents the boundary metric at $\rho=0$. Now substituting the form of the metric of eq.(\ref{FeffermanGraham}) into the gravity action (eq.(\ref{actionECG})),  $\bar{g}_{(2)ij}$ can be eliminated using the field equations. After writing the action in terms of $\bar{g}_{(0)ij\text{ }}$ and $~\bar{g}_{(1)ij}$ 
one can further express $\bar{g}_{(1)ij\text{ }}$ in terms of  $\bar{g}_{(0)ij\text{ }}$ and write down the action in terms of  $\bar{g}_{(0)ij\text{ }}$ only. 
Finally, one can read the anomaly coefficients by extracting the terms producing a log divergence,
\begin{equation}
S_{ln}\backsim \frac{1}{2}\ln \epsilon ~\int d^{4}x~\sqrt{\bar{g}_{(0)}}\langle
T_{\mu}^{\mu}\rangle ,  \label{Sln}
\end{equation}
where $\epsilon ~$is a short distance cut-off.
This procedure of finding the anomaly coefficients is straightforward but the calculation becomes complicated in higher derivative gravity theories. 
So, instead of following this approach, we follow \cite{Nemani}, \cite{Dehghani} in order to compute the central charges\footnote{The authors of \cite{Miao1} developed another beautiful and elegant method to simplify the computation of holographic Weyl anomaly and obtain the central charges for CFTs dual to higher derivative gravity. Using this approach, one does not need to solve any equations of motion. Instead, one can expand the action 
around a 'referenced curvature' and then derive the central charges from the coefficient in the expansion.}. We choose the following bulk metric,
\begin{equation}
ds^{2}=\frac{\tilde{L}^{2}}{4\rho ^{2}}d\rho ^{2}+\frac{1}{\rho }\left[
u(1+A \rho )\left( -R^{2}dt^{2}+\frac{dR^{2}}{R^{2}}\right) +v(1+B
\rho )(d\theta ^{2}+\sin ^{2}\theta d\phi ^{2})\right] .  
\label{bulkmetric}
\end{equation}
having a boundary $AdS_2 \times S^2$. After plugging this metric into the Lagrangian, we extract the coefficient of ${1 \over \rho}$ which gives rise 
to a log divergence. We call this term $L_{ln}$, given by,
\begin{eqnarray}
 L_{ln}&=&-\frac{1}{2 L^2 \tilde{L}^5 \ell _p^3} \bigg(-A^2 L^2 u v \tilde{L}^4+2 A B L^2 u v \tilde{L}^4-6 A B
   u v \tilde{L}^6-3 A \beta  L^6 u \tilde{L}^2\nonumber \\
   &-&A L^2 u\tilde{L}^6
   -B^2 L^2 u v \tilde{L}^4+3 \beta  B L^6 v
   \tilde{L}^2+B L^2 v \tilde{L}^6+6 \beta  L^6
   \tilde{L}^4-3 A^2 \beta  L^6 u v\nonumber \\
   &-&6 A \beta  B L^6 u
   v-3 \beta  B^2 L^6 u v\bigg)\sin \theta
\end{eqnarray}
The equations of motion for $A$ and $B$ are given by,
\begin{equation}
\frac{\partial }{\partial A }L_{\ln }=0\, , \ \ \mbox{and} \ \ \ \
\frac{\partial }{\partial B }L_{\ln }=0,  \label{ABeom}
\end{equation}
solving which we can determine $A$ and $B$. Now the four dimensional quantities $I_4$ and $E_4$ can be computed using the boundary metric  $\bar{g}_{0(ij)}$ as,
\begin{equation}
I_{4}=\frac{4(u-v)^2}{u\, v},\ \ \ \ \ \ \ \ \ \  E_{4}=-\frac{8}{u\, v}
\label{I4E4}
\end{equation}
Now, by noting the form of $I_4$ and $E_4$ in (eq.(\ref{I4E4})), it is easy to show that the central charges can be obtained by taking the following limits,
\begin{eqnarray}
c &=& \lim_{v\longrightarrow \infty }24\pi ^{2}\frac{L_{\ln }}{\sqrt{%
\bar{g}_{(0)}}}| _{u=1}, \nonumber \\
a &=& \lim_{v\longrightarrow 1 }4\pi ^{2}\frac{L_{\ln }}{\sqrt{%
\bar{g}_{(0)}}}| _{u=1}.
\label{caformula}
\end{eqnarray}
Finally, using eq.(\ref{caformula}) along with eq.(\ref{LtildeECG}), we determine the central charges up to linear order in $\beta$ as,
\begin{eqnarray}
c &=& \pi^2 \bigg({L\over \ell_p}\bigg)^3 \bigg(1+{9\over 2}\beta+\mathcal{O}(\beta^2)\bigg)\, , \label{cECG} \\
a &=&\pi^2 \bigg({L\over \ell_p}\bigg)^3 \bigg(1-{15\over 2}\beta+\mathcal{O}(\beta^2)\bigg)\, .
\label{aECG}
\end{eqnarray}
Note that the expressions of the entanglement entropy in section \ref{ERE} and the scaling dimension of twist operator in section \ref{Twist Operators} 
agree with the values for central charges obtained by the computation of holographic Weyl anomaly.

\subsection{Validity of R\'enyi Entropy Inequalities}

Since we work perturbatively up to linear order in the coupling constant $\beta$, we should consider relatively small values of $\beta$ in all our computations. 
However, to justify that our results are reliable (by taking into account the terms up to linear order in $\beta$), we should find the change in R\'enyi entropy 
due to the inclusion of next-to-leading order corrections and check that it is indeed small. For this purpose, we need to solve the Einstein equations at 
$\mathcal{O}(\beta^2)$. Following the same strategy as described earlier, we obtained a hyperbolic black hole solution 
at $\mathcal{O}(\beta^2)$ (see Appendix A). 

The numerical results for the ERE at different values of the cubic coupling with $\mathcal{O}(\beta)$ and $\mathcal{O}(\beta^2)$ corrections are shown 
in table \ref{tableqc}. To  note the deviation in the values with increasing $n$, we have chosen a large value of $n$ $(=1000)$ to produce the table. 
However, even with this large value of $n$, the second order correction in the entropy ratio is small and hence the first-order correction term is sufficiently 
reliable. It can be observed from the table that $\mathcal{O}(\beta^2)$ correction appears only at the third decimal place even when we increase the 
coupling $\beta$ to $0.04$.

\begin{table}[ht]
\begin{center}
 \begin{tabular}{|p{2.0cm}|p{3cm}|p{3cm}|}
\multicolumn{3}{c}
 {Numerical results for ${S_n\over S_1}$} \vspace{0.4cm}\\
\hline
\hspace{0.7cm}$\beta$ & \hspace{0.2cm}${S_n / S_1}$ at $\mathcal{O}(\beta)$  & \hspace{0.2cm}${S_n / S_1}$ at $\mathcal{O}(\beta^2)$  \\
\hline
\hspace{0.4cm}$0$ &\hspace{0.5cm}0.625272&\hspace{0.5cm}0.625272 \\
\hline
\hspace{0.4cm}$0.01$ &\hspace{0.5cm}0.617127&\hspace{0.5cm}0.617041 \\
\hline
\hspace{0.4cm}$0.02$ &\hspace{0.5cm}0.607545&\hspace{0.5cm}0.607182 \\
\hline
\hspace{0.4cm}$0.03$ &\hspace{0.5cm}0.596108&\hspace{0.5cm}0.59524 \\
\hline
\hspace{0.4cm}$0.04$ &\hspace{0.5cm}0.58222&\hspace{0.5cm}0.580575 \\
\hline
\end{tabular}
\caption{
Numerical results for ${S_n\over S_1}$ with different
values of the coupling $\beta$. The middle and the right part of the table show
results for ${S_n\over S_1}$ with $\mathcal{O}(\beta)$ and $\mathcal{O}(\beta^2)$ corrections, respectively.} 
\label{tableqc}
\end{center}
\end{table}
Here we should mention that the inequalities obeyed by R\'enyi entropy are satisfied in this regime of the coupling $\beta$ which are shown 
in figures \ref{SnbyS1ECG} and \ref{SecondInequalityECG}. 

Having studied ERE and its various features for Einsteinian cubic gravity, we will now move over to examples with gauge fields. Our next analysis
involves Born-Infeld gravity. 

\section{Charged R\'enyi Entropy with Born-Infeld Gravity}

We will start with the standard Einstein-Born-Infeld action in five dimensional AdS space which is given by,
\begin{equation}
S =  \frac{1}{ 2\ell _p ^3} \int d^5x \sqrt{-g} \left[ R + {12\over L^2}  
+ b^2 \ell _* ^2 \left( 1 - \sqrt{ 1 + \frac{ F^{\mu \nu}F_{\mu \nu}}{ 2 b^2}}\right) \right],
 \label{action}
\end{equation}
where, $b$ is the Born-Infeld parameter. $L$ is the AdS length, and is related to the cosmological constant $\Lambda$ by 
$\Lambda = -{6\over L^2}$. The factor $\ell _*$ in front of the 
Born-Infeld term is related to the five dimensional gauge coupling $g_5$ as, $g_5^2={ 2\ell _p ^3 \over \ell _*^2}$. As is well known, 
the Born-Infeld term reduces to the standard Maxwell term ($-{1\over 4}F_{\mu\nu}F^{\mu\nu}$), when we take the limit 
$b \rightarrow \infty$. On the other hand, with the limit $ b \rightarrow 0$, the Born-Infeld term vanishes and we are left with
Einstein's gravity with no gauge potential.

The variation of the action of eq.(\ref{action}) gives rise to the following Einstein's and Maxwell's equations:
\begin{eqnarray}
R_{\mu\nu}-{1\over2}g_{\mu\nu}R-{6\over L^2}g_{\mu\nu}=
{\ell _* ^2 \over 2}\left[g_{\mu\nu}b^2( 1 - \sqrt{ 1 + \frac{ F_{\rho \lambda}F^{\rho \lambda}}{ 2 b^2}})+
\frac{F_{\mu\alpha}F_{\nu}^{\alpha}}{\sqrt{ 1 + \frac{ F_{\rho \lambda}F^{\rho \lambda}}{ 2 b^2}}}\right] \,,
\label{EinsteinEOM}
\end{eqnarray}
\begin{eqnarray}
\partial_{\mu}\left
 (\frac{\sqrt{-g}F^{\mu\nu}}{\sqrt{1+\frac{F_{\mu\nu}F^{\mu\nu}}{2 b^{2}}}}\right
 )=0 \,.
 \label{MaxwellEOM}
\end{eqnarray}
The authors of \cite{Tanay} and particularly \cite{Cai} studied the black hole solutions in these theories, having horizons 
with positive(elliptic), zero (planar) and negative (hyperbolic) constant curvatures. Since we are interested here computing the 
charged R\'enyi entropy holographically, we only need to consider the charged hyperbolic black holes. The 
hyperbolic black hole solution of the Einstein-Born-Infeld gravity is given by,
 \begin{eqnarray}
 ds^2&=&-\bigg({r^2\over L^2}f(r)-1\bigg) \frac{L^2}{\mathcal{R}^2} dt^2 +{dr^2 \over \big({r^2\over L^2}f(r)-1\big)} + r^2 \ d\Sigma_{3} ^{2}  \,,
 \label{metric}
\end{eqnarray}
with the gauge field given by the expression
\begin{equation}
 A = \phi(r) \ dt \,.
 \label{gauge field}
\end{equation}
Here, $d\Sigma_{3} ^{2}$ is the line element for the three dimensional hyperbolic plane $H^3$. The function $f(r)$ 
has the form
\begin{eqnarray}
 f(r) = 1 - \frac{L^2 m}{3 r^4}  + \frac{b^2 \ell _* ^2 r^2}{12} \left( 1 - \sqrt{ 1 + \frac{q^2}{b^2 \ell _* ^2 r^6}} \right) 
 + \frac{ L^2 q^2}{ 8 r^6} \hspace{0.2cm} _2 {\cal F}_1 \left( \frac{1}{3}, \frac{1}{2}, \frac{4}{3}, -\frac{q^2}{ b^2 \ell _* ^2 r^6}\right)
\end{eqnarray}
where $m$ and $q$ are constants of integration and $_2 {\cal F}_1$ is the standard hypergeometric function. The constants $m$ and $q$
are related to the black hole mass $M$ and electric charge $Q$, respectively. Also, the position of the horizon $r_h$ is given by
$f(r_h)={L^2\over r_h^2}$. Further, the gauge field $\phi(r)$ is given by,
\begin{equation}
 \phi(r) = \frac{L \ q}{2 \ell _* \mathcal{R} \ r^2} \ 
 _2 {\cal F}_1 \left( \frac{1}{3}, \frac{1}{2}, \frac{4}{3}, -\frac{q^2}{ b^2 \ell _* ^2 r^6} \right) - {\mu_c \over 2\pi \mathcal{R}} \,,
 \label{gauge potential}
\end{equation}
where, $\mu_c$ is the chemical potential, given by 
\begin{eqnarray}
 \mu_c = \frac{\pi L \ q}{ \ell _* \ r_h^2} \ 
 _2 {\cal F}_1 \left( \frac{1}{3}, \frac{1}{2}, \frac{4}{3}, -\frac{q^2}{ b^2 \ell _* ^2 r_h^6} \right) \,.
 \label{chemical potential}
\end{eqnarray}

Notice that $\mu_c$ is chosen in such a way that the gauge field $\phi(r)$ vanishes at the horizon $r_h$ in order to prevent 
the appearance of a conical singularity. Another important point to note is that, compared to \cite{Cai}, we have inserted an extra factor of
$L^2\over \mathcal{R} ^2$ in the metric component $g_{tt}$. With this inclusion, we make sure that the boundary CFT dual to this black hole
geometry resides on $R \times H^3 $ where the hyperbolic spatial slice $H^3$ has curvature $\mathcal{R}$,
\begin{eqnarray}
 ds_{\infty}^2=-dt^2 + \mathcal{R}^2 \ d\Sigma_{3} ^{2} \,.
\end{eqnarray}
One can also compute the mass parameter $m$ in terms of the horizon radius $r_h$ from, $f(r_h)={L^2\over r_h^2}$ as
\begin{eqnarray}
 m &= \frac{3 r_h^4 }{L^2}-3r_h^2 + \frac{b^2 \ell _* ^2 r_h^4}{4} \left( 1 - \sqrt{ 1 + \frac{q^2}{b^2 \ell _* ^2 r_h^6}} \right) + \frac{ 3 q^2}{ 8 r_h^2} \hspace{0.2cm} _2 {\cal F}_1 \left( \frac{1}{3}, \frac{1}{2}, \frac{4}{3}, -\frac{q^2}{ b^2 \ell _* ^2 r_h^6}\right) \,.
\end{eqnarray}
The Hawking temperature of the black hole can be computed as,
\begin{eqnarray}
 T={\kappa \over 2\pi}={1\over 2\pi r_h}\big({L\over \mathcal{R}}+{r_h^3 \ f'(r_h)\over 2 \ L  \mathcal{R}}\big) \,,
 \label{BITemperature}
\end{eqnarray}
where $\kappa$ is the surface gravity of the black hole.
The black hole entropy computed using the Bekenstein-Hawking formula reads 
\begin{eqnarray}
 S=2\pi \bigg({r_h \over  \ell _p}\bigg)^3 V_{\Sigma} \,,
\end{eqnarray}
As was done in the previous section, we now express the chemical potential, temperature
and entropy of the black hole in terms of $x$ as,
\begin{eqnarray}
 \mu_c = \frac{\pi L \ q}{ \ell _* \ L^2 x^2} \ 
 _2 {\cal F}_1 \left( \frac{1}{3}, \frac{1}{2}, \frac{4}{3}, -\frac{q^2}{ b^2 \ell _* ^2 L^6 \ x^6} \right) \,.
 \label{chemicalpotentialnew}
\end{eqnarray}
\begin{eqnarray}
 T=\frac{2 x^2-1}{2 \pi  x \mathcal{R}} + \frac{b^2 \ell _*^2 L^2 }{12 \pi  \mathcal{R}} x 
 \bigg[1-\sqrt{1+{q^2\over b^2 \ell _*^2 L^2 x^6 }}\bigg] \,, 
\label{BITemperatureNew}
 \end{eqnarray}
\begin{eqnarray}
 S=2 \pi \bigg({L \over \ell _p}\bigg)^3 x^3 \ V_{\Sigma }\,. 
\end{eqnarray}

Here we should mention that since we want to compute the R\'enyi entropy in the grand canonical ensemble,
we should fix the chemical 
potential $\mu_c$. For this purpose, we need to express $q$ in terms of $\mu_c$ and substitute that in eq.(\ref{BITemperatureNew}) to
have an expression of temperature in terms of $\mu_c$. But from eq.(\ref{chemicalpotentialnew}), it is difficult to analytically express 
$q$ in terms of $\mu_c$ (this is because of the  factor of ${q^2\over x^6}$ inside the hypergeometric 
function). Hence, we numerically evaluate the R\'enyi entropy at a fixed chemical potential. 

At this point, we briefly describe the numerical routine for computing the R\'enyi entropy. First, we fix the chemical potential $\mu_c$ to a 
certain value and numerically solve eq.(\ref{chemicalpotentialnew}) to generate $q(x)$. Note that the entropy $S$ does not explicitly
depend on $q$, while the temperature $T$ does. But once we numerically generate the function $q(x)$, we can rewrite the function 
$T(q,x)$ as $T(x)$. In the process of computing the R\'enyi entropy $S_n$, we can express eq.(\ref{Sneqnmuc}) in terms of the dimensionless variable $x$ such that it can be rewritten in a form similar to eq.(\ref{RenyiEntropyFormula}) as,
\begin{eqnarray}
 S_n(\mu_c) &=& \frac{n}{n-1} \frac{1}{T_0}\, \int_{x_n}^{x_1} S(\mu_c,\, x)\,\frac{dT(\mu_c,\, x)}{dx}\,dx \nonumber \\
&=& \frac{n}{n-1} \frac{1}{T_0}\,\left[S(\mu_c,\, x)\,T(\mu_c,\, x)|^{x_1}_{x_n} -\int_{x_n}^{x_1} \frac{dS(\mu_c,\, x)}{dx}\,T(\mu_c,\, x)\,dx \right] \,.
\label{RenyiEntropyFormulamuc}
\end{eqnarray}
where $x_1$ and $x_n$ are the integration limits to be determined . Since we already have the function $q(x)$ we can straightforwardly 
determine those quantities by numerically solving the following equations,
\begin{eqnarray}
 T_0={1\over 2\pi \mathcal{R}}=\frac{2 x_1^2-1}{2 \pi  x_1 \mathcal{R}} + \frac{b^2 \ell _*^2 L^2 }{12 \pi  \mathcal{R}} x_1 
 \bigg[1-\sqrt{1+{q^2\over b^2 \ell _*^2 L^2 x_1^6 }} \bigg] \,,
\end{eqnarray}
\begin{eqnarray}
 {T_0\over n}={1\over 2\pi \mathcal{R} \, n}=\frac{2 x_n^2-1}{2 \pi  x_n \mathcal{R}} + \frac{b^2 \ell _*^2 L^2 }{12 \pi  \mathcal{R}} x_n 
 \bigg[1-\sqrt{1+{q^2\over b^2 \ell _*^2 L^2 x_n^6 }} \bigg] \,.
\end{eqnarray}

Now we are ready to compute the R\'enyi entropy $S_n$. Since, $S_n$ contains the volume factor $V_{\Sigma}$ involving
the short-distance cut-off of the boundary CFT, we plot the ratio of R\'enyi entropy $S_n$ to the entanglement entropy $S_1$,
instead of plotting $S_n$, as before. 
 \begin{figure}[h!]
 \centering
 \subfigure[$b=0.05$]{
 \includegraphics[width=2.5in,height=2.3in]{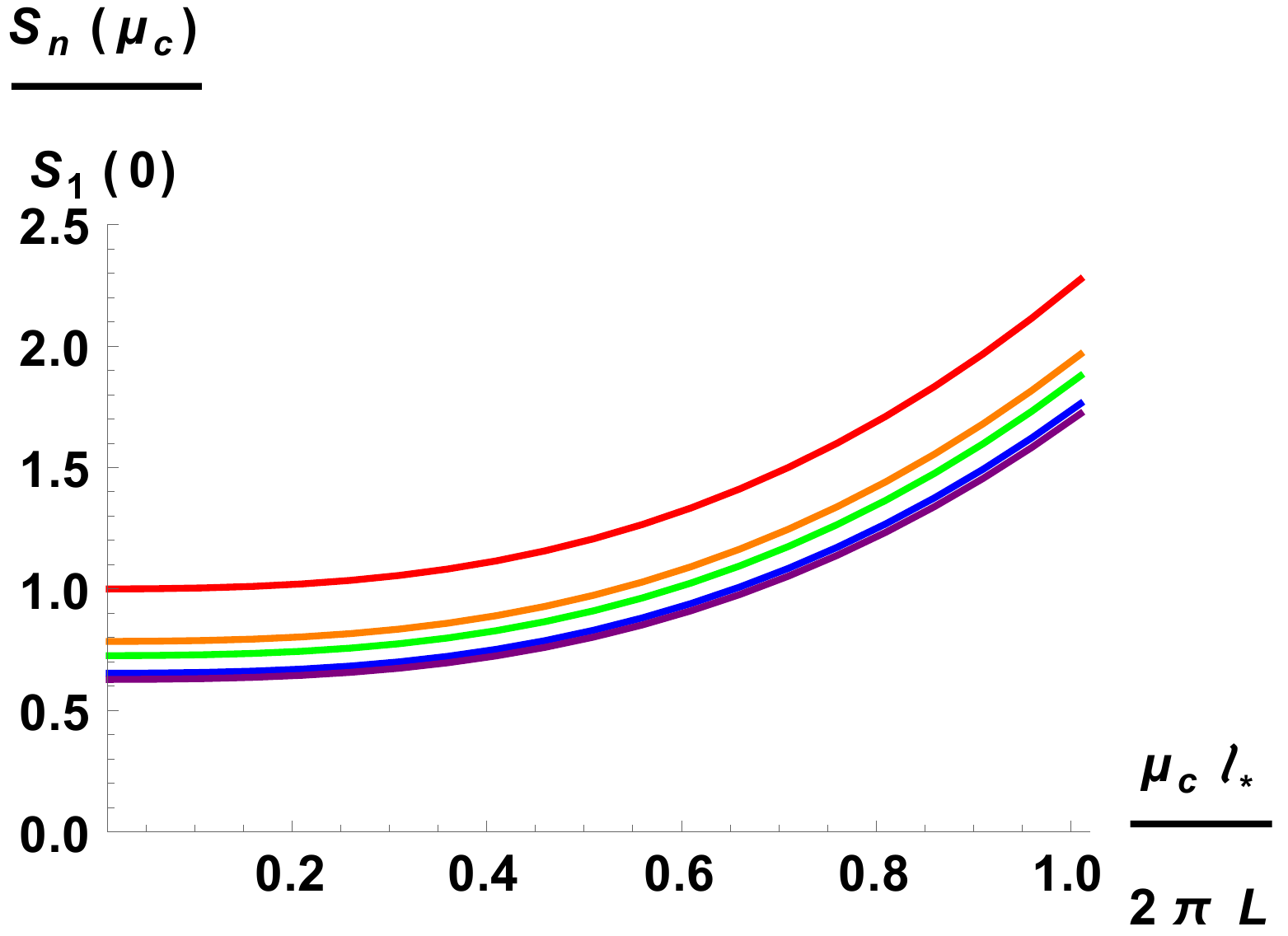}
 \label{b0Pt05Sqvsmuc} } 
 \subfigure[$b=0.1$]{
 \includegraphics[width=2.5in,height=2.3in]{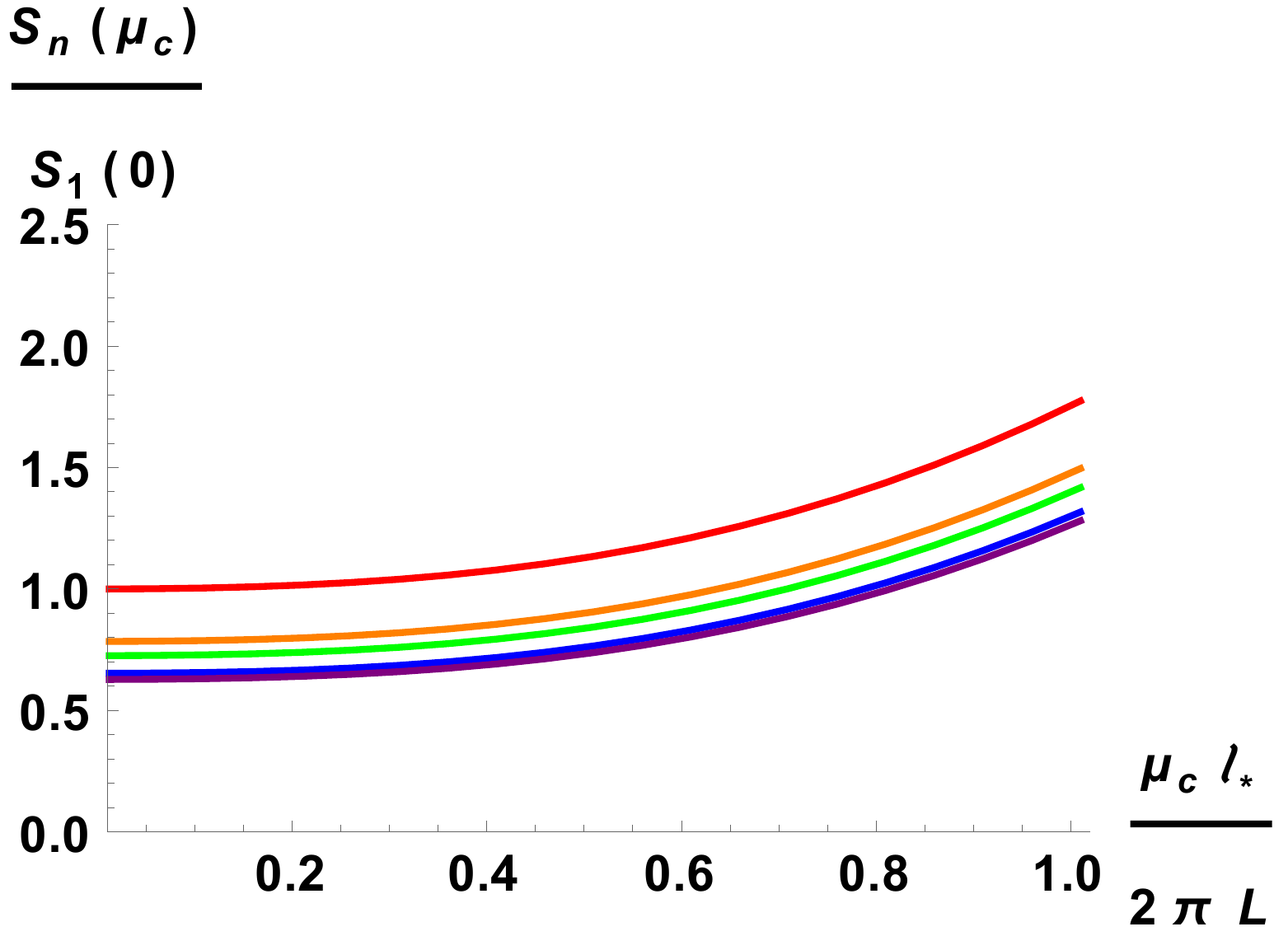}
 \label{b0Pt1Sqvsmuc} }
  \subfigure[$b=0.2$]{
  \includegraphics[width=2.5in,height=2.3in]{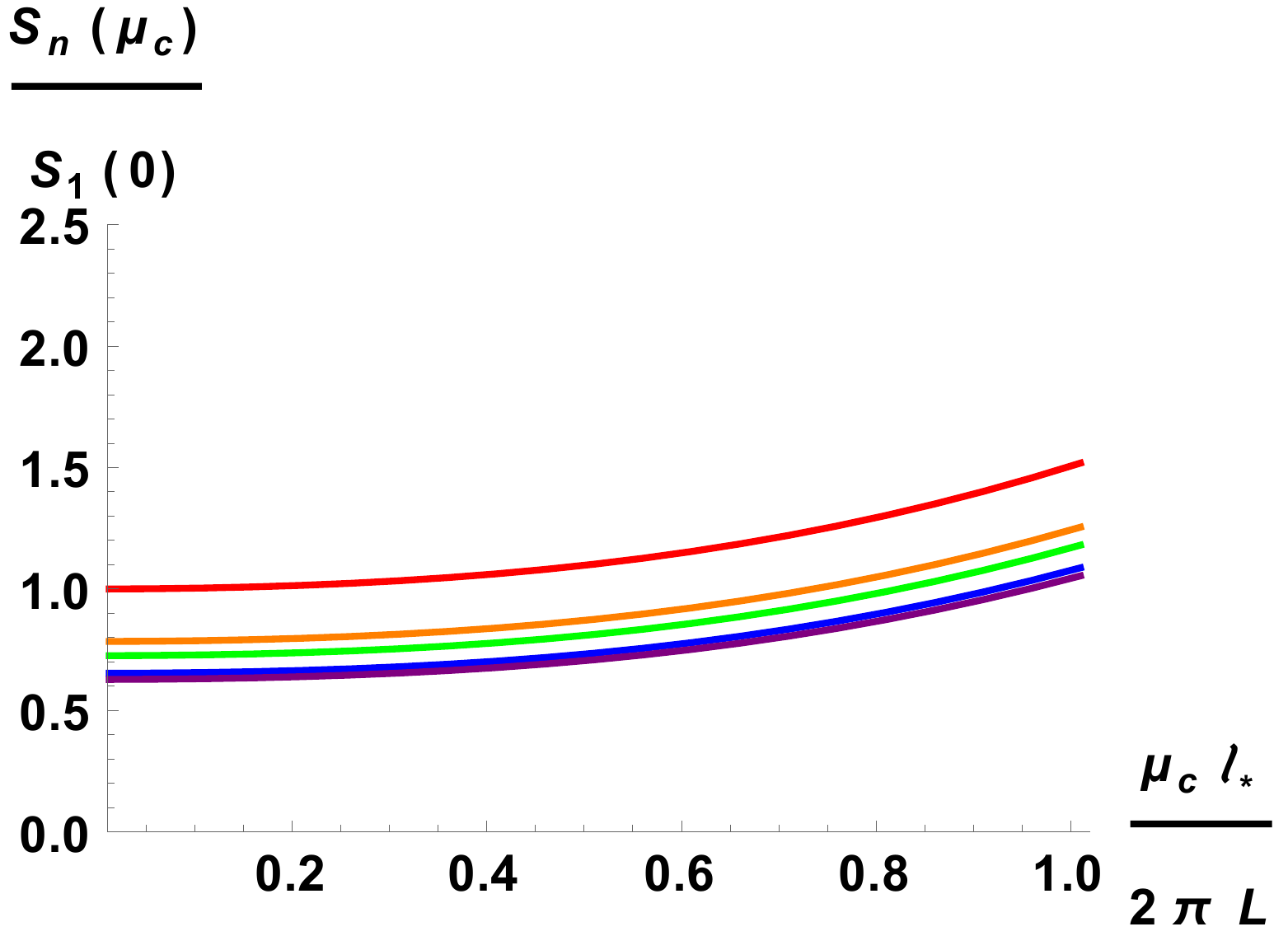}
  \label{b0Pt2Sqvsmuc} } 
  \subfigure[$b=0.5$]{
  \includegraphics[width=2.5in,height=2.3in]{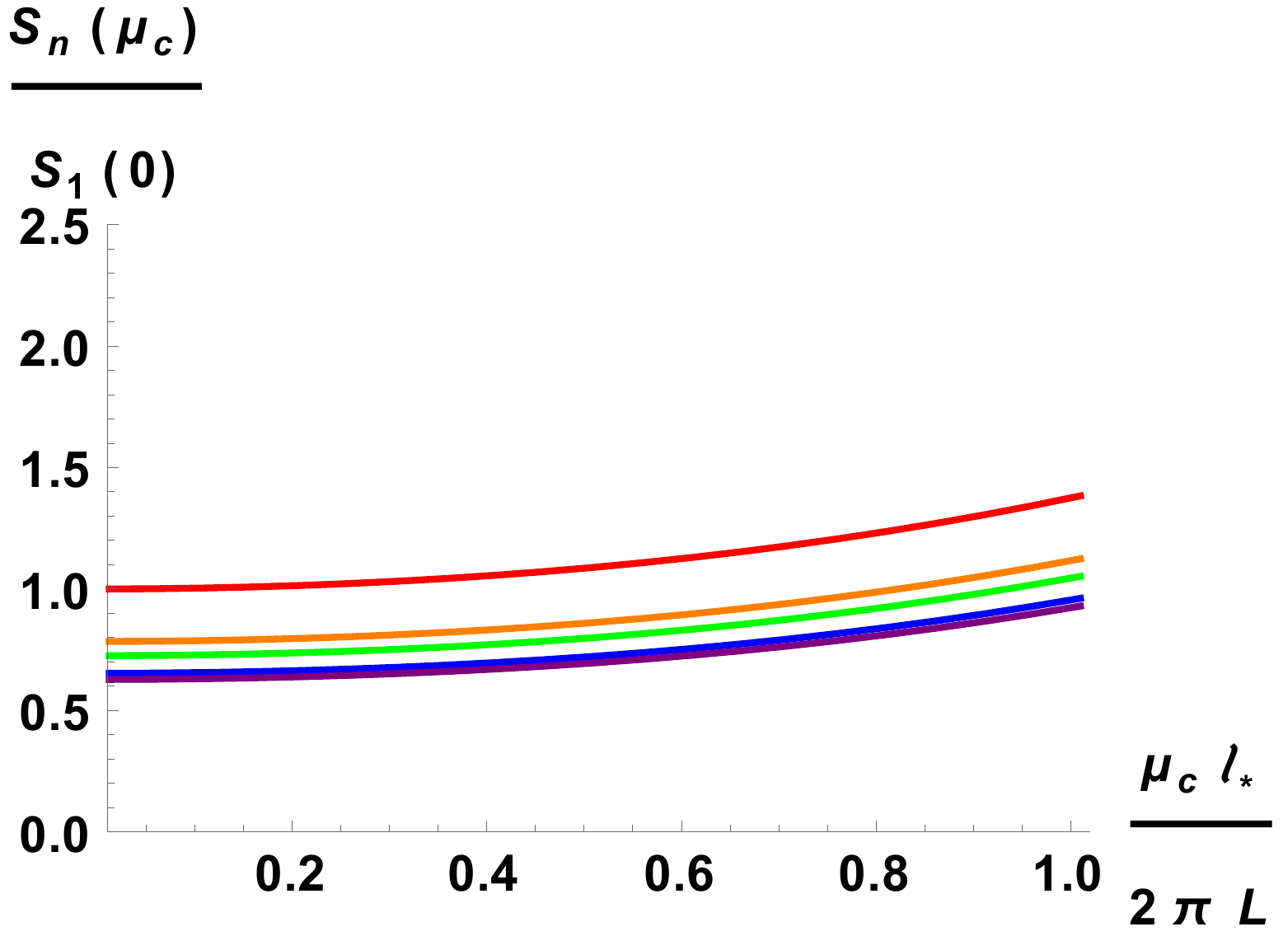}
  \label{b0Pt5Sqvsmuc} }
  \caption{\small{R\'enyi entropy $S_n(\mu_c)$ normalized by the entanglement entropy $S_1(0)$ is plotted as a function of the 
  chemical potential for different values of the Born-Infeld parameter $b$. In each subfigure the red, orange, green, blue and 
  purple lines (from top to bottom) denote $n=1$, $2$, $3$, $10$ and $100$ respectively.}}
 \label{SqvsmucDifferentb}
 \end{figure}

\subsection{Numerical Analysis and Results}

\begin{figure}[h!]
 \centering
 \subfigure[${\mu_c \ell _*\over 2\pi L}=0.75$]{
 \includegraphics[width=2.5in,height=2.3in]{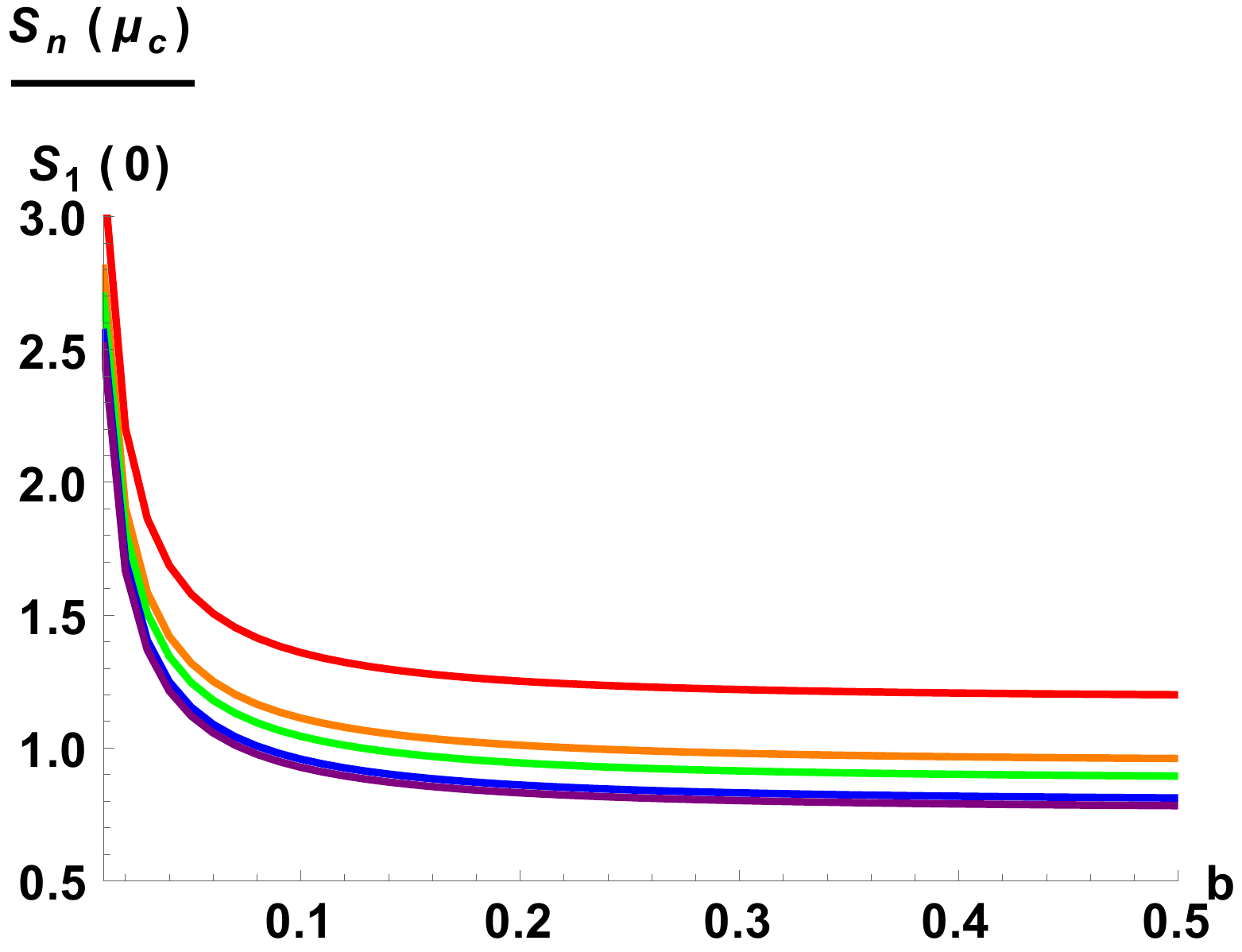}
 \label{ChemicalPotential0Pt75} } 
 \subfigure[${\mu_c \ell _*\over 2\pi L}=1$]{
 \includegraphics[width=2.5in,height=2.3in]{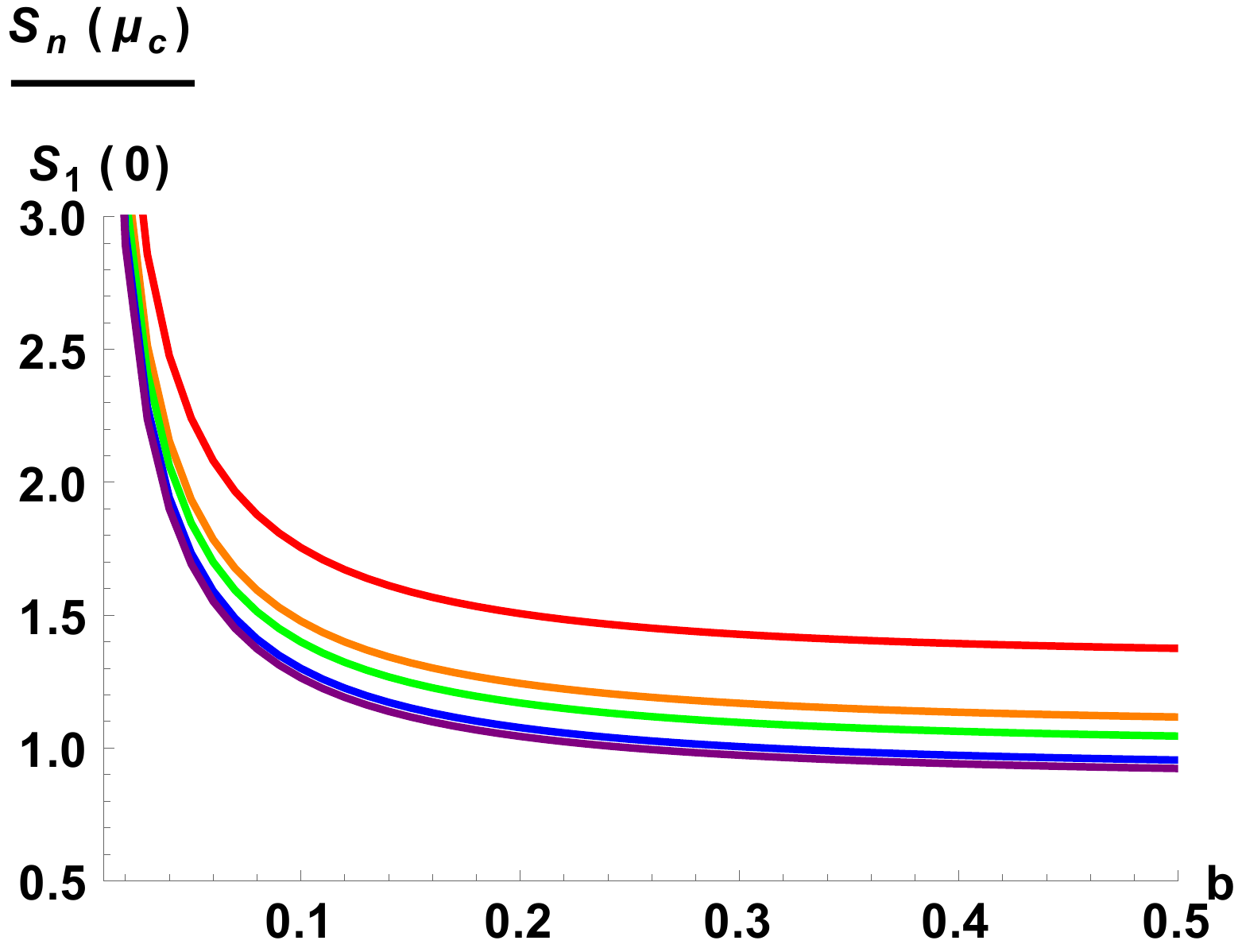}
 \label{ChemicalPotential1} }
  \caption{\small{R\'enyi entropy $S_n(\mu_c)$ normalized by the entanglement entropy $S_1(0)$ is plotted as a function of the 
  Born-Infeld parameter b for different values of the chemical potetial $\mu_c$. The red, orange, green, blue and purple lines (from
  top to bottom) denote $n=1$, $2$, $3$, $10$ and $100$ respectively.}}
 \label{SqvsbDifferentmuc}
 \end{figure}

In figure \ref{SqvsmucDifferentb}, we have shown the behavior of the R\'enyi entropy as a function of the chemical potential for 
different values of the Born-Infeld parameter $b$. In each figure, the red, orange, green, blue and purple lines correspond to
$n=1$, $2$, $3$, $10$ and $100$, respectively. The ratio of the R\'enyi entropy to entanglement entropy
${S_n(\mu_c)\over S_1(0)}$ always increases in a non-linear fashion as the chemical potential $\mu_c$ increases. Also, for 
any value of the Born-Infeld parameter $b$, the  R\'enyi entropy decreases as one increases the value of $n$. Note that, 
for a particular value of $n$ and $\mu_c$, ${S_n(\mu_c)\over S_1(0)}$ decreases as $b$ increases. This behavior is also
evident in figure \ref{SqvsbDifferentmuc} where we have plotted ${S_n(\mu_c)\over S_1(0)}$ as a function of the 
Born-Infeld parameter at fixed chemical potential. 

\begin{figure}[h!]
 \centering
 \subfigure[$b=0.05$]{
 \includegraphics[width=2.5in,height=2.3in]{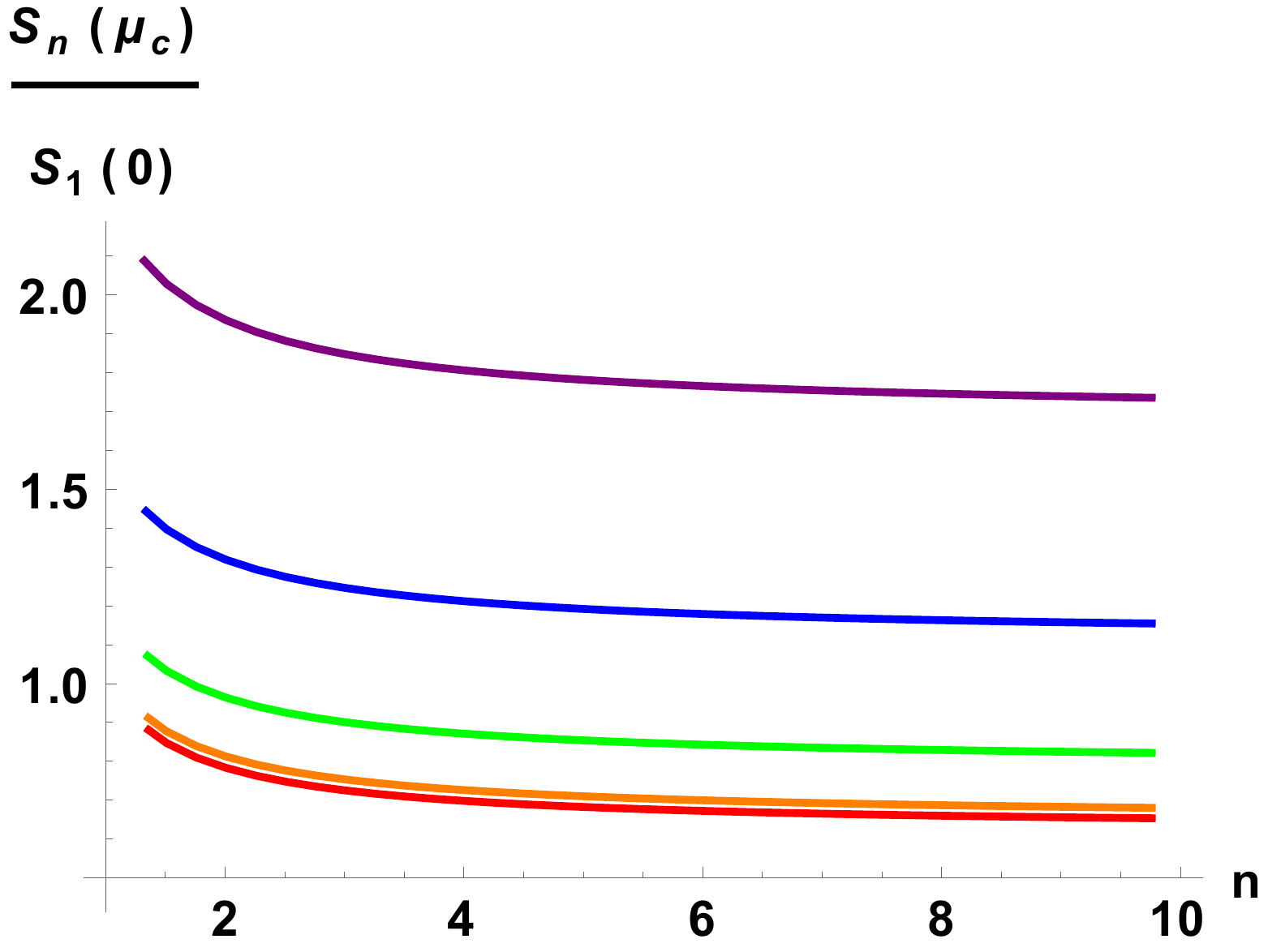}
 \label{FirstInequalityb0Pt05} } 
 \subfigure[${\mu_c \ell _*\over 2\pi L}=1$]{
 \includegraphics[width=2.5in,height=2.3in]{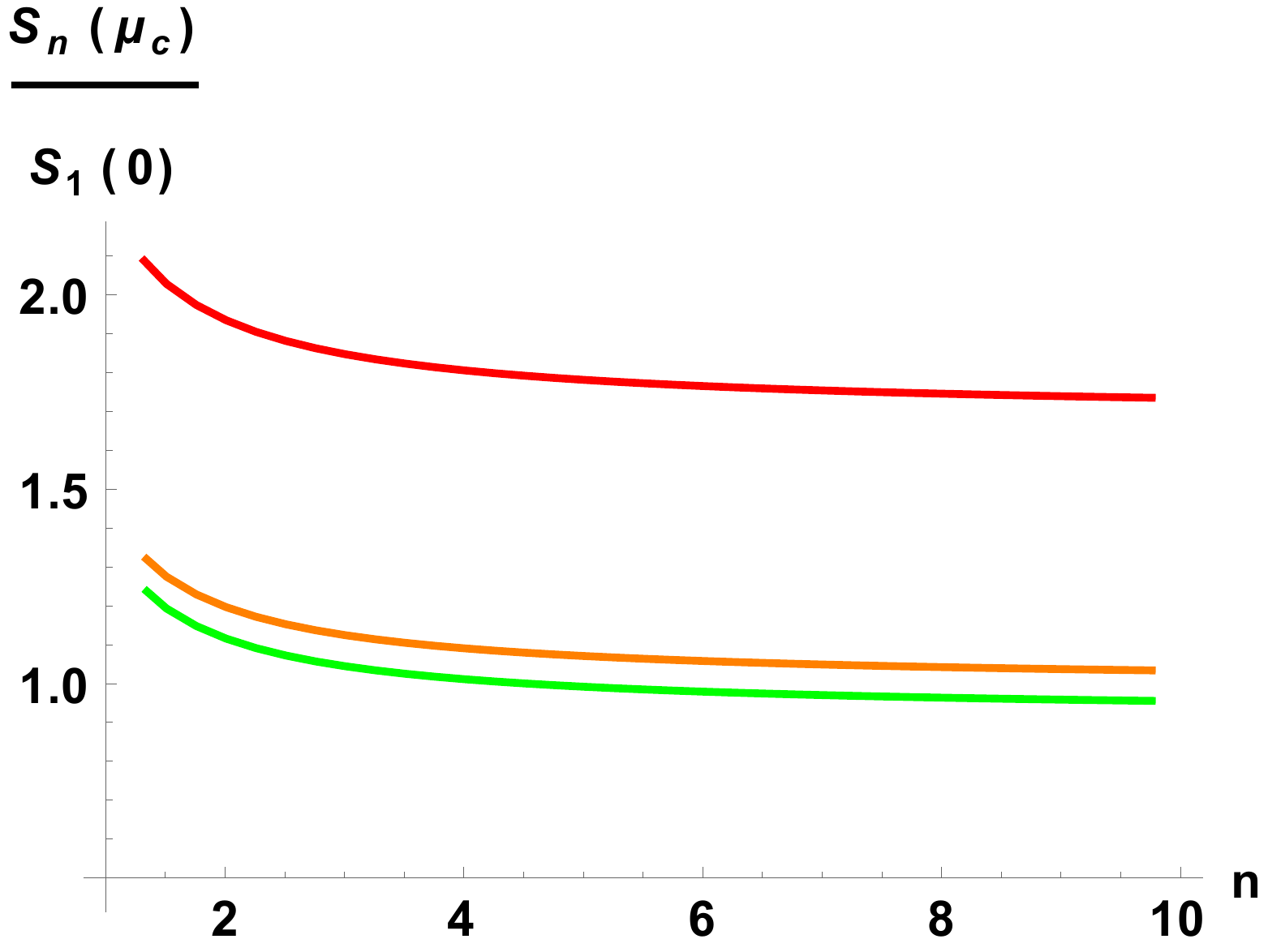}
 \label{FirstInequalityTestDifferentb} }
  \caption{\small {In (a) R\'enyi entropy $S_n(\mu_c)$ normalized by the entanglement entropy $S_1(0)$ is plotted as a function of 
  $n$ for fixed value of $b=0.05$. The red, orange, green, blue and purple lines (from bottom to top) denote 
  ${\mu_c \ell _*\over 2\pi L}=0.01$, $0.25$, $0.5$, $0.75$ and $1$ respectively. In (b) we have fixed ${\mu_c \ell _*\over 2\pi L}=1$
  and the red, orange and green curves correspond to $b=0.05$, $0.25$ and $0.5$, respectively.}}
 \label{FirstInequalityTest}
 \end{figure}

Next, in figure \ref{FirstInequalityTest} we plot the  R\'enyi entropy as a function of $n$. In 
figure \ref{FirstInequalityb0Pt05} we have fixed the Born-Infeld parameter $b=0.05$ while the red, orange, 
green, blue and purple lines (from bottom to top) correspond to ${\mu_c \ell _*\over 2\pi L}=0.01$, $0.25$, $0.5$, $0.75$ and $1$ 
respectively. In figure \ref{FirstInequalityTestDifferentb} we have fixed the chemical potential at ${\mu_c \ell _*\over 2\pi L}=1$
where the red, orange and green lines denote $b=0.05$, $0.25$ and $0.5$, respectively. In each figure it is evident that
the R\'enyi entropy has a negative slope with $n$, i.e., $\frac{\partial S_n(\mu_c)}{\partial n} \leq 0$.
 \begin{figure}[h!]
 \centering
 \subfigure[$b=0.05$]{
 \includegraphics[width=2.5in,height=2.3in]{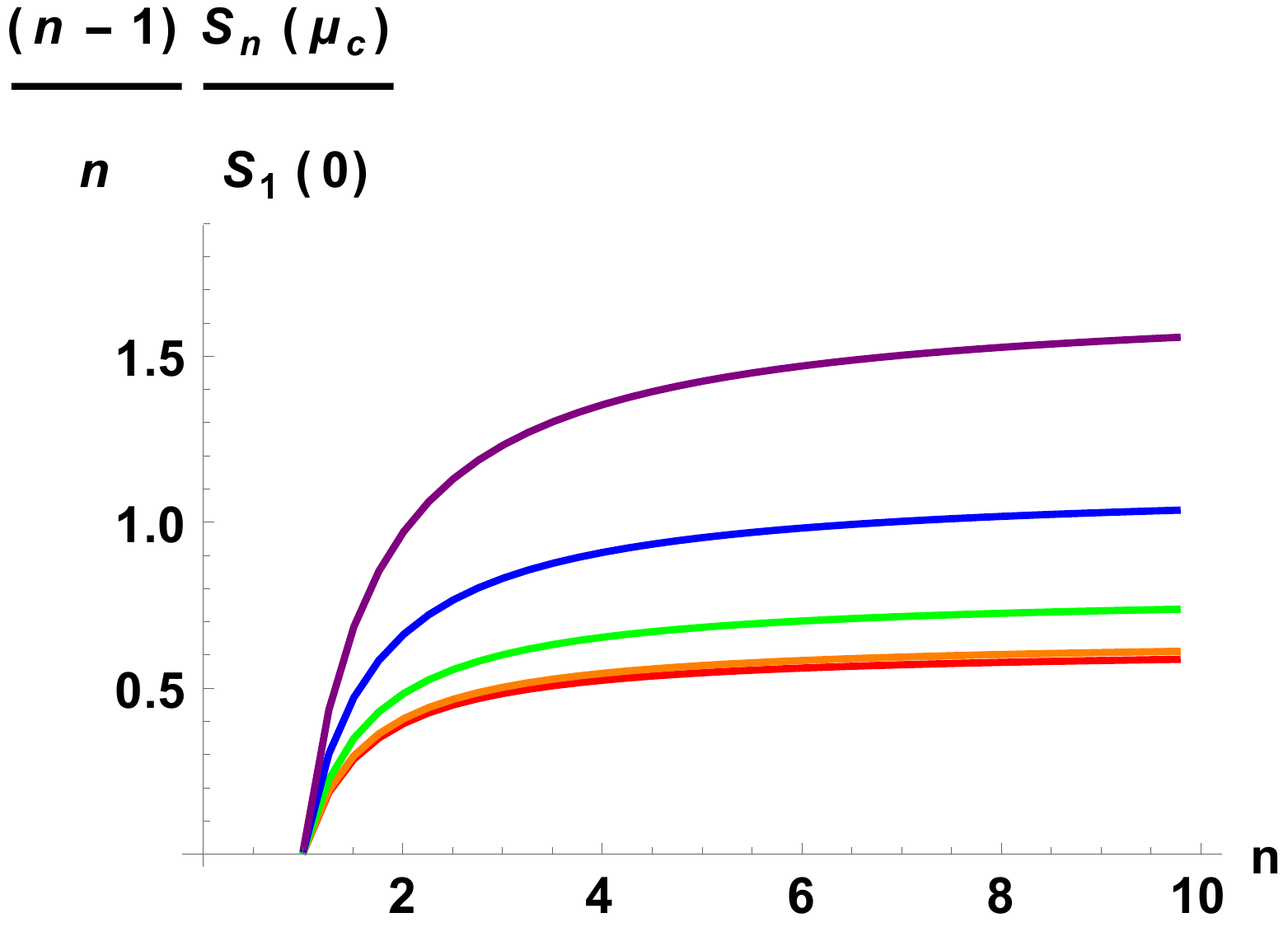}
 \label{SecondInequalityb0Pt05} } 
 \subfigure[${\mu_c \ell _*\over 2\pi L}=1$]{
 \includegraphics[width=2.5in,height=2.3in]{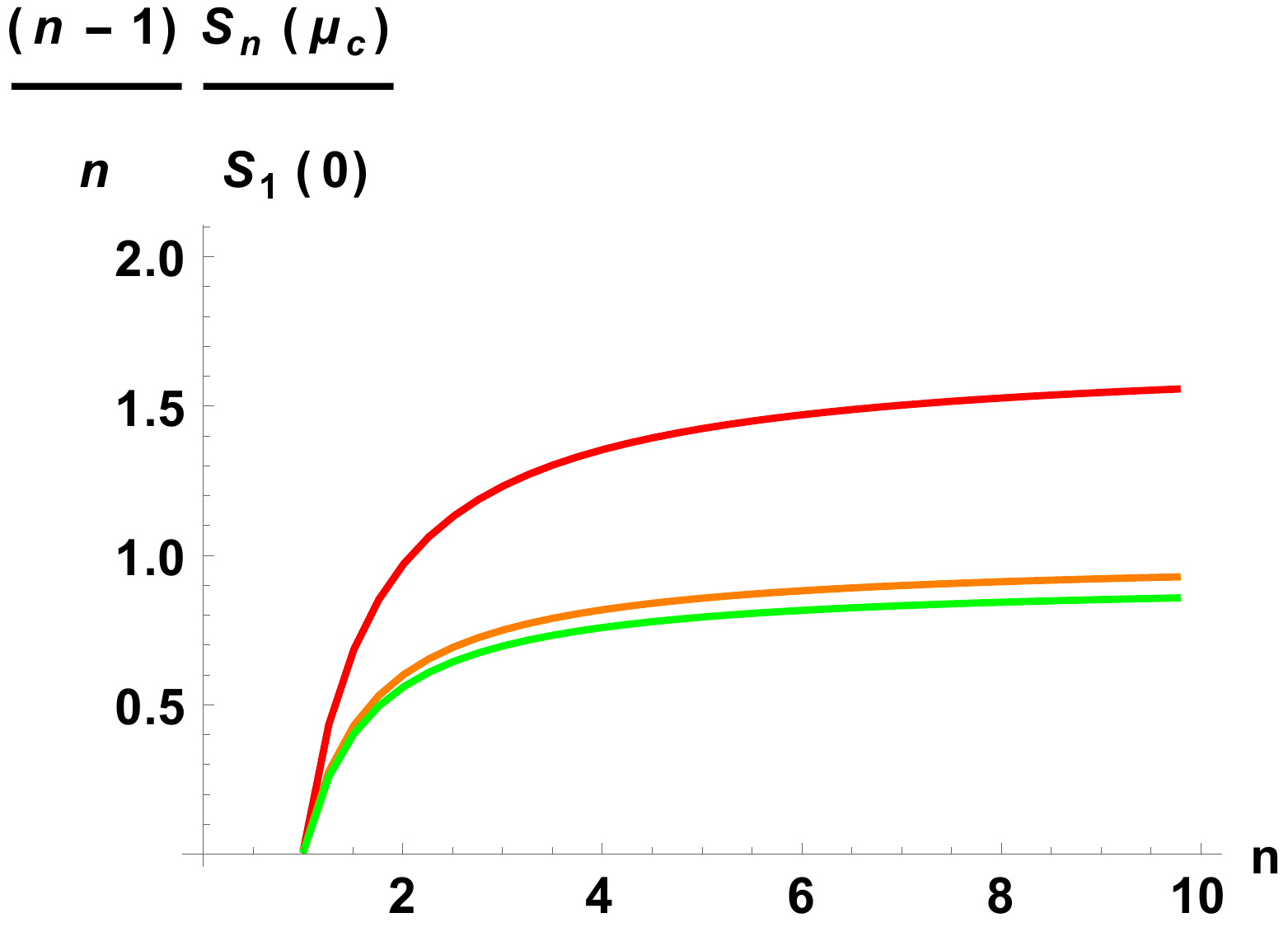}
 \label{SecondInequalityTestDifferentb} }
  \caption{\small{Second Inequality Test for R\'enyi entropy for Born-Infeld gravity. The same color coding as in figure \ref{FirstInequalityTest} has been used.}}
 \label{SecondInequalityTest}
 \end{figure}

Hence, we notice that the first inequality of eq.(\ref{EREinequalities}) is obeyed for any value of the Born-Infeld parameter in our case. In figure 
 \ref{SecondInequalityTest} we test the second inequality of eq.(\ref{EREinequalities}), which indicates that the quantity $\frac{n-1}{n} S_n(\mu_c) $ 
 should have a positive slope with $n$. We have shown this behavior with different values of chemical potential at fixed $b=0.05$ in
 figure \ref{SecondInequalityb0Pt05}, while in figure \ref{SecondInequalityTestDifferentb}, we have fixed the chemical potential
 at ${\mu_c \ell _*\over 2\pi L}=1$ and taken $b$ as a parameter. In both cases we got positive slope $\frac{n-1}{n} S_n(\mu_c) $
 with $n$, i.e., $\frac{\partial}{\partial n}\left(\frac{n-1}{n} S_n \right) \geq  0$.

\subsection{Scaling Dimension of Twist Operators}

Following the same method as described in section 2, we compute the scaling dimension of the twist operators in a CFT dual to the Born-Infeld gravity. 
Recall that the scaling dimension is given by,
\begin{eqnarray}
 h_n(\mu_c)={n\over 3}{\mathcal{R}^3\over T_0} \big(\mathcal{E}(T_0,\, 0)-\mathcal{E}({T_0\over n},\, \mu_c)\big)
 \label{hnformula1}
\end{eqnarray}
where, $\mathcal{E}$ is the energy density given by,
\begin{equation}
\mathcal{E} = \frac{m}{2 \ell_p^{3} \mathcal{R}^{3}}
\label{epsilon}
\end{equation}
Here, the mass $m$ is given by,
\begin{equation}
m(\frac{T_0}{n},\mu_c) = \frac{r_h^4}{L^2}-3 r_h^2 + \frac{b^2 \ell_{*}^2 r_h^4}{4} \bigg(1-\sqrt{1+\frac{q^2}{b^2 \ell_{*}^2 r_h^6}} \bigg) + \frac{3 q^2}{8 r_h^2} \ {}_2 \mathcal{F}_1 (\frac{1}{3},\frac{1}{2},\frac{4}{3},-\frac{q^2}{b^2 \ell_{*}^2 r_h^6})
\label{mBI}
\end{equation}
Since $m(T_0,0)=0$, finally we have the expression of the scaling dimension,
\begin{equation}
h_n = -\frac{n \pi L}{3\ell_p^3} m(\frac{T_0}{n},\mu_c)
\end{equation}
However, from eq.(\ref{chemicalpotentialnew}) it is evident that expressing $q$ in terms of $\mu_c$ is difficult because of the presence of the hypergeometric 
function. Hence, it is difficult to obtain an analytical expression of the expansion of the scaling dimension, i.e $h_{10}$ or $h_{02}$ are difficult to calculate. 
The same difficulty also holds for the calculation of the coefficients of magnetic response.  Of course one could have considered various limits of
the parameters in order to (perturbatively) expand the hypergeometric function, but such an analysis may be of limited interest and we will not belabor 
upon this here. Alternatively, it is possible to do the computation in a canonical (fixed charge) ensemble. However we will not undertake such an
analysis here. 

We now move to our second example of ERE with a gauge field, namely charged quasi-topological gravity. 

\section{Charged R\'enyi Entropy with Quasi-Topological Gravity}

In this section, we will discuss the properties of the charged  R\'enyi Entropy of a four dimensional CFT, dual to five dimensional charged
quasi-topological gravity (QTG). The action of the five dimensional QTG coupled to a $U(1)$ gauge field $A$ is given by,
\begin{eqnarray}
 S = \frac{1}{2\ell_p^3} \int \mathrm{d}^5x \, \sqrt{-g}\, \left[R+\frac{12}{L^2} 
 + \frac{\lambda }{2}L^2 \mathcal{\chi}_4 +\frac{7\mu}{4}L^4\mathcal{Z}_5-{\ell_*^2\over 4} F_{\mu\nu}F^{\mu\nu} \right]\, ,
 \label{QTGaction}
\end{eqnarray}
where $\mathcal{\chi}_4$ is the standard four-derivative Gauss-Bonnet term,
\begin{eqnarray}
 \mathcal{\chi}_4 = R_{\mu\nu\rho\lambda}R^{\mu\nu\rho\lambda}-4R_{\mu\nu}R^{\mu\nu}+R^2\, ,
 \label{GBterm}
\end{eqnarray}
while $\mathcal{Z}_5$ is a six-derivative interaction term \cite{Robinson},\cite{Sinha} given by,
 \begin{eqnarray}
&& \mathcal{Z}_5=   \tensor{R}{_\mu ^\rho _\nu ^\lambda} \tensor{R}{_\rho ^\alpha _\lambda ^\beta} \tensor{R}{_\alpha ^\mu _\beta ^\nu}  
+{1\over 56}(21 R_{\mu\nu\rho\lambda}R^{\mu\nu\rho\lambda}R-72 \tensor{R}{_{\mu\nu\rho\lambda}}\tensor{R}{^{\mu\nu\rho} _\alpha} \tensor{R}{^{\lambda \alpha}} \nonumber \\
&& +120 R_{\mu\nu\rho\lambda} R^{\mu\rho} R^{\nu\lambda}+144 \tensor{R}{_\mu ^\nu} \tensor{R}{_\nu ^\rho} \tensor{R}{_\rho ^\mu} 
-132 \tensor{R}{_\mu  ^\nu} \tensor{R}{_\nu ^\mu} R +15R^3 )\, .
\label{QTGterm}
\end{eqnarray}

In order to compute the holographic charged R\'enyi Entropy dual to this bulk gravity theory, we need to construct charged hyperbolic black hole 
solutions in this \cite{Mann}. This can be done by choosing the following metric and gauge field ansatz,
 \begin{eqnarray}
 ds^2&=&-\bigg({r^2\over L^2}f(r)-1\bigg) \mathcal{N}(r)^2 dt^2 +{dr^2 \over \bigg({r^2\over L^2}f(r)-1\bigg)} + r^2 \ d\Sigma_{3} ^{2}  \,, \nonumber \\
 \label{metricQTG}
 A &=& \phi(r) \ dt \,.
\end{eqnarray}
Substituting the ansatz of eq.(\ref{metricQTG}) in the action of eq.(\ref{QTGaction}), and varying the action with respect to $f(r)$, 
we have $ \mathcal{N}'(r)=0$, yielding 
\begin{eqnarray}
 \mathcal{N}(r)=\frac{\tilde{L}}{\mathcal{R}}\, ,
 \label{NeomQTG}
\end{eqnarray}
where $\tilde{L}$ is the AdS curvature scale to be determined. Also, note that we have chosen the ${1\over \mathcal{R}^2}$ factor in the same 
spirit as we did earlier.
Now varying the action with respect to $\phi(r)$ and solving that equation, we get 
\begin{eqnarray}
\phi(r)=\frac{q \tilde{L}}{2 r^2 \ell _* \mathcal{R}}-\frac{\mu_c}{2\pi \mathcal{R}} 
\label{phiQTG}
\end{eqnarray}
Here, the chemical potential is given by 
\begin{eqnarray} 
\mu_c=\frac{\pi \tilde{L} q}{\ell _* r_h^2}\, . 
\label{mucQTG}
\end{eqnarray}
In the above expression, $q$ is an integration constant, related to the charge of the black hole and $r_h$ is the position of the horizon. Note that,
we have again chosen the chemical potential $\mu_c$ in such a way that the gauge field $\phi(r)$ vanishes at the horizon.

The variation of the action with respect to $\mathcal{N}(r)$ yields a first order differential equation for $f(r)$, solving which we get a cubic 
equation for $f(r)$,
\begin{eqnarray}
 1-f(r)+\lambda f(r)^2+\mu f(r)^3={L^2 m \over 3 r^4}-{L^2 q^2\over 12 r^6}\, ,
 \label{fQTG}
\end{eqnarray}
where $m$ is an integration constant, related to the mass of the black hole. 
Using the fact that $f(r_h)={L^2\over r_h^2}$, one can also express the mass parameter in terms of the horizon radius $r_h$ from 
eq.(\ref{fQTG}),
\begin{eqnarray}
 m &=& \frac{3 r_h^4}{L^2}-3 r_h^2+\frac{q^2}{4 r_h^2}+3 \lambda L^2+\frac{3 \mu  L^4}{r_h^2}\,.
 \label{mQTG}
\end{eqnarray}
Here we should mention that unlike the Einstein gravity, the AdS curvature scale $\tilde{L}$ is not equal to the length scale $L$ 
related to the cosmological constant in QTG. These two are related by $\tilde{L}={L\over \sqrt{f_{\infty}}}$, where
$f_{\infty}$ is the asymptotic value of $f(r)$ and can be determined by solving eq.(\ref{fQTG}) at $r\rightarrow \infty$,
\begin{eqnarray}
 1-f_{\infty}+\lambda f_{\infty}^2+\mu f_{\infty}^3=0\, ,
 \label{finfQTG}
\end{eqnarray}
As discussed in \cite{Hofman}, \cite{Sinha}, the $3+1$ dimensional boundary CFT dual to this cubic gravity, can be characterized by two central charges
$c$ and $a$, and an extra parameter $t_4$ which is required to describe certain scattering phenomena. Further, they can be expressed in terms of the
coupling constants in the theory as,
\begin{eqnarray}
t_4 &=& \frac{2 \mu f_\infty ^2}{1 - 2 \lambda  f_\infty - 3\mu f_\infty^2} \,,  \nonumber\\
c &=& \pi^2 \frac{\tilde{L}^3}{\ell_p^3} ( 1 - 2 \lambda f_\infty - 3 \mu f_\infty^2) \,, \\
\label{t4centralcharges}
a &=& \pi^2 \frac{\tilde{L}^3}{\ell_p^3} ( 1 - 6 \lambda f_\infty + 9 \mu f_\infty^2)\, . \nonumber
\end{eqnarray}
It is to be noted that setting $\mu=0$ yields $t_4=0$, while $c$ and $a$ reduce to the form of the central charges in Gauss-Bonnet gravity.

From the above equations one can write down the coupling constants in terms of $t_4$ and the central charges as,
\begin{eqnarray}
 \frac{\tilde{L}^3}{\ell_p^3}&=&\frac{a}{2\pi^2} \big[3 \frac{c}{a} (1 + 3 t_4 ) - 1\big] \,,\nonumber\\
 \lambda f_{\infty}&=&\frac{1}{2}\, \frac{\frac{c}{a}\left(1 + 6 t_4 \right)-1}{3\frac{c}{a}\left(1 + 3 t_4 \right) -1}\,,\\
 \label{couplings}
\mu f_{\infty}^2&=&\frac{\frac{c}{a} t_4}{3 \frac{c}{a}\left(1 + 3 t_4 \right) - 1 }\, .\nonumber
\end{eqnarray}
Using the above equations, one can further write down the form of $f_{\infty}$ from eq.(\ref{finfQTG}), in terms of $t_4$ and the central charges as,
\begin{eqnarray}
 f_{\infty}= 2\, \frac{3 \frac{c}{a} ( 1 + 3 t_4) -1}{5 \frac{c}{a} ( 1 + 2 t_4 )-1}
\end{eqnarray}

The two coupling constants and hence the central charges along with $t_4$ can be constrained to avoid negative energy excitations in the 
boundary CFT \cite{Sinha} :
 \begin{eqnarray}
 \frac{c}{a} ( 1 - 84\, t_4 )  & \leq & 2 \, , \nonumber \\
 \frac{c}{a} ( 1 - 210\, t_4)  & \geq & \frac{1}{2}\, , 
 \label{t4centralchargesinequalities} \\
 \frac{c}{a} ( 1 + 224\, t_4) & \geq & \frac{2}{3}  \,. \nonumber
\end{eqnarray}
As was done earlier with earlier examples, here we first define $r_h=\tilde{L}\, x$ and compute the Hawking temperature as,
\begin{eqnarray}
 T=\frac{1}{2 \pi  x \mathcal{R}}\bigg[1+\frac{x^4 \ell _*^2 \mu _c^2 f_{\infty }^2-24 \pi ^2 L^2 \left(\mu  f_{\infty
   }^3-x^4 f_{\infty }+\lambda  x^2 f_{\infty }^2+x^6\right)}{12 \pi ^2 L^2
   f_{\infty } \left(3 \mu  f_{\infty }^2+2 \lambda  x^2 f_{\infty }-x^4\right)}\bigg]
 \label{tempQTG}
\end{eqnarray}
Unlike the Born-Infeld case, here it is straightforward to analytically express the charge parameter $q$ in terms of the 
chemical potential $\mu_c$ and hence in eq.(\ref{tempQTG}) we have written down the final expression of $T$ in terms of $\mu_c$.

Also, the thermal entropy of the black hole can be computed using Wald's prescription \cite{Wald},
\begin{eqnarray}
S= 2 \pi  \bigg(\frac{\tilde{L}}{\ell _p}\bigg)^3 V_{\Sigma }\, x^3 \bigg(1-\frac{6 \lambda  f_{\infty }}{x^2}+\frac{9 \mu  f_{\infty }^2}{x^4}\bigg)
\label{SQTG}
\end{eqnarray}
Finally, we compute the charged R\'enyi entropy,
\begin{eqnarray}
 S_n(\mu_c)&=&\bigg(\frac{\tilde{L}}{\ell _p}\bigg)^3 \pi V_{\Sigma } {n\over n-1}
 \left[3 \left(x_1^2-x_n^2\right) \bigg(1+\frac{\ell _*^2 \mu _c^2 f_{\infty }}{12 \pi ^2 L^2}
 +\frac{\mu  f_{\infty }^2}{x_1^2 x_n^2}\right)-\frac{3}{f_{\infty }}(x_1^4-x_n^4)\nonumber \\
 &+&2(P_1 Q_1-P_n Q_n)\bigg]
 \label{SnQTG}
\end{eqnarray}
where, 
\begin{eqnarray}
P_1= \frac{9 \mu  f_{\infty }^2-6 \lambda  x_1^2 f_{\infty }+x_1^4}{x_1^2 \left(3 \mu 
   f_{\infty }^2+2 \lambda  x_1^2 f_{\infty }-x_1^4\right)}\, , \ \ Q_1= x_1^4 \left(1+\frac{\ell _*^2 \mu _c^2 f_{\infty }}{12 \pi ^2 L^2}\right)+\mu 
   f_{\infty }^2-\frac{2 x_1^6}{f_{\infty }} \nonumber \\ 
    P_n=\frac{9 \mu  f_{\infty }^2-6 \lambda  f_{\infty } x_n^2+x_n^4}{x_n^2 \left(3 \mu 
   f_{\infty }^2+2 \lambda  f_{\infty } x_n^2-x_n^4\right)}\, , \ \ Q_n=x_n^4 \left(1+\frac{\ell _*^2 \mu _c^2 f_{\infty }}{12 \pi ^2 L^2}\right)+\mu 
   f_{\infty }^2-\frac{2 x_n^6}{f_{\infty }}
   \label{P1Q1PnQn}
\end{eqnarray}
As with the Born-Infeld gravity, here we define $x_n$ through $T(\mu_c,\, x)={T_0\over n}$, where with $n=1$ we can obtain $x_1$. Hence, the only 
remaining task here is to solve the following sixth order algebraic equation,
\begin{eqnarray}
 \frac{4 x_n^6}{f_{\infty }}-\frac{2 x_n^5}{n}-2 x_n^4 \left(1+\frac{\ell _*^2 \mu _c^2 f_{\infty }}{12 \pi ^2 L^2}\right)
 +\frac{4 \lambda  f_{\infty } x_n^3}{n}+\frac{6 \mu  f_{\infty }^2
   x_n}{n}-2 \mu f_{\infty }^2=0
   \label{xneqnQTG}
\end{eqnarray}
The largest real solution of the above equation determines the value of $x_n$. The analytical solution for $x_n$ is very difficult to obtain,
hence we would numerically solve this equation and compute the R\'enyi entropy.
Here, by considering the limit $n \rightarrow 1$, one can obtain the entanglement entropy, $S_1(\mu_c)$. Like we did earlier, 
we will show here the variation of the R\'enyi entropy $S_n(\mu_c)$ normalized by the entanglement entropy $S_1(0)$, instead of plotting $S_n(\mu_c)$.

\subsection{Numerical Analysis and Results}

\begin{figure}[h!]
 \centering
 \subfigure[${c\over a}=1$, ${\mu_c \ell _*\over 2\pi \tilde{L}}=2$]{
 \includegraphics[width=2.5in,height=2.3in]{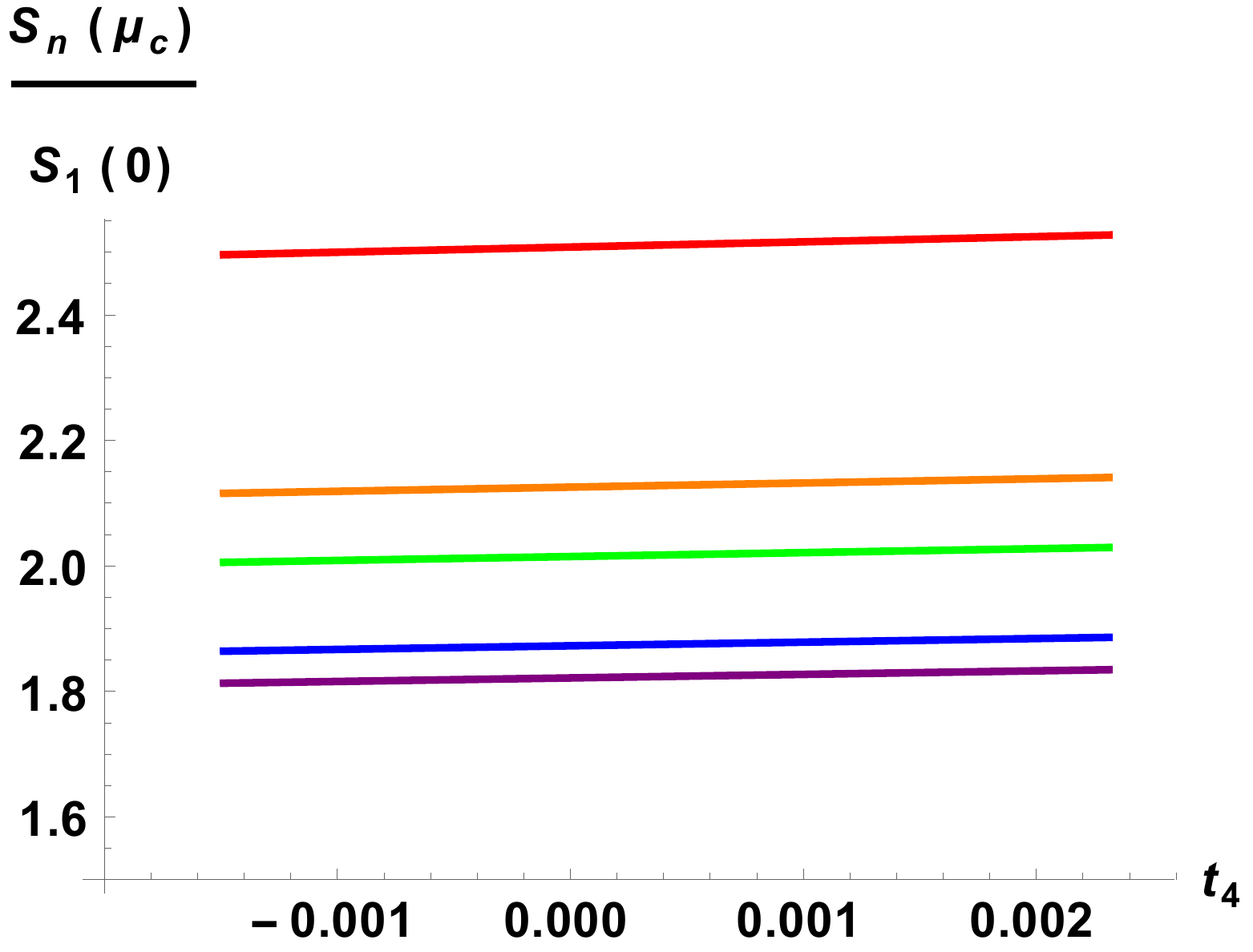}
 \label{muc2cbya1Snvst4} } 
 \subfigure[${c\over a}=1$, ${\mu_c \ell _*\over 2\pi \tilde{L}}=20$]{
 \includegraphics[width=2.5in,height=2.3in]{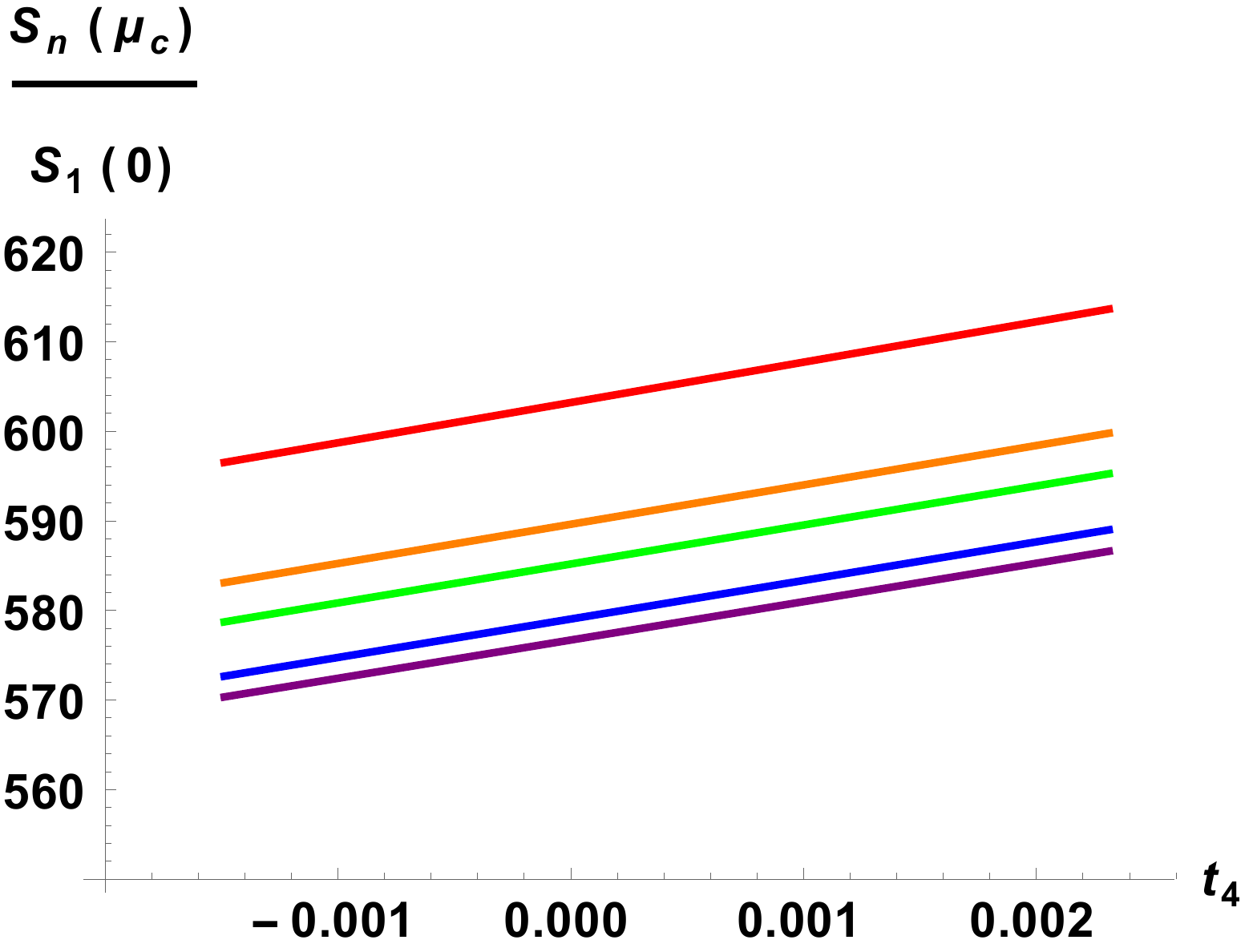}
 \label{muc20cbya1Snvst4} }
\caption{\small{${S_n(\mu_c)\over S_1(0)}$ as a function of $t_4$ at fixed value of ${c \over a}$ with different fixed 
values of the chemical potential.}}
 \label{Snvst4fixedcByaDifferentmuc}
 \end{figure}

First, fixing the value of the ratio of central charges ${c\over a}$, we plot the behavior of R\'enyi entropy ratio ${S_n(\mu_c) \over S_1(0)}$
with $t_4$ for different values of the chemical potential ($\mu_c$), as shown in figure \ref{Snvst4fixedcByaDifferentmuc}. 
We set ${c\over a}=1$, for which $t_4$ must be in the following physically allowed regime from eq.(\ref{t4centralchargesinequalities}):
\begin{eqnarray}
-0.00149\leq t_4 \leq 0.00238\, . \nonumber
\end{eqnarray}

In both figures we seem to get a linear behavior. Notice that for a relatively small value of chemical potential (figure \ref{muc2cbya1Snvst4}),
where ${\mu_c \ell _*\over 2\pi \tilde{L}}=2$, the slope of the lines are very small. In fact, the authors of \cite{Myers} considered the 
uncharged quasi-topological gravity and concluded that the ratio ${S_n\over S_1}$ is almost independent of $t_4$ for a fixed value
of ${c\over a}$ in the physically allowed regime and the dependence comes into play when we are well outside the physical regime. However, by coupling the
quasi-topological gravity to a $U(1)$ gauge field, it is evident in figure \ref{muc20cbya1Snvst4} that with a higher value of chemical potential
the R\'enyi entropy ratio increases as $t_4$ increases, where we have fixed ${c\over a}$ to the same value as in figure \ref{muc2cbya1Snvst4},
but chosen a higher value of chemical potential ${\mu_c \ell _*\over 2\pi \tilde{L}}=20$. In both figures, the red, orange, green, blue and purple
lines correspond to the values $n=1$, $2$, $3$, $10$ and $100$, respectively. Notice that, as we increase chemical potential, the ratio 
${S_n(\mu_c)\over S_1(0)}$ reaches its maximum value for the maximum bound of $t_4$ (i.e., for $t_4=0.00238$ with ${c\over a}=1$). 
Another important observation is that, although the slope of the entropy ratio increases as the chemical potential increases, the slope is 
independent of $n$ for a fixed value of chemical potential with fixed ${c\over a}$.

 \begin{figure}[h!]
 \centering
 \subfigure[${c\over a}=1$, $t_4=-0.00149$]{
 \includegraphics[width=2.5in,height=2.3in]{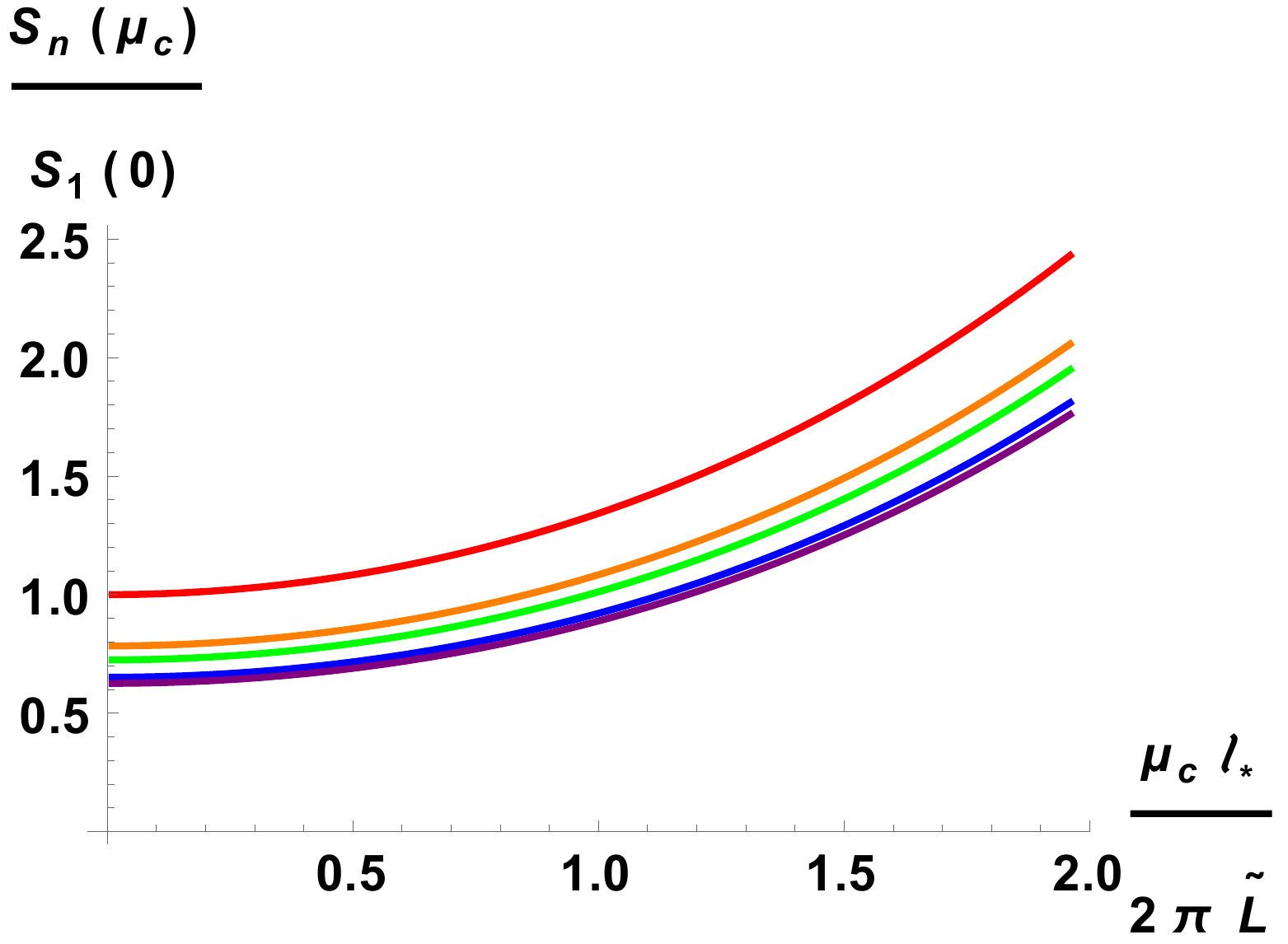}
 \label{Snvsmuct4N0Pt00149cBya1} } 
 \subfigure[${c\over a}=1$, $t_4=0.00238$]{
 \includegraphics[width=2.5in,height=2.3in]{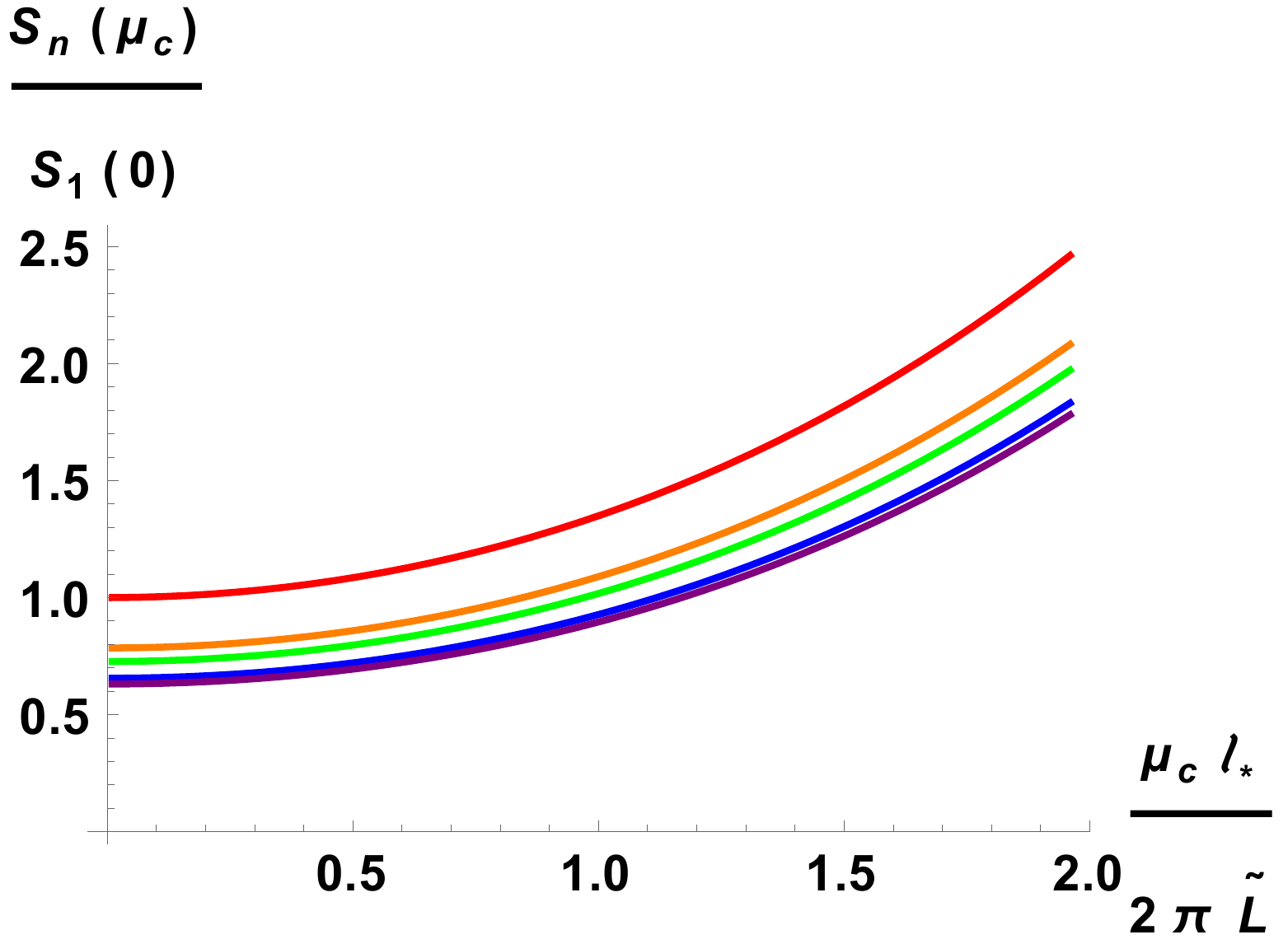}
 \label{Snvsmuct40Pt00238cBya1} }
  \caption{\small{${S_n(\mu_c)\over S_1(0)}$ as a function of chemical potential at fixed value of ${c \over a}=1$ with different fixed values of $t_4$.}}
 \label{SnvsmucfixedcByaDifferentt4}
 \end{figure}
Now we proceed to show the behavior of the entropy ratio with the chemical potential, keeping ${c\over a}$ fixed.
Figure \ref{Snvsmuct4N0Pt00149cBya1} shows this variation with $t_4=-0.00149$ and 
figure \ref{Snvsmuct40Pt00238cBya1} shows the same with $t_4=0.00238$, where we have fixed ${c\over a}=1$. Recall that, these two chosen
values of $t_4$ are the two extreme bounds (i.e., end points of the physically allowed parameter space for $t_4$) with ${c\over a}=1$. In each
figure, the entropy ratio increases monotonically and in a non-linear fashion, as $\mu_c$ increases. Here we follow the same color coding for $n$,
i.e., $n=1$, $2$, $3$, $10$ and $100$ correspond to the red, orange, green, blue and purple curves. Note that for any value of $n$, the behavior
of the entropy with chemical potential is the same. The curve with $n=1$ has the highest value of entropy ratio at a fixed chemical potential, while
$n=100$ has the lowest.  

 \begin{figure}[h!]
 \centering
 \subfigure[${c\over a}=2$, $t_4=0$]{
 \includegraphics[width=2.5in,height=2.3in]{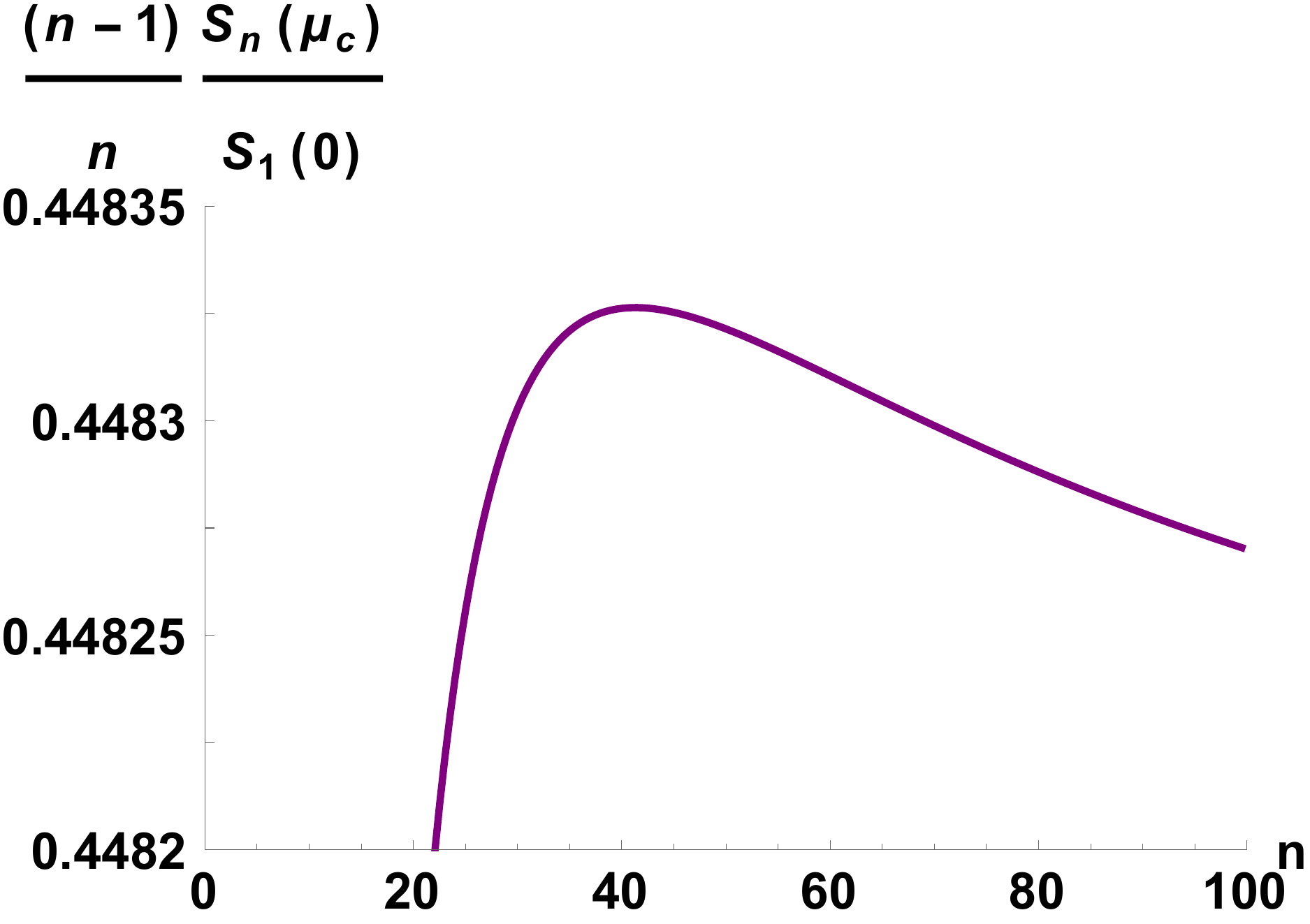}
 \label{muc0Pt45cBya2t40} } 
 \subfigure[${c\over a}=2$, $t_4=0.0036$]{
 \includegraphics[width=2.5in,height=2.3in]{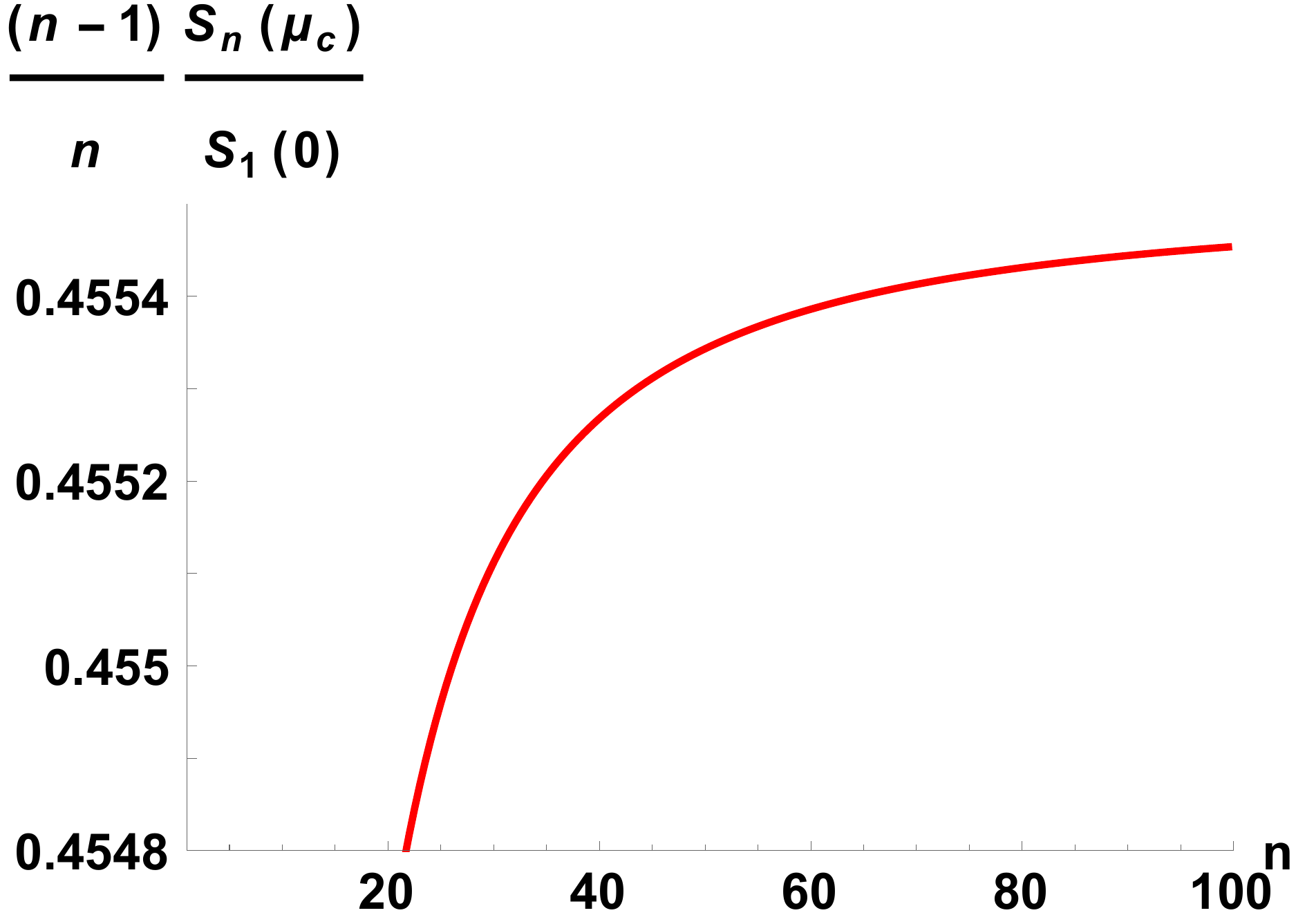}
 \label{muc0Pt45cBya2t40Pt0036} }
  \caption{\small{(a) Second inequality is violated with $t_4=0$, (b) Second inequality is obeyed with $t_4=0.0036$. In both figures
  we have fixed the chemical potential, ${\mu_c \ell _*\over 2\pi \tilde{L}}=0.45$.}}
 \label{SecondInequalitymuc0Pt45cBya2}
 \end{figure}
\begin{figure}[h!]
 \centering
 \subfigure[${c\over a}=2$, ${\mu_c \ell _*\over 2\pi \tilde{L}}=0.25$]{
 \includegraphics[width=2.5in,height=2.3in]{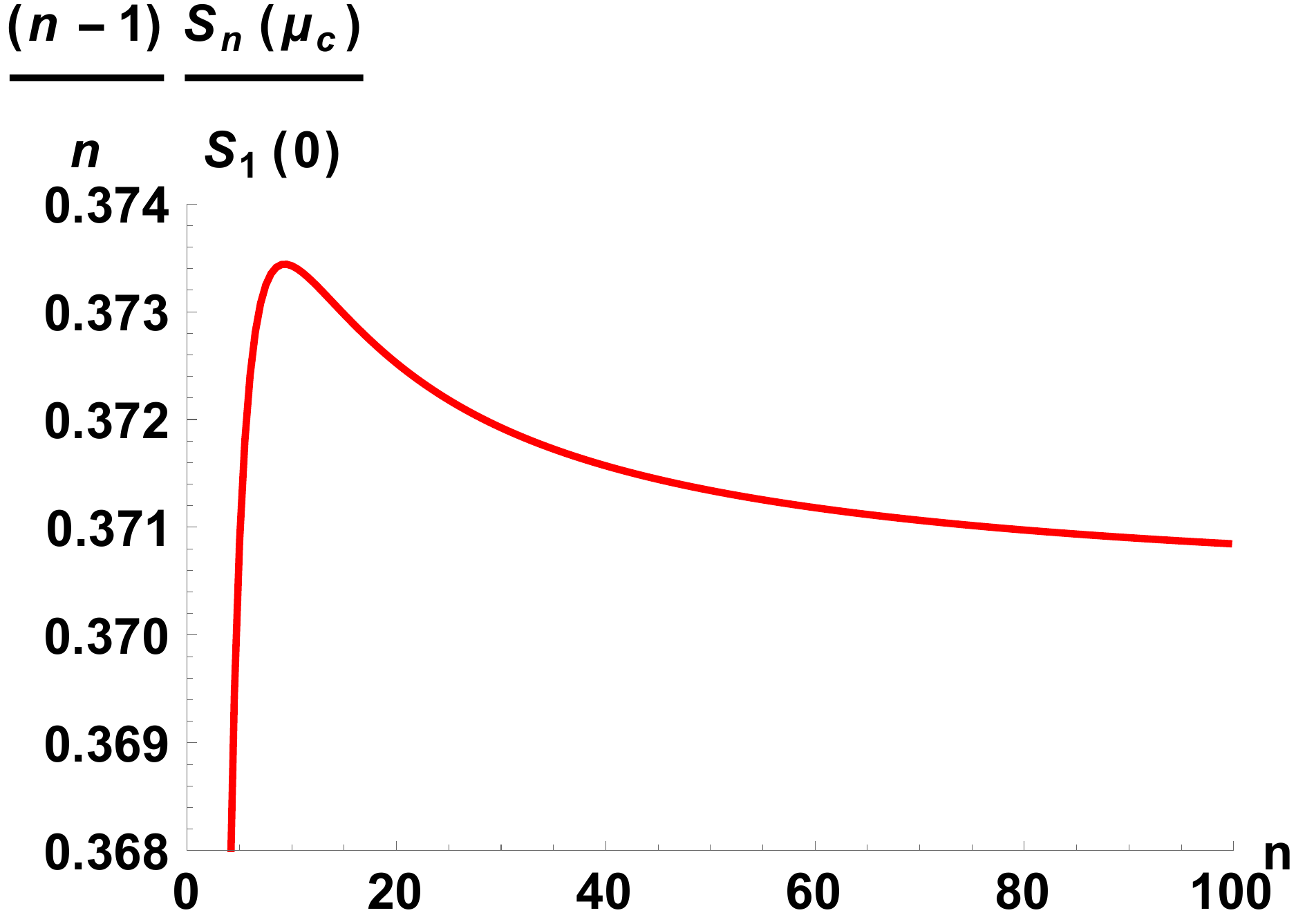}
 \label{cbya2t40Pt001muc0Pt25} } 
 \subfigure[${c\over a}=2$, ${\mu_c \ell _*\over 2\pi \tilde{L}}=0.75$]{
 \includegraphics[width=2.5in,height=2.3in]{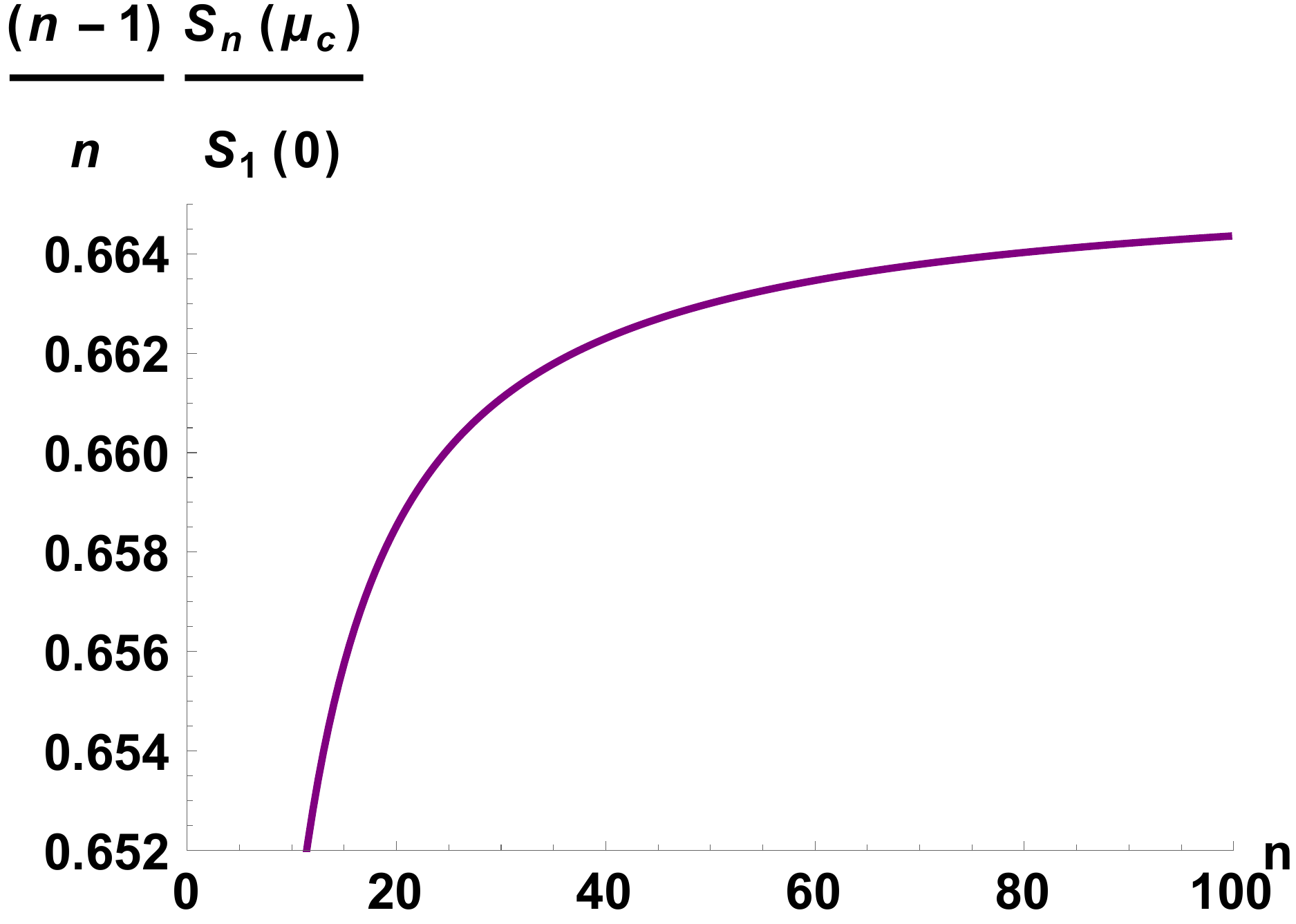}
 \label{cbya2t40Pt001muc0Pt75} }
  \caption{\small{(a) second inequality is violated with ${\mu_c \ell _*\over 2\pi \tilde{L}}=0.25$,
  (b) Second inequality is obeyed with ${\mu_c \ell _*\over 2\pi \tilde{L}}=0.75$. In both figures we have fixed, $t_4=0.001$.}}
 \label{SecondInequalityt40Pt001cBya2}
 \end{figure}

Now we go ahead to verify the R\'enyi entropy inequalities as we did in the Born-Infeld gravity. First, we verified that the first inequality of
eq.(\ref{EREinequalities}) holds for any values of $t_4$ and chemical potential for a fixed value of ${c\over a}$. We just mention this result here 
and do not show the relevant graphs since they are similar to the Born-Infeld case. The results become more interesting when we test the second 
inequality as shown in figures (\ref{SecondInequalitymuc0Pt45cBya2}) and (\ref{SecondInequalityt40Pt001cBya2}). 
Figure (\ref{SecondInequalitymuc0Pt45cBya2}) shows the variation
of ${(n-1)\over n} {S_n(\mu_c)\over S_1(0})$ with $n$ for fixed value of the chemical potential ${\mu_c \ell _*\over 2\pi \tilde{L}}=0.45$ and 
the central charge ratio ${c\over a}=2$. For this chosen value of ${c\over a}$, it is clear from (\ref{t4centralchargesinequalities}) that $t_4$ 
must be in the following physically allowed regime,
\begin{eqnarray}
  0\leq t_4 \leq 0.0036\, . \nonumber
\end{eqnarray}
Note that with the upper bound $t_4=0.0036$, the R\'enyi entropy obeys the second inequality, but with the lower bound $t_4=0$, it violates
the inequality although $t_4=0$ is a well-accepted physical regime from causality considerations. This is interesting and let us discuss this further. 
In this context, we should mention here that this kind of violation was obtained earlier
by the authors of \cite{Pastras} while studying the charged R\'enyi entropies in CFTs dual to Gauss-Bonnet gravity. They showed that this violation
can occur in CFTs dual to a hyperbolic black hole geometry when the black hole possesses negative thermal entropy. Now, the negative thermal entropy
of a topological black hole is a typical behavior of higher derivative gravity theories and is controlled by the parameters of the higher derivative
terms \cite{Cvetic}, \cite{Anninos}.

It is clear from eq.(\ref{SQTG}) that the negative entropy black holes must satisfy,
\begin{eqnarray}
 x < \sqrt{3 f_{\infty}(\lambda+\sqrt{\lambda^2-\mu})}.
\end{eqnarray}
Now from eq.(\ref{tempQTG}), the negative entropy black holes would appear when the chemical potential,
\begin{eqnarray}
 \bigg({\mu_c \ell _*\over 2\pi \tilde{L}}\bigg)^2 < \frac{2}{3 \mu } \bigg[\lambda  \left(\sqrt{\lambda ^2-\mu }+27 \mu \right)+\mu 
   \left(27 \sqrt{\lambda ^2-\mu }-4\right)-\lambda ^2\bigg].
   \label{MaxChemicalPotentialQTG}
\end{eqnarray}
with
\begin{eqnarray}
 \lambda  \left(\sqrt{\lambda ^2-\mu }+27 \mu \right)+\mu 
   \left(27 \sqrt{\lambda ^2-\mu }-4\right)-\lambda ^2 > 0.
   \label{MinlambdamuQTG}
\end{eqnarray}

Eq.(\ref{MaxChemicalPotentialQTG}) can be rewritten in terms of $t_4$ and the central charges of the theory as,
\begin{eqnarray}
 \bigg({\mu_c \ell _*\over 2\pi \tilde{L}}\bigg)^2 &<& {1\over 6{c\over a}\, t_4\, \alpha_2\, \alpha_3} \bigg[-\alpha_1^2\, \alpha_2 
 +{c\over a}\, t_4\, {\alpha_2\over \alpha_3}(-16 \alpha_3^2+27\alpha_2\, \sqrt{\alpha_4}) \nonumber \\
 &+&{\alpha_1\over \alpha_3}(\sqrt{\alpha_4}\alpha_2 \alpha_3+27{c\over a}\, t_4\, \alpha_2^2)\bigg]
 \label{MaxChemicalPotentialQTGcByat4}
\end{eqnarray}
where, 
\begin{eqnarray}
 \alpha_1&=&-1+{c\over a}(1+6t_4)\,, \ \ \ \alpha_2=-1+5{c\over a}(1+2t_4)\nonumber \\
 \alpha_3&=&-1+3{c\over a}(1+3t_4)\,, \ \ \alpha_4=(-1+{c\over a})^2-8{c\over a}\, t_4\, .
 \label{alphas}
\end{eqnarray}
Now with ${c\over a}=2$ and $t_4=0$, from eq.(\ref{MaxChemicalPotentialQTGcByat4}), one can see that
negative entropy black holes appear when ${\mu_c \ell _*\over 2\pi \tilde{L}} < 0.49$. 
In figure \ref{muc0Pt45cBya2t40}, since the chosen chemical potential (i.e., ${\mu_c \ell _*\over 2\pi \tilde{L}} =0.45$) is less than 
the aforementioned value, negative entropy black holes would appear in this case and the CFT dual to this negative entropy hyperbolic black hole 
would exhibit a violation in the second inequality obeyed by R\'enyi entropy. However, with $t_4=0.0036$,  negative entropy black holes would appear
when ${\mu_c \ell _*\over 2\pi \tilde{L}} < 0.44$. Since the chosen potential is greater than this value, the black hole can not have negative entropy
and hence, the corresponding CFT obeys the second inequality as shown in  figure \ref{muc0Pt45cBya2t40Pt0036}. It can be checked that, with 
${c\over a}=2$ and ${\mu_c \ell _*\over 2\pi \tilde{L}} =0.45$, the physically allowed regime of the $t_4$ parameter space becomes narrower
in order to obey the inequalities: 
\begin{eqnarray}
 0.0029\leq t_4\leq 0.0036\, .
\end{eqnarray}

In figure \ref{SecondInequalityt40Pt001cBya2}, we have shown how the second inequality is controlled by the chemical potential for a fixed value 
of ${c\over a}=2$ and $t_4=0.001$. With these CFT parameters, the negative entropy black hole can appear only when 
${\mu_c \ell _*\over 2\pi \tilde{L}} =0.476$. Since in figure \ref{cbya2t40Pt001muc0Pt25}, ${\mu_c \ell _*\over 2\pi \tilde{L}} =0.25$, black holes
would have negative thermal entropy and hence the second inequality of R\'enyi entropy would be violated. On the other hand, 
in figure \ref{cbya2t40Pt001muc0Pt75}, ${\mu_c \ell _*\over 2\pi \tilde{L}} =0.75$ which can not produce negative thermal entropy and hence there 
would be no violation of the inequality in this case.

After analyzing the bound on $t_4$ for a fixed value of the ratio of the central charges, it is important to examine what bound the R\'enyi entropy
imposes on ${c\over a}$ when we fix $t_4$. As we mentioned before, the authors of \cite{Pastras} analyzed this bound on the ratio of 
central charges in the context of Gauss-Bonnet gravity. Since, we have an extra parameter $\mu$ in QTG and hence $t_4$, this bound needs to be
reexamined in the context of QTG. Note that, if we set $\mu=0$, we should recover the result of Gauss-Bonnet gravity. 

 \begin{figure}[h!]
 \centering
 \subfigure[$t_4=0.001$, ${\mu_c \ell _*\over 2\pi \tilde{L}}=0$]{
 \includegraphics[width=2.5in,height=2.3in]{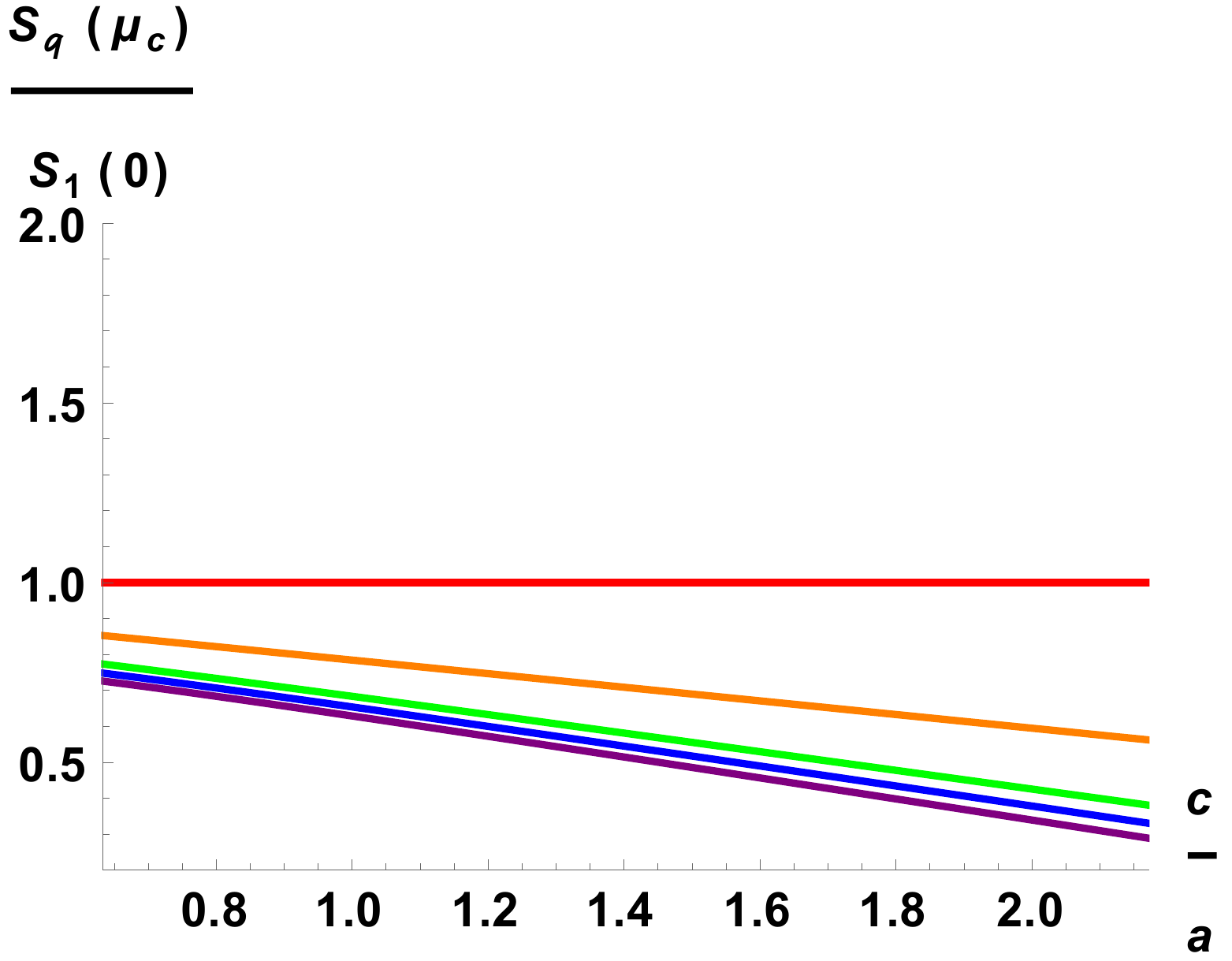}
 \label{QTGSnVscByat40Pt001muc0} } 
 \subfigure[$t_4=0.001$, ${\mu_c \ell _*\over 2\pi \tilde{L}}=1$]{
 \includegraphics[width=2.5in,height=2.3in]{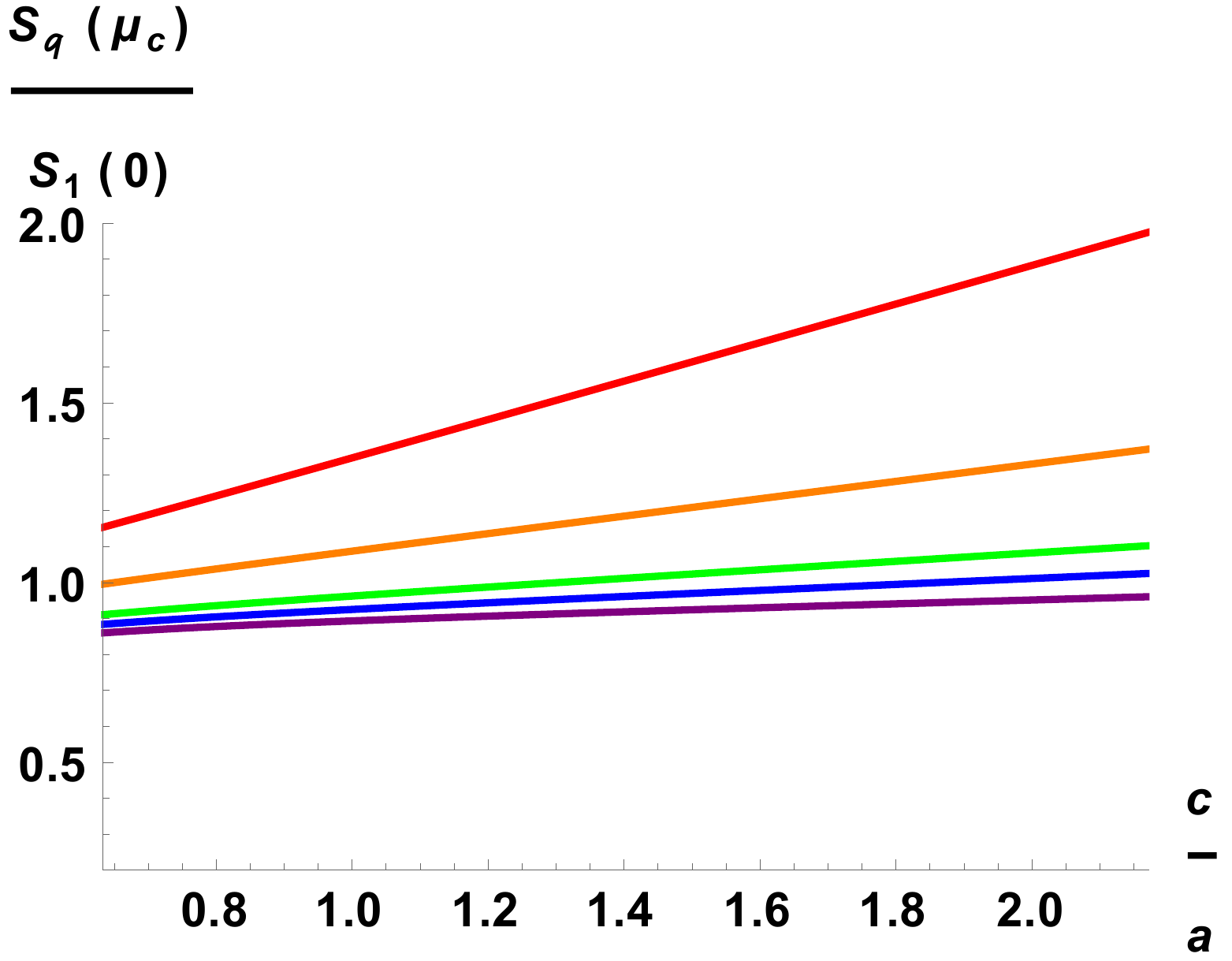}
 \label{QTGSnVscByat40Pt001muc1} }
\caption{\small{${S_n(\mu_c)\over S_1(0)}$ as a function of ${c\over a}$ at fixed value of $t_4=0.001$ with different fixed values of the 
chemical potential.}}
\label{QTGSnVscByat40Pt001Differentmuc}
\end{figure}
However, before analyzing the bound on central charges we briefly describe the behavior of the entropy ratio ${S_n(\mu_c)\over S_1(0)}$ 
as a function of ${c\over a}$ and the chemical potential for a particular value of $t_4$. The plots are similar to those obtained with Gauss-Bonnet gravity 
\cite{Pastras}. Figure \ref{QTGSnVscByat40Pt001muc0} and \ref{QTGSnVscByat40Pt001muc1} shows the variation of  ${S_n(\mu_c)\over S_1(0)}$ 
with the central charge ratio with $t_4=0.001$ for ${\mu_c \ell _*\over 2\pi \tilde{L}}=0$ and $1$, respectively. 
Like figure \ref{Snvst4fixedcByaDifferentmuc}, this behavior is almost linear as long as we are inside the physical regime determined by 
eq.(\ref{t4centralchargesinequalities}). But unlike figure \ref{Snvst4fixedcByaDifferentmuc}, here the slopes of the lines are different for 
different $n$. With ${\mu_c \ell _*\over 2\pi \tilde{L}}=0$, the lower bound of ${c\over a}$ has the maximum value of ${S_n(\mu_c)\over S_1(0)}$,
while with ${\mu_c \ell _*\over 2\pi \tilde{L}}=1$, the maximum value of  ${S_n(\mu_c)\over S_1(0)}$ is obtained for the upper bound of ${c\over a}$.
Hence, as we increase the chemical potential, the entropy ratio increases and there exists a critical value of the chemical potential beyond which
${S_n(\mu_c)\over S_1(0)}$ attains its maximum value at the upper bound of ${c\over a}$ for any value of $n$.
 \begin{figure}[h!]
 \centering
 \subfigure[$t_4=0.001$, ${c\over a}=0.633$]{
 \includegraphics[width=2.5in,height=2.3in]{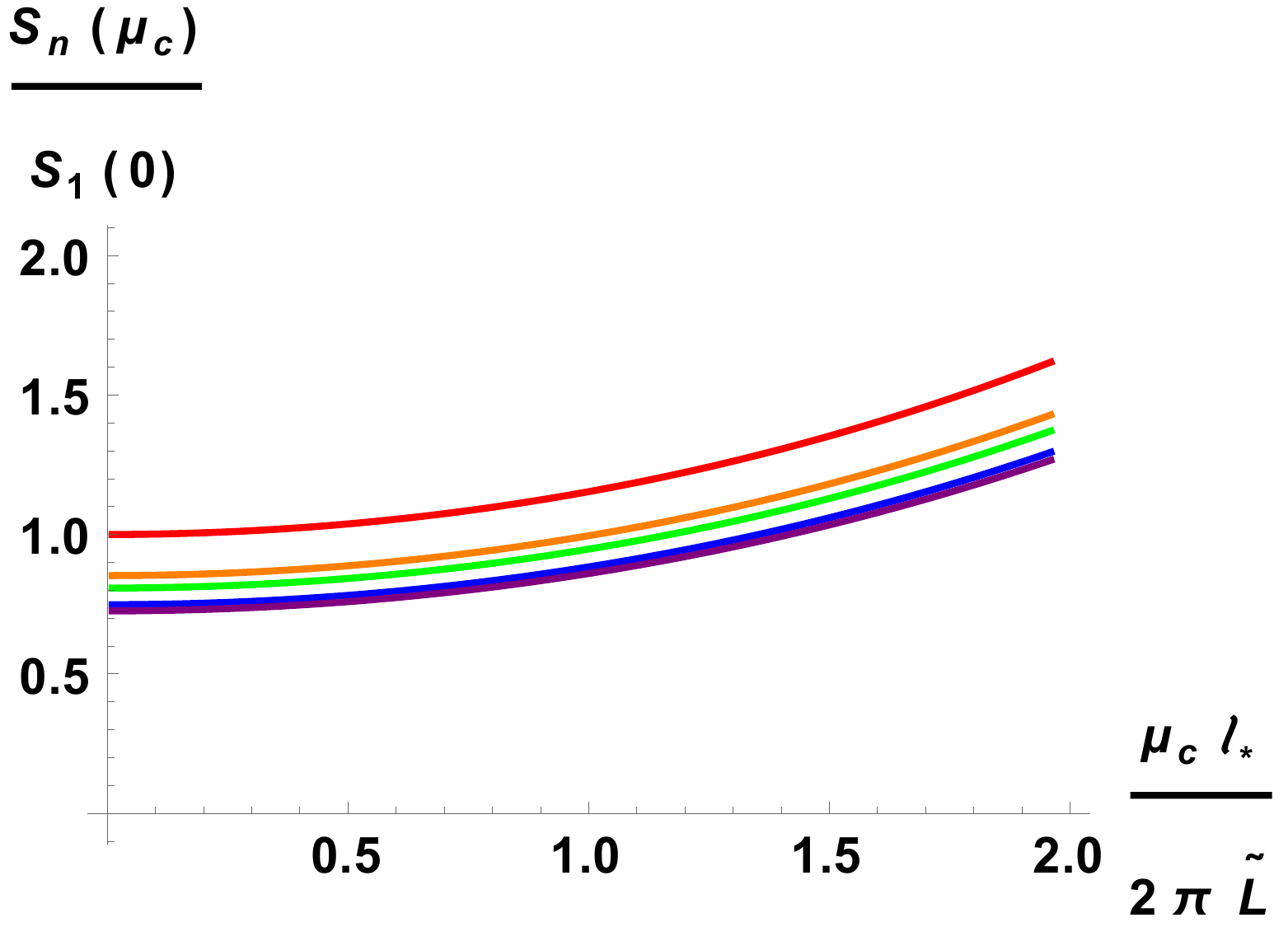}
 \label{QTGt40Pt001cBya0Pt633} } 
 \subfigure[$t_4=0.001$, ${c\over a}=2.183$]{
 \includegraphics[width=2.5in,height=2.3in]{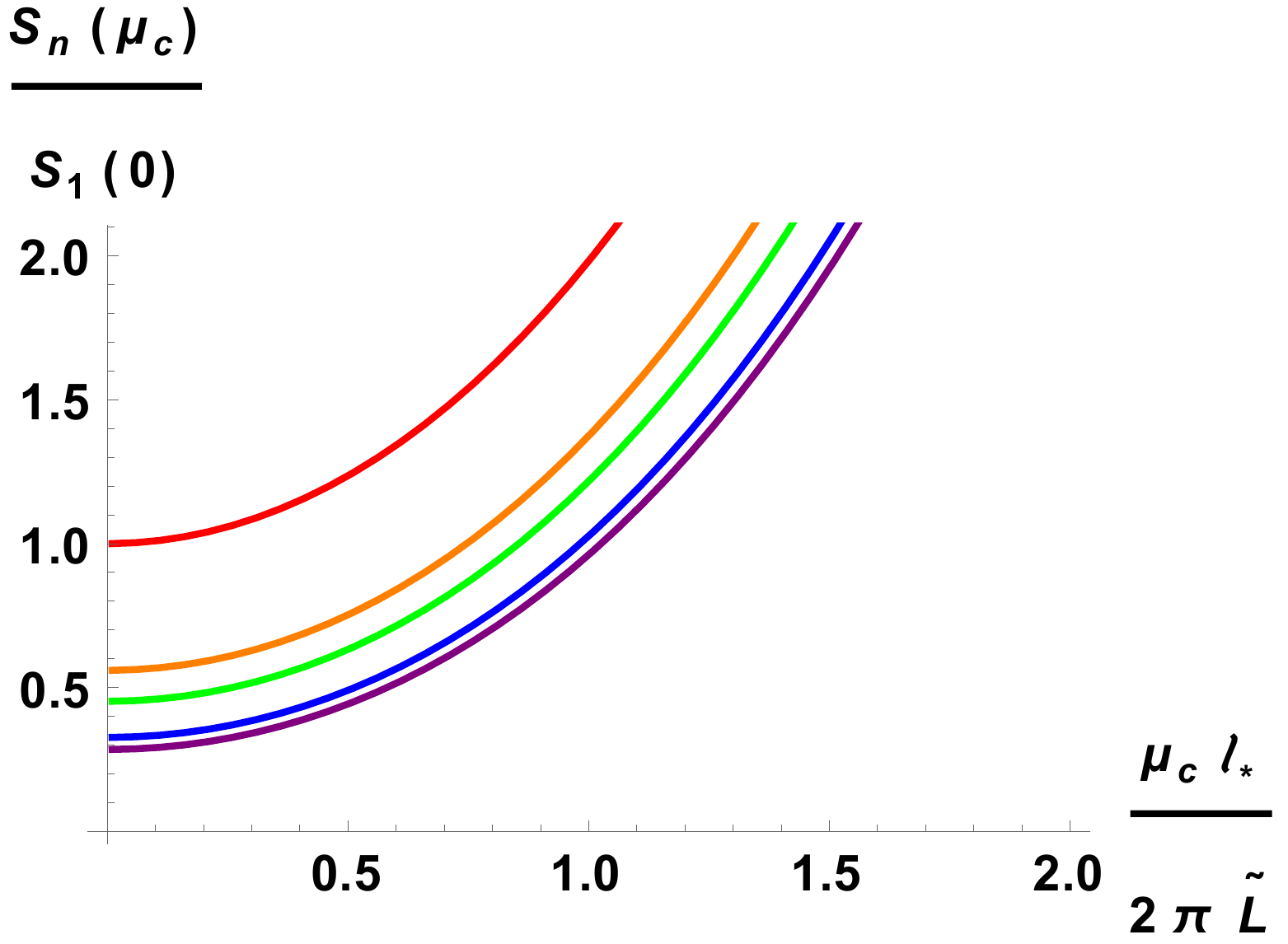}
 \label{QTGt40Pt001cBya2Pt183} }
  \caption{\small{${S_n(\mu_c)\over S_1(0)}$ as a function of chemical potential at fixed value of $t_4=0.001$ with different fixed 
  values of ${c\over a}$.}}
 \label{QTGSnVsmuct40Pt001DifferentcBya}
 \end{figure}

For completeness, we also show the plots of the entropy ratio as a function of the chemical potential at fixed value of $t_4=0.001$ for different values of ${c\over a}$ in
figure \ref{QTGSnVsmuct40Pt001DifferentcBya}. In both figures \ref{QTGSnVscByat40Pt001Differentmuc} and \ref{QTGSnVsmuct40Pt001DifferentcBya},
the red, orange, green, blue and purple lines correspond to $n=1$, $2$, $3$, $10$ and $100$ respectively. Note that for higher values of
${c\over a}$, the entropy ratio increases more rapidly with chemical potential as compared to smaller values of ${c\over a}$. 

 \begin{figure}[h!]
 \centering
 \subfigure[$t_4=0.002$, ${c\over a}=0.862$]{
 \includegraphics[width=2.5in,height=2.3in]{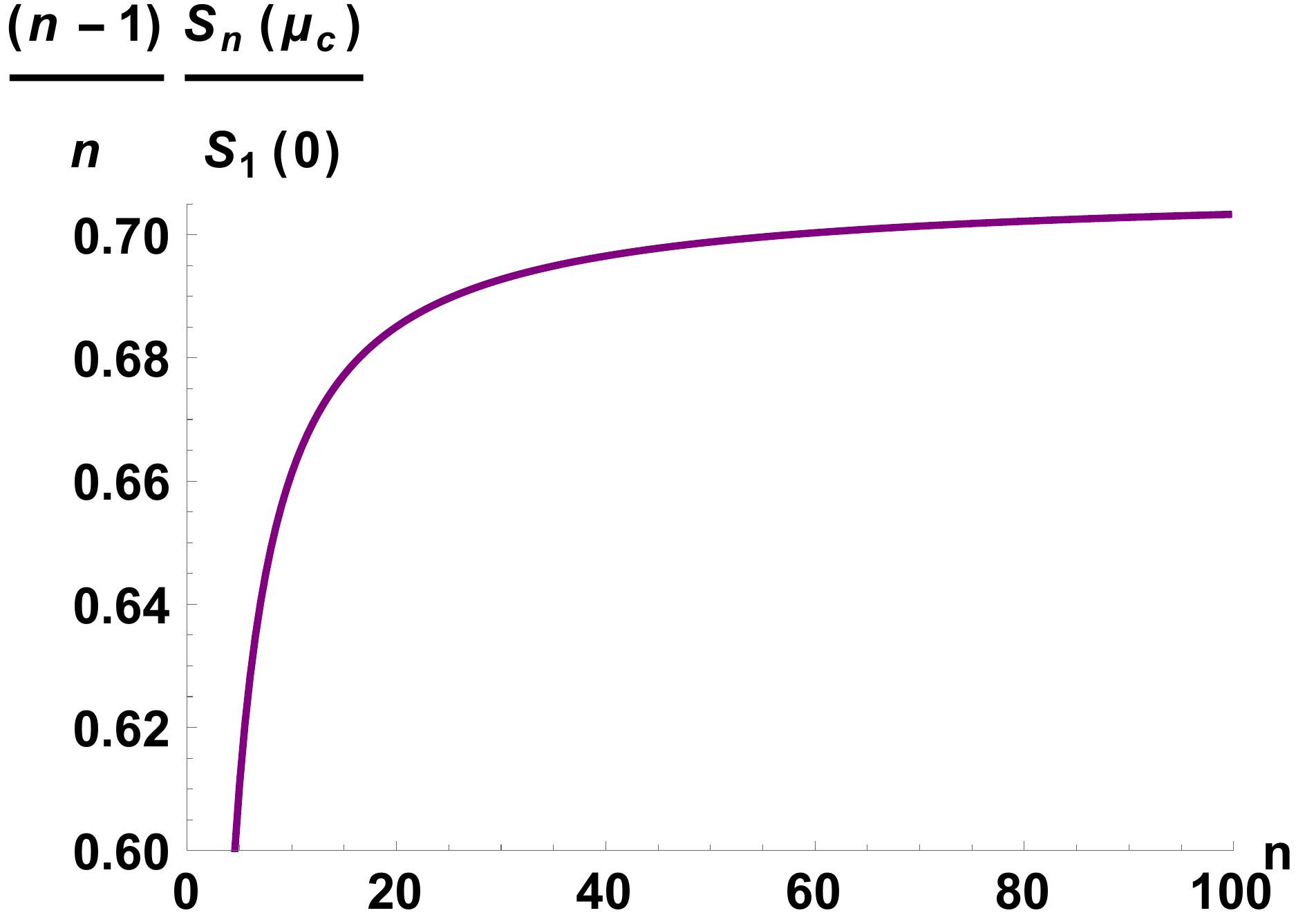}
 \label{QTGt40Pt002muc0Pt45cbya0Pt86} } 
 \subfigure[$t_4=0.002$, ${c\over a}=2.404$]{
 \includegraphics[width=2.5in,height=2.3in]{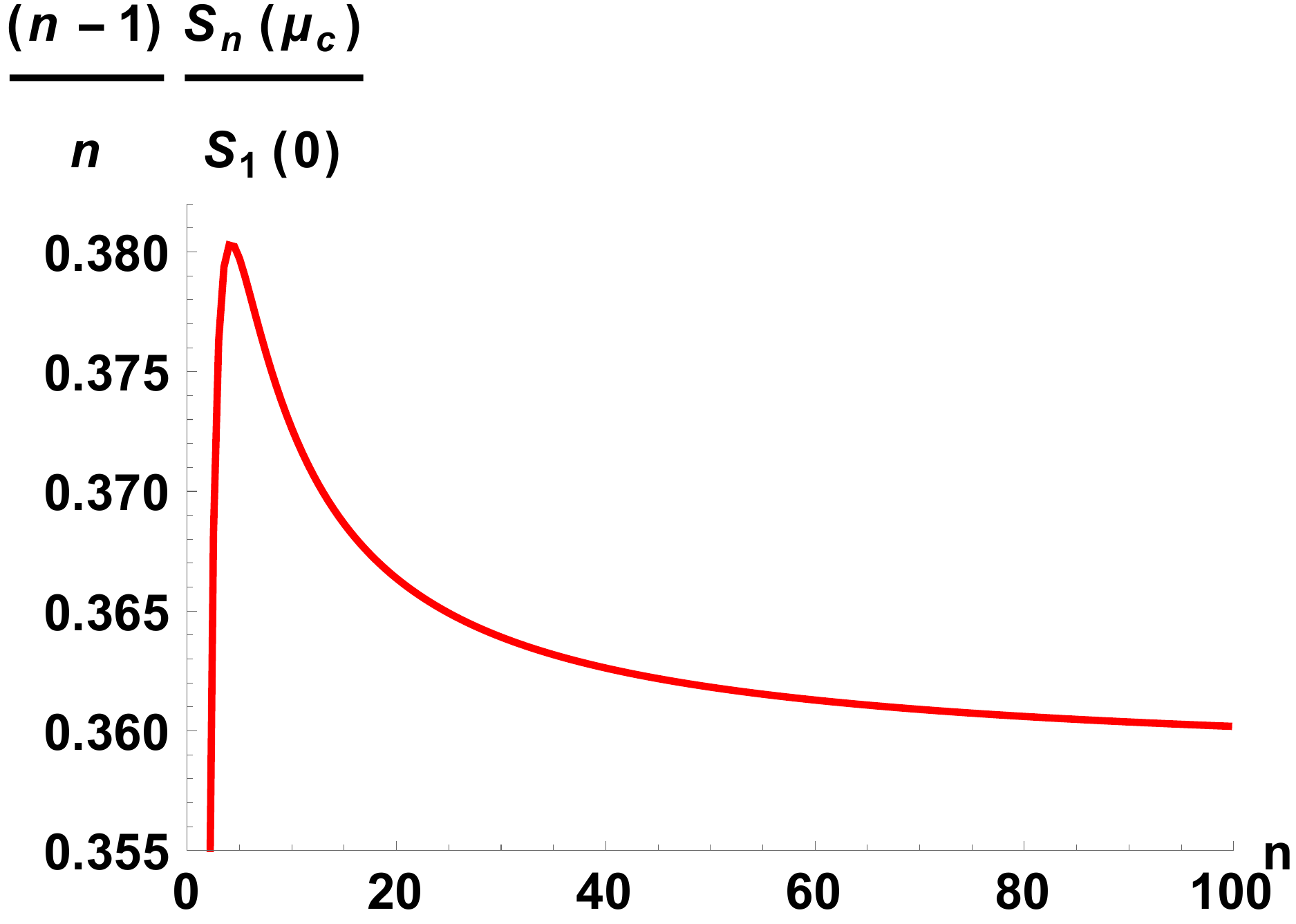}
 \label{QTGt40Pt002muc0Pt45cbya2Pt4} }
  \caption{\small{(a) The second inequality is obeyed with ${c\over a}=0.862$, (b) Second inequality is violated with ${c\over a}=2.404$. In both figures
  we have fixed the chemical potential, ${\mu_c \ell _*\over 2\pi \tilde{L}}=0.45$.}}
 \label{SecondInequalitymuc0Pt45t40Pt002}
 \end{figure}
Now we verify the second inequality obeyed by R\'enyi entropy for fixed values of $t_4$, and analyze the bound on the central charge ratio
${c\over a}$. We set $t_4=0.002$ and the chemical potential ${\mu_c \ell _*\over 2\pi \tilde{L}}=0.45$. With this chosen value of $t_4$, using
eq.(\ref{t4centralchargesinequalities}) one can easily determine the physically allowed regime for ${c\over a}$ which is, 
$0.862 \leq {c\over a} \leq 2.404$. Now with the lower bound of ${c\over a}$, figure \ref{QTGt40Pt002muc0Pt45cbya0Pt86} shows that the second inequality
of R\'enyi entropy is obeyed while with the upper bound, figure \ref{QTGt40Pt002muc0Pt45cbya2Pt4} shows that the inequality is
violated. This is expected if we connect this result with the negative entropy condition of the dual black hole geometry. Following the 
same logic as we elaborated upon with fixed values of ${c\over a}$, here one can explain these results with fixed value of $t_4$ using the same arguments 
based on eq.(\ref{MaxChemicalPotentialQTGcByat4}) and we will not go into the details. However, for this example, we mention the bound on the central charge
ratio imposed by the inequality of R\'enyi entropy is
\begin{eqnarray}
 0.862 \leq {c\over a} < 1.990\, .
\end{eqnarray}

\subsection{Scaling Dimension of Twist Operators}

Let us compute the scaling dimension $h_n(\mu_c)$ and the magnetic flux response $k_n(\mu_c)$ of the generalized twist operators in the four 
dimensional CFT dual to charged quasi-topological gravity. Here, we will use eq.(\ref{hnformula1}) with the mass $m$ being given in eq.(\ref{mQTG}),
\begin{equation}
m = 3 \tilde{L}^2 \bigg( \frac{x^4}{f_{\infty}}-x^2+\frac{\mu_c^2 \ell_{*}^2}{12\pi^2 \tilde{L}^2}x^2+\lambda f_{\infty}+\frac{\mu}{x^2}f_{\infty}^2 \bigg)
\label{massQTG2}
\end{equation}
Using eqs.(\ref{hnformula}), (\ref{epsilon}) and (\ref{massQTG2}), it is straightforward to show that the scaling dimension $h_n$ is given by,
\begin{equation}
h_n(\mu_c) = - \frac{\pi \tilde{L}^3 n}{\ell_p^3} \bigg( \frac{x_n^4}{f_{\infty}}-x_n^2+\frac{\mu_c^2 
\ell_{*}^2}{12\pi^2 \tilde{L}^2}x_n^2 + \lambda f_{\infty} + \frac{\mu}{x_n^2}f_{\infty}^2 \bigg)
\label{hmfinal}
\end{equation}
Note that with $\mu_c=0$ and $\mu=0$ we recover the dimension of the twist operator in a CFT dual to Einstein Gauss-Bonnet gravity, 
while with $\mu_c=0$ and $\mu=0$ and $\lambda=0$, the results match with the same in pure Einstein gravity, as expected. We also analyze the first order 
coefficients of the expansion of the scaling dimension $h_{10}$ and $h_{02}$ (see section 2) which are given by 
\begin{equation}
h_{1 0} = \frac{2\pi}{3}\bigg(\frac{\tilde{L}}{\ell_p}\bigg)^3 (1-2\lambda f_{\infty}-3\mu f_{\infty}^2) = \frac{2}{3\pi}c
\end{equation}
and
\begin{eqnarray}
h_{0 2} = -\frac{5}{18\pi} \bigg( \frac{\tilde{L}}{\ell_p} \bigg)^3 \bigg(\frac{\ell_{*}^2}{\tilde{L}^2} \bigg)
\end{eqnarray}
In appropriate limits, the results agree with the corresponding ones in pure Einstein and Gauss-Bonnet gravity.

On the other hand, the magnetic response can be calculated by computing the charge density,
\begin{eqnarray}
 \rho (x,\mu_c)=\frac{2}{\mathcal{R}^3} \bigg( \frac{\tilde{L}}{\ell_p}\bigg)^3 \bigg(\frac{\ell_{*}}{\tilde{L}} \bigg)^2 \frac{\mu_c}{4\pi} x^2
\end{eqnarray}
This yields the result
\begin{equation}
k_n (\mu_c)= 2\pi n \mathcal{R}^{3} \rho (x,\mu_c) = 2 n \bigg(\frac{\tilde{L}}{\ell_p} \bigg)^3 \bigg(\frac{\ell_{*}}{\tilde{L}} \bigg)^2 \frac{\mu}{2} x_n^{2} 
\end{equation}
Note that this result of magnetic response depends on the couplings $\lambda$ and $\mu$ through $x_n$. One can also extract the expansion coefficients as,
\begin{eqnarray}
 k_{01}(\mu, \lambda)=\bigg({\tilde{L}\over \ell_p}\bigg)^3\bigg({\ell_{*}\over \tilde{L}}\bigg)^2\, \  \text{and}\, \ \ 
 k_{11}(\mu, \lambda)={1\over 3} \bigg({\tilde{L}\over \ell_p}\bigg)^3\bigg({\ell_{*}\over \tilde{L}}\bigg)^2
\end{eqnarray}
The expressions for magnetic response are again in agreement with the case of pure Einstein ($\lambda=0$, $\mu=0$, $\mu_c=0$) and the 
Gauss-Bonnet gravity ( $\mu=0$, $\mu_c=0$).

Having elucidated the nature of ERE in charged quasi-topological gravity, we will now analyse our final example, ERE in Weyl corrected gravity. 

\section{Charged R\'enyi Entropy in Weyl Corrected Gravity}

In this section, we consider the Einstein-Maxwell action along with a Weyl coupling. Since the analysis is
similar to the ones considered in the previous two sections, we will be brief here. We will start with the action 
\begin{equation}
\textit{S} = \frac{1}{2\ell_p ^3}\int \mathrm{d^5}x \sqrt{-g} \ \ \bigl(R+\frac{12}{L^{2}}
-\frac{\ell _*^2}{4}\textit{F}_{\mu\nu}\textit{F}^{\mu\nu}+\gamma \ell _*^2 L^2 \textit{C}_{\mu \nu\rho\lambda}
\textit{F}^{\mu\nu}\textit{F}^{\rho\lambda}\bigr) ,
\label{actionWeyl}
\end{equation}
where $\textit{C}_{\mu \nu\rho\lambda}$ is the five dimensional Weyl-tensor and $\gamma$ is the Weyl coupling. The motivation for
this action has been provided in \cite{Dey}, \cite{Sarkar} (in which black holes in this theory with planar and spherical horizon were studied), to which we refer the reader 
for further details. 

The hyperbolic black hole solutions here can be obtained by choosing the following metric and gauge field ansatz,
 \begin{eqnarray}
 ds^2&=&-\bigg({r^2\over L^2}f_0(r)-1\bigg)\big(1+\gamma \mathcal{F}_1(r)\big) \mathcal{N}(r)^2 dt^2 
 +{dr^2 \over \big({r^2\over L^2}f_0(r)-1\big)\big(1+\gamma \mathcal{F}_2(r)\big)} \nonumber \\
 &+& r^2 \ d\Sigma_{3} ^{2}  \,, \nonumber \\
 \label{metricWeyl}
\end{eqnarray}
and
\begin{equation}
 A = (\phi_0(r)+\gamma \mathcal{H}(r)) \ dt \,.
 \label{gauge field QTG}
\end{equation}
where $f_{0}(r)$ and $\phi_{0}(r)$ are the solutions to linear order in $\gamma$ representing a hyperbolic Reissner-Nordstr\"{o}m black hole solution 
in five dimensional AdS space, with
\begin{eqnarray}
 f_0(r) &=& 1-\frac{L^2 m}{3 r^4}+\frac{L^2 q^2}{12 r^6} \,,\nonumber\\
\phi_0(r) &=& \frac{q \tilde{L}}{2 r^2 \ell _* \mathcal{R}}-\frac{q \tilde{L}}{2 \ell _* \mathcal{R}
   r_h^2}.
\label{zeroth order soln}
\end{eqnarray}

Solving Einstein's and Maxwell's equation up to linear order in $\gamma$, one can determine the form of $\mathcal{F}(r)$, $\mathcal{H}(r)$ 
and $\mathcal{N}(r)$ as 
\begin{eqnarray}
\mathcal{F}_1(r)&=& \frac{L^2 q^2 \left(L^2 \left(32 r^2 r_h^8 \left(m+3
   r^2\right)-q^2 \left(7 r_h^8+r^8\right)\right)+96 r^6
   \left(r^2-2 r_h^2\right) r_h^6\right)}{12 r^6 r_h^8
   \left(L^2 \left(q^2-4 r^2 \left(m+3
   r^2\right)\right)+12 r^6\right)}\,,\nonumber\\
 \mathcal{F}_2(r)&=& \frac{L^2 q^2 \left(L^2 \left(64 r^2 r_h^8 \left(m+3
   r^2\right)-q^2 \left(15 r_h^8+r^8\right)\right)+96 r^6
   \left(r^2-3 r_h^2\right) r_h^6\right)}{12 r^6 r_h^8
   \left(L^2 \left(q^2-4 r^2 \left(m+3
   r^2\right)\right)+12 r^6\right)}\,,\nonumber\\  
\mathcal{H}(r)&=&  \frac{L q \left(L^2 \left(r_h^8 \left(16 m r^2-7
   q^2\right)+48 r^8 r_h^4+3 q^2 r^8\right)-48 r^8
   r_h^6\right)}{12 r^8 \ell _* \mathcal{R} r_h^8} .\ \nonumber\\
\mathcal{N}(r) &=& \frac{\tilde{L}}{\mathcal{R}}
\label{perturbations}
\end{eqnarray}

In the above equations, $\tilde{L}$ is the effective AdS curvature scale. It is straightforward to show that for Weyl-corrected gravity, 
$\tilde{L}=L$. The Hawking temperature of the Weyl-corrected black hole can be computed up to linear order in $\gamma$ as,
\begin{eqnarray}
 T= \frac{2 x^2-1}{2 \pi  x \mathcal{R}}-\frac{q^2}{24 \pi  L^4 x^5 \mathcal{R}} + \gamma \left(\frac{q^4}{36 \pi  L^8 x^{11} \mathcal{R}}+\frac{q^2}{2 \pi  L^4 x^7 \mathcal{R}}-\frac{2 q^2}{3 \pi  L^4 x^5 \mathcal{R}}\right)
  \label{Weyl Hawking Temp}
\end{eqnarray}
where again we have set $r_h=L \, x$. 
Note that, in the limit $\gamma \rightarrow 0$, we get back the temperature of a hyperbolic RNAdS black hole.
Using Wald's prescription one can compute the linear order correction in entropy as \cite{Sarkar},
\begin{eqnarray}
 S=2 \pi  \bigg(\frac{L}{\ell _p}\bigg)^3 V_{\Sigma }\,\bigg( x^3 -\gamma{q^2\over L^4 x^3}\bigg).
  \label{Weyl Entropy}
\end{eqnarray}

Since the R\'enyi entropy can be calculated at this stage by exactly the same methods and numerical routines given in the previous sections, 
we do not repeat the description of the procedures here. We only study the variation of the R\'enyi entropy $S_n (\mu_c)$ normalized by $S_1 (0)$ as 
a function of the ratio $\frac{\mu_c l_{*}}{2\pi L}$.

 \begin{figure}[h!]
 \centering
 \subfigure[$\gamma=-0.005$]{
 \includegraphics[width=2.5in,height=2.3in]{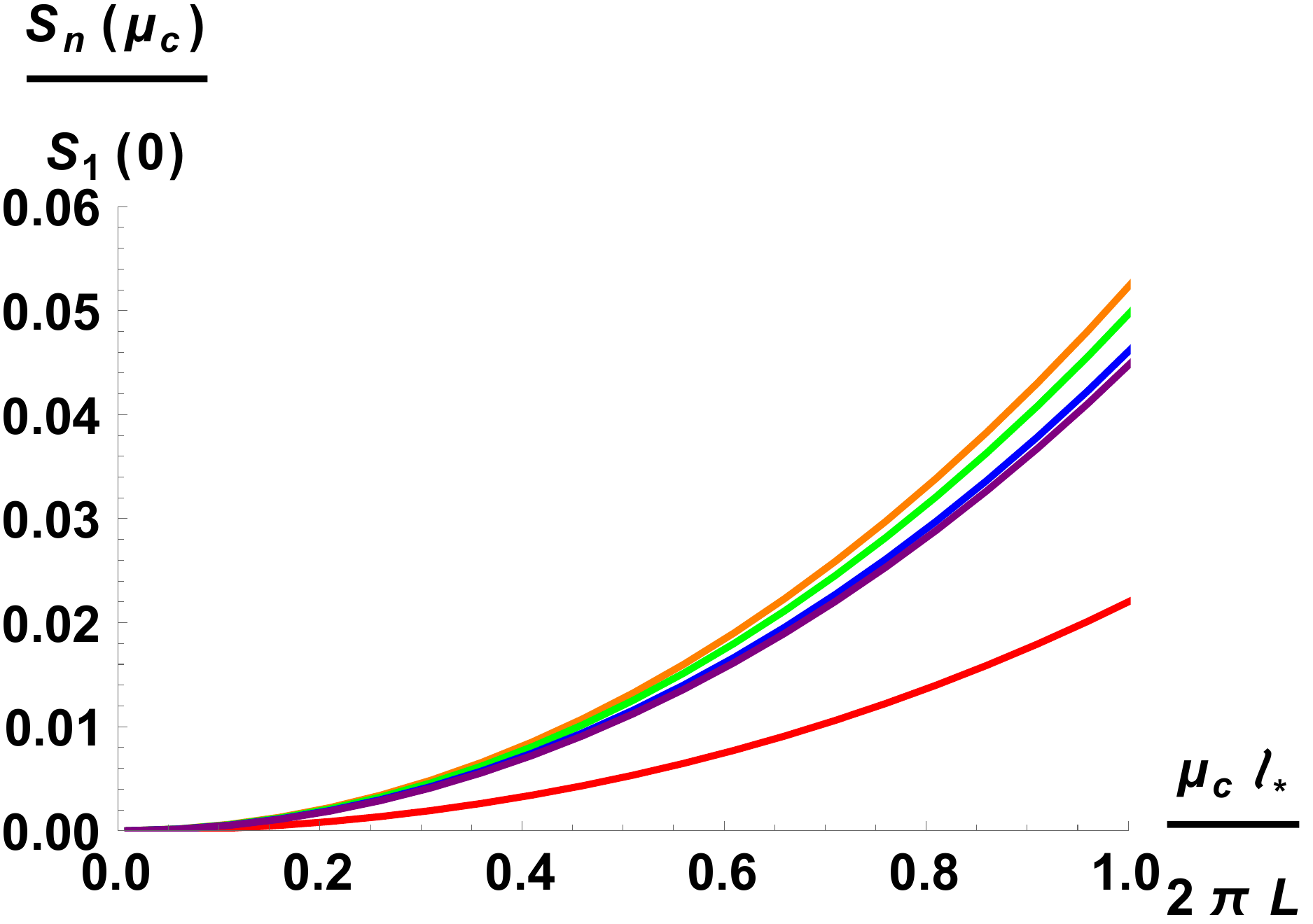}
 \label{Snbymu} } 
 \subfigure[${\mu_c \ell _*\over 2\pi L}=0.5$]{
 \includegraphics[width=2.5in,height=2.3in]{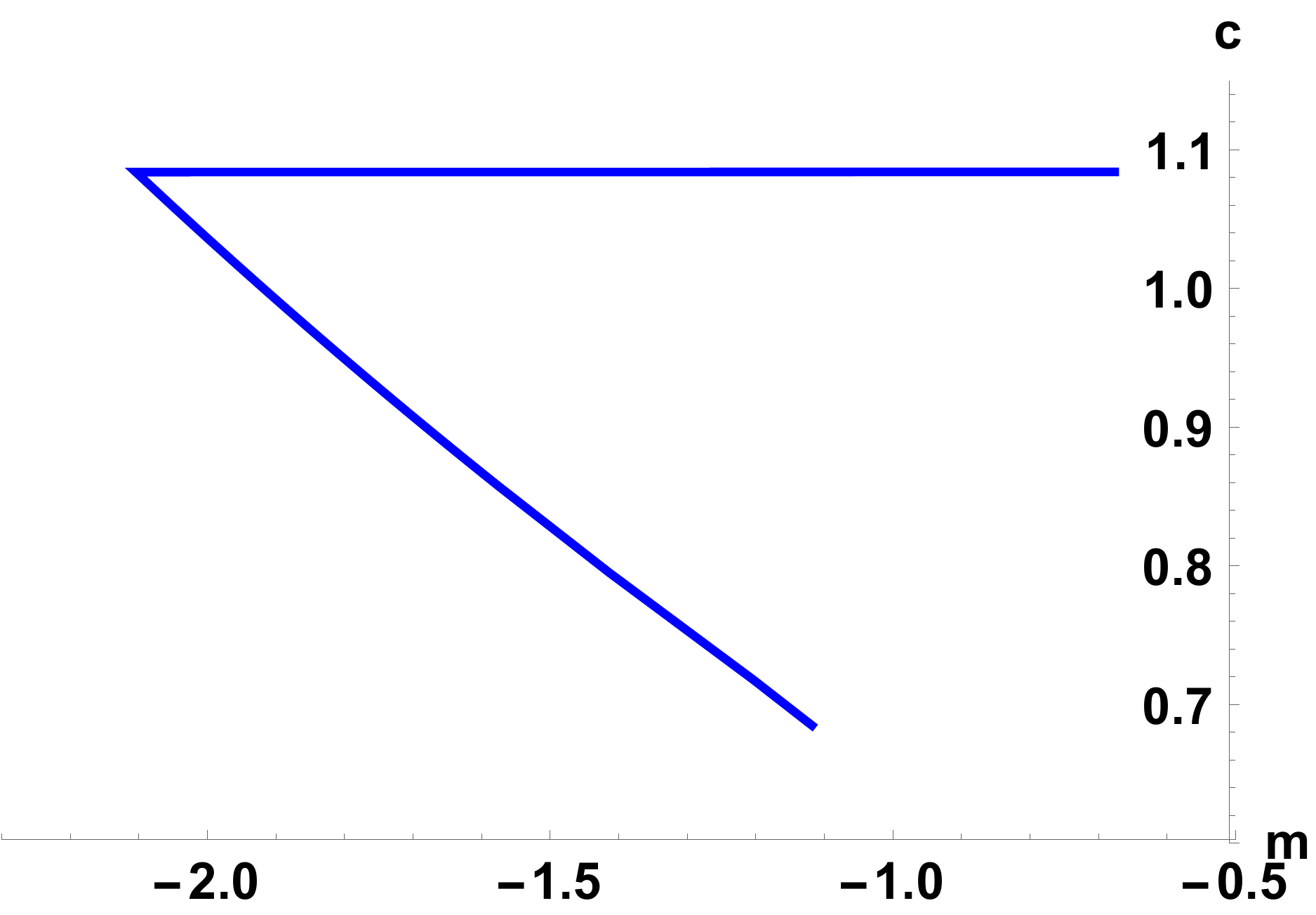}
 \label{Snbygamma} }
  \caption{\small{(a) ${S_n(\mu_c)\over S_1(0)}$ is plotted against $\mu_c \ell _*\over 2\pi L$ with fixed $\gamma=-0.005$ where only the
  Weyl corrected contribution is shown, (b) ${S_n(\mu_c)\over S_1(0)}$ is 
  fitted to a curve $m\gamma + c$ and we show the $c$ vs $m$ plot. Here, different points on the curve correspond to different values of $n$, and
we have fixed $\frac{\mu_c \ell _*}{ 2\pi L}=0.5$.}}
 \label{Sn}
 \end{figure}

In figure \ref{Snbymu} the curve correspond to $n=1$ is the bottom-most, while those with $n=2,3,10,100$, are the others
from the top to the bottom, respectively. The graphs are similar to the cases considered in 
previous sections, and show that the R\'enyi entropy increases monotonically with increasing $\frac{\mu_c l_{*}}{2\pi L}$.
In figure \ref{Snbygamma}, we have investigated the behaviour of $\frac{S_n(\mu_c)}{S_1 (0)}$ vs $\gamma$. We find that 
the behaviour of $\frac{S_n(\mu_c)}{S_1 (0)}$ is almost linear in $\gamma$, and hence we fit it to a curve of the form $m\gamma + c$ 
and show the behaviour of $c$ vs $m$. Each point on the curve corresponds to a different value of $n$, and
$\frac{\mu_c \ell _*}{ 2\pi L}=0.5$ is fixed. 

 \begin{figure}[h!]
 \centering
 \subfigure[]{
 \includegraphics[width=2.5in,height=2.3in]{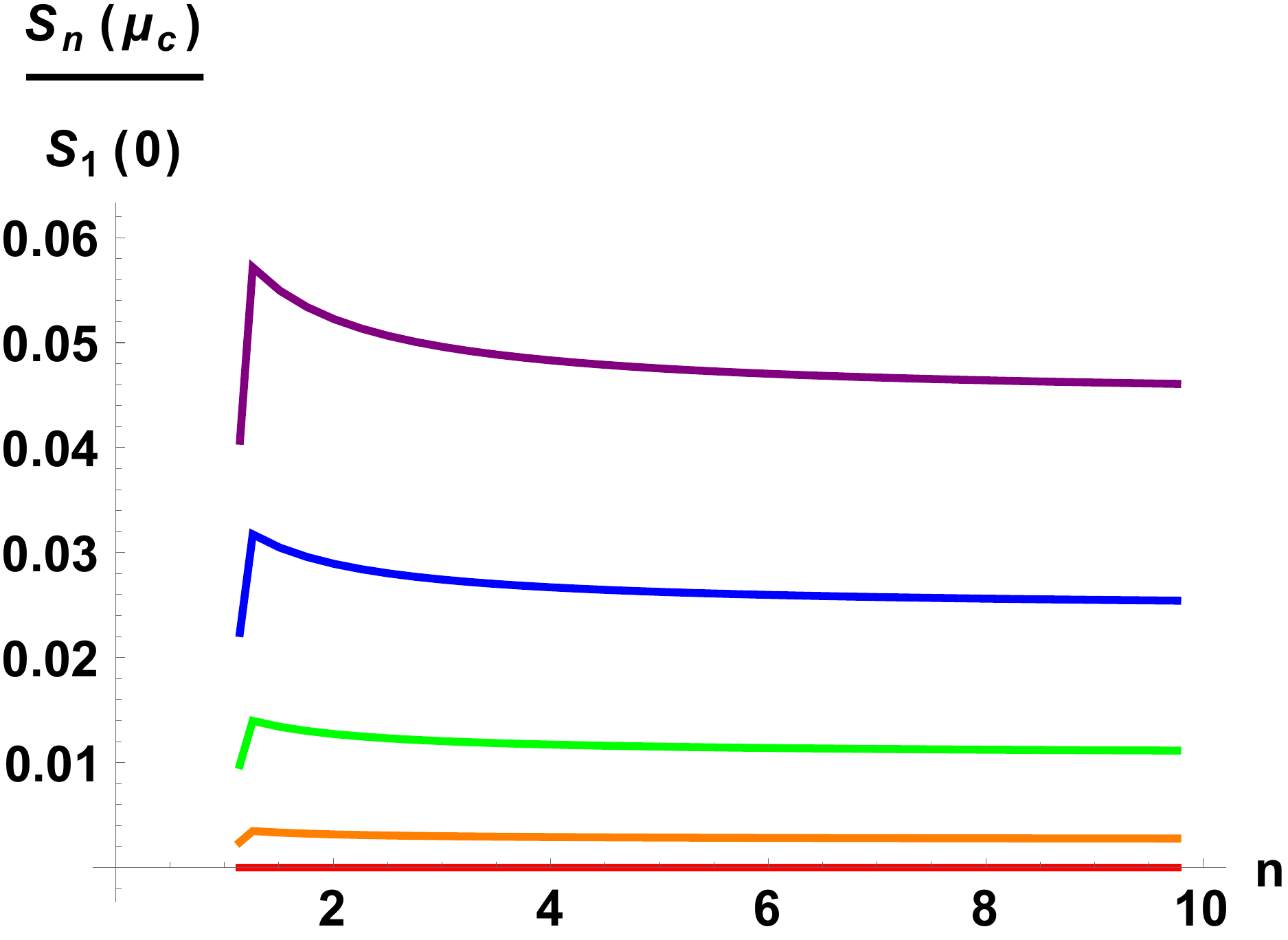}
 \label{FirstInequalityWeyl} } 
 \subfigure[]{
 \includegraphics[width=2.5in,height=2.3in]{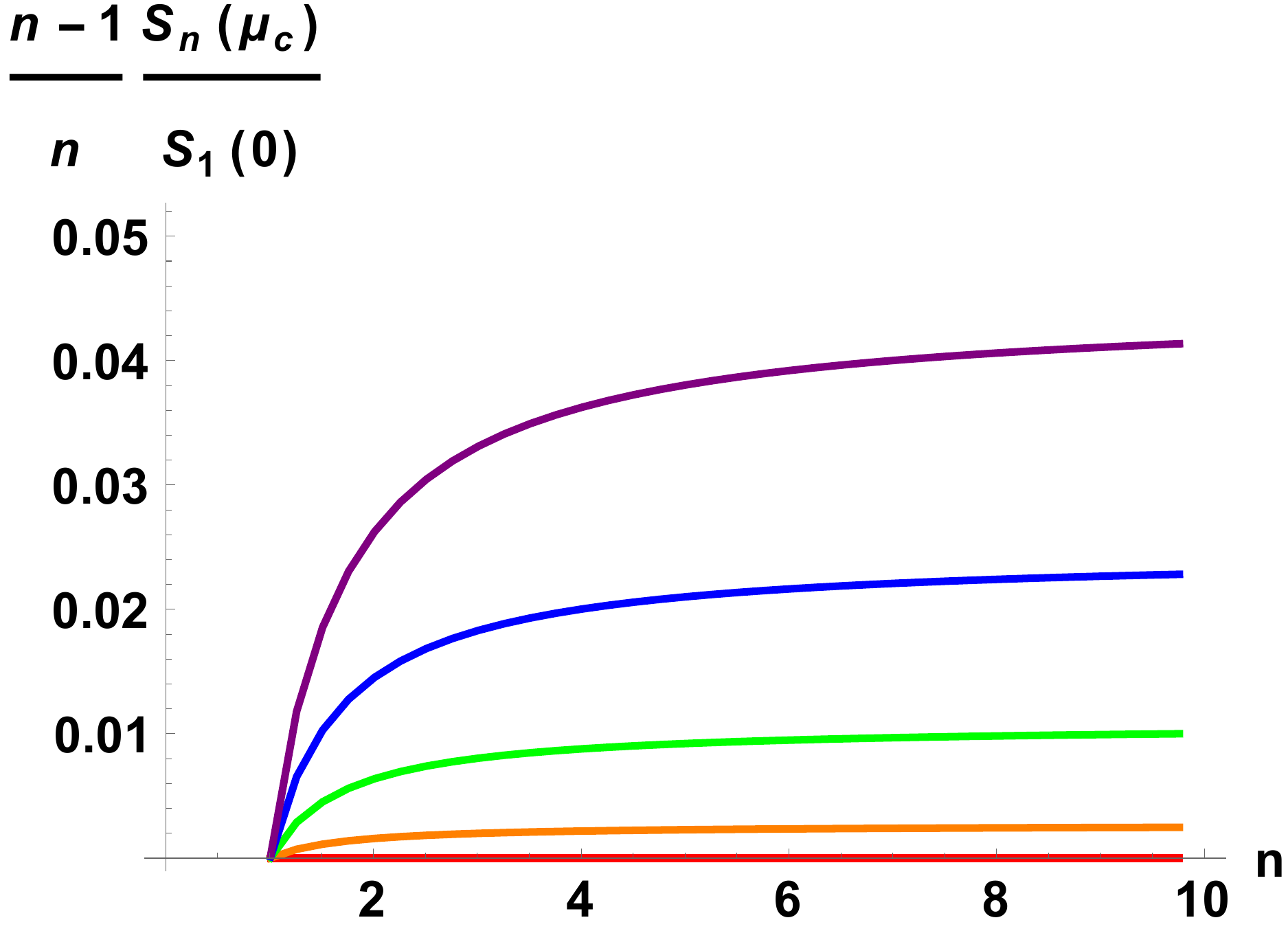}
 \label{SecondInequalityWeyl} }
  \caption{\small{(a) ${S_n(\mu_c)\over S_1(0)}$ is plotted against $n$ with fixed $\gamma=-0.005$, (b) ${n-1\over n}{S_n(\mu_c)\over S_1(0)}$ is plotted 
  against $n$ with fixed $\gamma=-0.005$. In both these figures, the corresponding Einstein-Maxwell part has been subtracted and the figures
  show the quantities only from the Weyl corrected part.}}
 \label{SnInequality}
 \end{figure}
Next, in figure \ref{FirstInequalityWeyl}, we have plotted ${S_n(\mu_c)\over S_1(0)}$ as a function of $n$, while in figure \ref{SecondInequalityWeyl},
we have plotted ${n-1\over n}{S_n(\mu_c)\over S_1(0)}$ as a function of $n$. In both these figures, we have fixed $\gamma=-0.005$, where the red, orange, green, 
blue and purple lines correspond to ${\mu_c \ell _*\over 2\pi L}=0$, $0.25$, $0.5$, $0.75$ and $1.0$, respectively. In both figures, we have
isolated the correction to the Einstein-Maxwell result, and the figures depict the correction resulting from the Weyl term only. Notice that in 
fig.(\ref{FirstInequalityWeyl}), the inequality obeyed by the R\'enyi entropy becomes difficult to satisfy if we consider only the Weyl term. However,
it can be checked that the Einstein-Maxwell part modifies the total contribution, such that the first inequality of eq.(\ref{EREinequalities})
is indeed valid. 
 
A word regarding the upper limit of the Weyl coupling $\gamma$ is in order. Ideally, the maximum magnitude of $\gamma$ for which our results are 
trustworthy should be determined by computing the corrections to the R\'enyi entropy that arise by expanding the metric perturbatively to second order in 
$\gamma$ and by ensuring that changes in the value of the R\'enyi entropy are small for the value of the Weyl coupling chosen. This is what we
had done for the case of Einstein cubic gravity in section 3. In this case, however, due to the complicated nature of the 
expressions involved (details can be found in Appendix B), such an analysis could not be performed. This is a caveat in our analysis. 
Specifically, to compute the contribution of the second order correction terms to R\'enyi entropy, we need to evaluate the lower
limit ($x_n$) and upper limit ($x_1$) (see eq.(\ref{RenyiEntropyFormulamuc})) numerically, using
$T(x_n)={1\over 2\pi \mathcal{R} n}$, where the temperature $T(x)$ is given in eq.(\ref{HawkingTempWeyl2}). 
For this computation, one needs to generate $q(x)$ at a certain value of the chemical potential $\mu_c$. Unfortunately, unlike
the case with first order correction, we were unable to determe $q(x)$ precisely, due to some issues with numerical stability. 
However, we should mention that such an analysis was performed in \cite{Sarkar} for black holes with spherical horizons, where it was checked 
that small values of $\gamma$ like the ones chosen in this paper were indeed trustable. 

For completeness, using the methods described earlier, we also computed the scaling dimension $h_n(\mu_c)$ of the twist operators. Since the form of $h_n$ is not particularly illuminating we do not write it here. However, we should mention that the expansion coefficient $h_{10}={2\over 3}\pi\big({L\over \ell_p}\big)^3$ turns out to be the same as with pure Einstein gravity. It is not difficult to convince oneself that this is due to the fact that in this case the AdS curvature scale $\tilde{L}=L$. 

\section{Summary and Conclusions}

In this paper, we have undertaken a detailed study of holographic entanglement Ren\'yi entropies in various extended theories of gravity. The results of
this paper complements the ones currently available in the literature, and provides novel examples of the computation of EREs in strongly coupled quantum 
field theories in four dimensions. As mentioned in the introduction, all the theories that we consider have tuneable parameters, which might be of interest in
understanding field theories for realistic systems. Let us now summarise the main findings of this paper. 

We first considered the example of Einsteinian cubic gravity recently proposed in \cite{Bueno}. Here, we constructed a black hole solution of the theory
perturbatively, up to the second order in a particular parameter. We computed the ERE for this case and verified the relevant inequalities satisfied by the 
same. We then computed the central charges of the dual field theory, and checked that it matched with a corresponding calculation of the Weyl 
anomaly on the gravity side. The validity of our perturbative analysis was also justified. As a curiosity, we point out here that the calculation of the 
free energy using the standard counterterm method seems to be somewhat problematic in this case. This will be commented upon elsewhere. 

We next considered examples of extended theories of gravity with a chemical potential. In this context, we studied the ERE for the Born-Infeld and 
the charged quasi-topological theories of gravity in five dimensions. For both these, the ERE was computed. The complicated nature of the solution for 
Born-Infeld gravity prevented us from obtaining an analytic computation of the scaling dimensions of the twist operators, but this was not the 
case for quasi-topological gravity. There, we obtained novel bounds on the field theory parameters. Finally, we briefly considered an example of a class of Weyl 
corrected gravity theories in five dimensions which was treated perturbatively, and similar analyses as above was carried out. 

In this paper, we have presented numerical analysis and plots of relevant quantities throughout, in order to quantify our results. We note here that it might be
interesting to investigate some aspects of these further. For example, one could envisage a scaling form for the curves in figs.(\ref{SqvsmucDifferentb}),
(\ref{SqvsbDifferentmuc}), (\ref{FirstInequalityTest}), (\ref{SecondInequalityTest}) and for the ones in figs.(\ref{Snvst4fixedcByaDifferentmuc}), 
(\ref{SnvsmucfixedcByaDifferentt4}), (\ref{QTGSnVscByat40Pt001Differentmuc}) and (\ref{QTGSnVsmuct40Pt001DifferentcBya}). Our initial attempts at such
an analysis seems to suggest that this is somewhat difficult, but if this can be done, one might glean further analytical insights into the results predicted
in these figures. In general, it might be a good idea to attempt such analysis, although we have found it difficult in this work. 

It might be interesting to investigate the R\'enyi entropy with an imaginary chemical potential in the higher derivative charged modified gravity theories considered 
in this paper, following the line of \cite{Belin}, \cite{Pastras}. It will also be interesting to study the charged ERE in canonical ensembles where the charge 
parameter is kept fixed, instead of the chemical potential. Further, instead of a purturbative analysis, one could numerically solve the Einstein equations 
to obtain numerical black hole solutions in Einsteinian cubic gravity and Einstein gravity with a Weyl-corrected gauge field. This way of studying the ERE would 
help understand better the entropy bounds, and hence the bounds on the coupling constants of the theories, from the validation of the R\'enyi entropy 
inequalities.
\begin{center}
{\bf Acknowledgements}
\end{center}

We sincerely thank P. Bueno and P. Cano for pointing out an error in an ansatz, in a previous version of this paper.   
It is also a pleasure to thank Subhash Mahapatra for helpful discussions. Some of the computations of this paper have been performed with the 
MATHEMATICA packages xAct \cite{xAct} and RGTC \cite{RGTC}. The authors of these packages are gratefully acknowledged. 

\appendix
\renewcommand{\theequation}{A.\arabic{equation}}
\setcounter{equation}{0}
\section{Correction up to  $\mathcal{O}(\beta^2)$ in Einsteinian Cubic Gravity}
Here we note down the hyperbolic black hole solution up to $\mathcal{O}(\beta^2)$ in Einsteinian cubic gravity, considering the following metric ansatz ,
\begin{eqnarray}
ds^2 &=& -\big(\frac{r^2}{L^2}f_0(r) -1\big)(1+\beta \mathcal{F}_1(r)+ \beta^2 \mathcal{G}_1(r))\mathcal{N}(r)^2 dt^2 \nonumber \\
&+& \frac{1}{\big(\frac{r^2}{L^2}f_0(r) -1\big)(1+\beta \mathcal{F}_2(r)+ \beta^2 \mathcal{G}_2(r))}dr^2 
+ r^2 d\Sigma_3^2\, .
\end{eqnarray}

Here, $f_0(r)$ is the zeroth order solution while $\mathcal{F}_1(r)$ and $\mathcal{F}_2(r)$ are the $\mathcal{O}(\beta)$ terms given in section \ref{ERE}. 
$\mathcal{G}_1(r)$ and $\mathcal{G}_2(r)$ are the $\mathcal{O}(\beta^2)$ corrections to be determined. Since, the recipe to find out the solution is already discussed in detail in section \ref{ERE}, we just mention the functions second order in $\beta$,

\begin{eqnarray}
\mathcal{G}_1(r) = \frac{3 \ \text{A}}{7 r^{16} r_h^6 \big(-L^2 r^2+r^4-\omega ^4\big)}\,, \ \ \ \
\mathcal{G}_2(r) = \frac{3 \ \text{B}}{7 r^{16} r_h^6 \big(-L^2 r^2+r^4-\omega ^4\big)}\
\end{eqnarray}
where,
\begin{eqnarray}
\text{A} &=& 7 r^2 \omega ^{16} r_h^6 \big(14447 r^2-13314 
   L^2\big)+r^4 \omega ^{12} r_h^6 \big(-48000 L^4 +104336 L^2 r^2 -57883 r^4\big)\nonumber \\
   &+&14 r^{10} \omega^4 r_h^4 \big(12 L^2-19 r^2\big) \big(-54 L^4
   r_h^2+75 L^2 r_h^4-31 r_h^6+11 L^6\big) -44926 \omega ^{20}r_h^6 \nonumber \\ 
   &+& 7 r^8 \omega^8 r_h^4 \big(1875 L^2 r_h^4-2 r_h^2 \big(675 L^4-18
   L^2 r^2+38 r^4\big)-775 r_h^6+275 L^6\big)+r^{16}\nonumber \\
   &&\big(-1501 L^8 r_h^2
   + 4313 L^6 r_h^4-5610 L^4
   r_h^6+3212 L^2 r_h^8+7 r_h^6 \big(r^4-88
   r_h^4\big)+195 L^{10}\big) \nonumber
\end{eqnarray}
\begin{eqnarray}
\text{B} &=& 27 r^2 \omega ^{16} r_h^6 \big(5285 r^2-5104
   L^2\big)-7 r^4 \omega ^{12} r_h^6 \big(9600 L^4-19872 L^2 r^2+10481 r^4\big)\nonumber \\
   &+&14 r^{10} \omega ^4 r_h^4 \big(16 L^2-23 r^2\big) \big(-54 L^4
   r_h^2+75 L^2 r_h^4-31 r_h^6+11 L^6\big)-70028 \omega ^{20} r_h^6 \nonumber \\
   &+&7 r^8 \omega ^8 r_h^4 \big(2775 L^2 r_h^4-2 r_h^2 \big(999 L^4-24
   L^2 r^2+46 r^4\big)-1147 r_h^6+407 L^6\big)+r^{16}\nonumber \\ 
   &&\big(-1501 L^8 r_h^2 
   +4313 L^6 r_h^4-5610 L^4
   r_h^6+3212 L^2 r_h^8+7 r_h^6 \big(r^4-88
   r_h^4\big)+195 L^{10}\big) \nonumber
\end{eqnarray}

Also, the function $\mathcal{N}(r)$ at $\mathcal{O}(\beta^2)$ is given by,
\begin{eqnarray}
 \mathcal{N}(r) = \frac{\tilde{L}}{\mathcal{R}}=\frac{L}{\mathcal{R}\sqrt{f_{\infty}}}
\end{eqnarray}
where $f_{\infty}$ is determined as, $f_{\infty}=1-\beta+3\beta^2$. 

For completeness, here we mention the Hawking temperature and the Wald entropy at $\mathcal{O}(\beta^2)$,
\begin{equation}
T = \frac{2 x^2-1}{2 \pi  x \mathcal{R}}-\beta \frac{ \left(x^2-1\right)^3}{\pi 
   x^5 \mathcal{R}}+\beta^2 \frac{\left(x^2-1\right)^3 \left(1834 x^4-459
   x^2-201\right)}{14 \pi  x^9 \mathcal{R}}
\label{TempECG2}
\end{equation}
and
\begin{eqnarray}
S_{Wald}&=& 2 \pi \big({L\over \ell_p}\big)^3\, V_{\Sigma}\big[x^3+{3\over 2}\beta \big(25 x^3-48 x+\frac{18}{x}\big)+{3\over 8}\beta^2 \big(905 x^3-3168 x  \nonumber \\
&+&\frac{3996}{x}-\frac{2112}{x^3}
+\frac{384}{x^5}\big)\big]\, .
\label{WaldEntropyECG2}
\end{eqnarray}

We also computed the R\'enyi entropy correction at $\mathcal{O}(\beta^2)$, but do not write it explicitly here, since the expression is unwieldy. Instead we give the expression for the entanglement entropy which is obtained by taking the $n\rightarrow 1$ limit. 
\begin{equation}
S_1 = 2 \pi  \bigg(\frac{L}{\ell_p}\bigg)^3 V_{\Sigma} \big(1-{15\over 2}\beta+{15\over 8}\beta^2\big)\, .
\label{EEECG2}
\end{equation}

We also calculated the scaling dimension of the twist operator at $\mathcal{O}(\beta^2)$ and computed the quantity $\partial_n h_n |_{n=1}$, in a similar way as we did with the leading order solution,
\begin{eqnarray}
 \partial_n h_n |_{n=1}=\frac{2}{3} \pi \bigg({L\over \ell_p}\bigg)^3 \big(1+{9\over 2}\beta-{33\over 8}\beta^2\big)\, .
 \label{SDECG2}
\end{eqnarray}

As a further check for our computations with the second order solution, we calculated the $\mathcal{O}(\beta^2)$ correction to the Weyl anomaly, i.e., the correction to the central charges $c$ and $a$ using the methods outlined in section \ref{Anomaly}. We find that the coefficients match exactly with those that can be read off from the above expressions (\ref{EEECG2}) and (\ref{SDECG2}),
\begin{eqnarray}
 c &=&\pi ^2 \big({L\over \ell_p}\big)^3 \big(1+{9\over 2}\beta-{33\over 8}\beta^2\big) \nonumber \\
 a &=&\pi^2 \bigg({L\over \ell_p}\bigg)^3 \bigg(1-{15\over 2}\beta+{15\over 8}\beta^2\bigg)\, .
\end{eqnarray}

\renewcommand{\theequation}{B.\arabic{equation}}
\setcounter{equation}{0}
\section{Correction up to  $\mathcal{O}(\gamma^2)$ in Weyl-Corrected Gravity}

Here we note down the Weyl-corrected hyperbolic black hole solution up to $\mathcal{O}(\gamma^2)$, considering the following ansatz for the metric and gauge field ansatz,
 \begin{eqnarray}
 ds^2&=&-\big({r^2\over L^2}f_0(r)-1\big)\big(1+\gamma \mathcal{F}_1(r)+\gamma^2 \mathcal{G}_1(r)\big) \mathcal{N}(r)^2 dt^2 \nonumber \\
 &+&{dr^2 \over \big({r^2\over L^2}f_0(r)-1\big)\big(1+\gamma \mathcal{F}_2(r)+\gamma^2 \mathcal{G}_2(r)\big)} 
 + r^2 \ d\Sigma_{3} ^{2}  \,, \nonumber \\
 \label{metricWeyl2}
\end{eqnarray}
and
\begin{equation}
 A = \phi_0(r)+\gamma \mathcal{H}(r)+\gamma^2 \mathcal{J}(r) \ dt \,,
 \label{gauge field Weyl2}
\end{equation}
where $f_0(r)$ is the zeroth order solution, $\mathcal{F}_1(r)$ and $\mathcal{F}_2(r)$ are the first order solutions of the metric, while $\mathcal{G}_1(r)$ and $\mathcal{G}_2(r)$ represent the $\mathcal{O}(\gamma^2)$ corrections to the metric. On the other hand, $\phi_0(r)$ and $ \mathcal{H}(r)$ are respectively the zeroth order and first order corrections to the gauge field, while $\mathcal{J}(r)$ stands for the second order correction to the gauge field. Since, the procedure to find the solution has been described earlier with $\mathcal{O}(\gamma)$ correction, here we simply write down the functions at $\mathcal{O}(\gamma^2)$,
\begin{eqnarray}
 \mathcal{G}_1(r)&=&{1\over 90 r^{12} r_h^{14} (12 r^6-4 L^2 m r^2+L^2 q^2-12 L^2 r^4)}\bigg[L^2 q^2 \big(L^4 \big(8 q^2 r^2 r_h^4 \big(672
   r^{12}\nonumber \\ &-& 11 r_h^{10} \big(41 m+171
   r^2\big)\big)+384 r^4 r_h^8 \big(m r_h^6 \big(11
   m+54 r^2\big)+ 63 r^{10}\big)+q^4 \big(5 r^8
   r_h^6 \nonumber \\ &+& 365 r_h^{14}+268 r^{14}\big)\big)-12 L^2 r^6
   r_h^6 \big(384 \big(7 m r^2 r_h^8+18 r^8
   r_h^4\big) +q^2 \big(40 r^2 r_h^6 \nonumber \\ &-& 1895 r_h^8+809
   r^8\big)\big)+58752 r^{14} r_h^{12}\big)\bigg]\, ,
\end{eqnarray}
\begin{eqnarray}
 \mathcal{G}_2(r)&=&{1\over 45 r^{12} r_h^{14} (12 r^6-4 L^2 m r^2+L^2 q^2-12 L^2 r^4)}\bigg[L^2 q^2 \big(L^4 \big(32 q^2 \big(84 r^{14} r_h^4-r^2
   r_h^{14} \nonumber \\ && \big(350 m+927 r^2\big)\big)+192 r^4
   r_h^8 \big(m r_h^6 \big(53 m+180 r^2\big)+63
   r^{10}\big)+q^4 \big(5 r^8 r_h^6 \nonumber \\ &+& 2025 r_h^{14}+134
   r^{14}\big)\big)- 6 L^2 r^6 r_h^6 \big(192 r^2
   r_h^4 \big(35 m r_h^4+36 r^6\big)+q^2 \big(80 r^2
   r_h^6-5625 r_h^8 \nonumber \\ &+& 809 r^8\big)\big) + 29376 r^{14}
   r_h^{12}\big)\bigg]\, ,
\end{eqnarray}
and
\begin{eqnarray}
 \mathcal{J}(r)&=& \frac{1}{180 \ell _* \mathcal{R} r^{14} r_h^{14}}\bigg[L q \big(L^4 \big(1152 \big(m^2 r^4 r_h^{14}-9 r^{14}
   r_h^8\big)-48 q^2 \big(r^2 r_h^{14} \big(33 m+32
   r^2\big) \nonumber \\ &+&31 r^{14} r_h^4\big)+q^4 \big(5 r^8
   r_h^6+275 r_h^{14}+44 r^{14}\big)\big)+24 L^2 r^6
   r_h^6 \big(q^2 \big(-20 r^2 r_h^6+75 r_h^8+71
   r^8\big) \nonumber \\ &+& 864 r^8 r_h^4\big)-10368 r^{14}
   r_h^{12}\big)\bigg]\, .
\end{eqnarray}
Also, $\mathcal{N}(r)$ is simply given by, $\mathcal{N}(r)={L \over \mathcal{R}}$.

For completeness, we also note down the Hawking temperature and Wald entropy computed at $\mathcal{O}(\gamma^2)$,
\begin{eqnarray}
 T(x)&=& \frac{2 x^2-1}{2 \pi  x \mathcal{R}}-\frac{q^2}{24 \pi  L^4 x^5 \mathcal{R}} + \gamma \left(\frac{q^4}{36 \pi  L^8 x^{11} \mathcal{R}}+\frac{q^2}{2 \pi  L^4 x^7 \mathcal{R}}-\frac{2 q^2}{3 \pi  L^4 x^5 \mathcal{R}}\right) \nonumber \\
 &+&\gamma^2 \left(\frac{49 q^6}{540 \pi  L^{12} x^{17} \mathcal{R}}-\frac{7 q^4}{18 \pi  L^8 x^{13}
   \mathcal{R}}+\frac{121 q^4}{45 \pi  L^8 x^{11} \mathcal{R}}-\frac{32 q^2}{5 \pi 
   L^4 x^9 \mathcal{R}}+\frac{136 q^2}{5 \pi  L^4 x^7 \mathcal{R}}-\frac{104 q^2}{5
   \pi  L^4 x^5 \mathcal{R}}\right)\, , \nonumber \\
  \label{HawkingTempWeyl2}
\end{eqnarray}
and
\begin{eqnarray}
 S_{Wald}=2 \pi  \bigg(\frac{L}{\ell _p}\bigg)^3 V_{\Sigma }\,\bigg( x^3 -\gamma{q^2\over L^4 x^3}+\gamma^2 \frac{6 q^4-48 L^4 q^2 x^6+48 L^4 q^2 x^4}{L^8 x^9}\bigg)\, .
   \label{WaldEntropyWeyl2}
\end{eqnarray}

\end{document}